\documentclass[preprint,1p,times,authoryear]{elsarticle}

\usepackage{amssymb}

\usepackage{functan}

\usepackage{dsfont}
\usepackage{mathtools}
\usepackage{color}
\usepackage{tikz}
\usepackage[hidelinks]{hyperref}

\newcommand{\bbN}{\mathbb{N}}

\newcommand{\bbR}{\mathbb{R}}
\newcommand{\bbS}{\mathbb{S}}

\newcommand{\cB}{\mathcal{B}}

\newcommand{\cD}{\mathcal{D}}

\newcommand{\cL}{\mathcal{L}}

\newcommand{\cO}{\mathcal{O}}

\newcommand{\cQ}{\mathcal{Q}}
\newcommand{\cR}{\mathcal{R}}

\newcommand{\cW}{\mathcal{W}}

\newcommand{\cY}{\mathcal{Y}}

\newcommand{\scal}[3][]{\scalprod*{#1}{#2}{#3}}

\newcommand{\rd}{\mathrm{d}}
\newcommand{\mmd}{\rd}
\newcommand{\md}{\,\rd}

\newcommand{\mv}[1]{{\boldsymbol{\mathrm{#1}}}}
\newcommand{\trsp}{\ensuremath{\top}}

\newcommand{\proper}{\textsf}
\newcommand{\pN}{\proper{N}}

\newcommand{\pPo}{\proper{Poisson}}
\newcommand{\pGRF}{\proper{GRF}}

\newcommand{\ol}{\overline}
\newcommand{\wt}{\widetilde}

\newcommand{\adj}[1]{#1^{*}}

\newcommand{\mat}[1]{\begin{bmatrix}#1\end{bmatrix}}

\newcommand{\s}{\mv{s}}

\newcommand{\pproper}{\mathsf}
\newcommand{\pE}{\pproper{E}}
\newcommand{\pP}{\pproper{P}}

\newcommand{\pCov}{\pproper{Cov}}
\newcommand{\pPrec}{\pproper{Prec}}
\newcommand{\cov}{\varrho}

\newcommand{\RRR}{\textsf{R}}
\newcommand{\Rpkg}[1]{\texttt{#1}}
\newcommand{\RINLA}{\Rpkg{R-INLA}}
\newcommand{\inlabru}{\Rpkg{inlabru}}
\newcommand{\rspde}{\Rpkg{rSPDE}}

\journal{Spatial Statistics}
\makeatletter
\def\ps@pprintTitle{%
 \let\@oddhead\@empty
 \let\@evenhead\@empty
 \def\@oddfoot{\centerline{\thepage}}%
 \let\@evenfoot\@oddfoot}
\makeatother

\begin{document}

\begin{frontmatter}



\title{The SPDE approach for Gaussian and non-Gaussian fields:\\ 10
    years and still running}


\author[label1]{Finn Lindgren\corref{cor1}}
\affiliation[label1]{
    organization={The University of Edinburgh},
             addressline={James Clerk Maxwell Building, Peter Guthrie Tait Rd},
             city={Edinburgh},
             postcode={EH9 3FD},
             country={Scotland}}
\ead{Finn.Lindgren@ed.ac.uk}

\author[label2]{David Bolin}
\author[label2]{H{\aa}vard Rue}
\affiliation[label2]{
    organization={King Abdullah University of Science and Technology (KAUST)},
    city={Thuwal},
    postcode={23955-6900},
    country={Saudi Arabia}}
\cortext[cor1]{Corresponding author:}

\begin{abstract}

    Gaussian processes and random fields have a long history, covering
    multiple approaches to representing spatial and spatio-temporal
    dependence structures, such as covariance functions, spectral
    representations, reproducing kernel Hilbert spaces, and graph
    based models. This article describes how the stochastic partial
    differential equation approach to generalising Mat\'ern covariance
    models via Hilbert space projections connects with several of
    these approaches, with each connection being useful in different
    situations. In addition to an overview of the main ideas, some
    important extensions, theory, applications, and other recent
    developments are discussed. The methods include both Markovian and
    non-Markovian models, non-Gaussian random fields, non-stationary
    fields and space-time fields on arbitrary manifolds, and practical
    computational considerations.

\end{abstract}



\begin{keyword}

    Random fields \sep
    Gaussian Markov random fields \sep
    Mat{\'e}rn covariances \sep
    stochastic partial differential equations \sep
    computational efficiency \sep
    INLA



\end{keyword}

\end{frontmatter}


\section{Introduction}\label{sec:introduction}

It has been 10 years since the publication of the paper
\citep{lindgren11} that introduced the \emph{Stochastic Partial
    Differential Equation} (SPDE) approach for Gaussian fields. We
will use this opportunity to put the method in perspective in relation
to different ways of characterising Gaussian random field
distributions, review recent developments, and show some main
applications of the approach. We will also discuss the closely related
construction of non-Gaussian fields based on a generalisation of SPDE
approach, that was initially proposed in David Bolin's PhD-thesis that
appeared a year later \citep{phd11}, and which since then has been
developed in a series of papers starting with \citet{bolin14}.

Although most of the following (rather technical) discussion in this
paper will focus on the properties of, and opportunities with, the
SPDEs and their finite dimensional (Hilbert space) representations, it
is important to keep in mind the immediate \emph{practical} relevance
of this approach. An important aim for our research is to construct
models that also have good computational properties so we and other
non-specialists can make practical use of them. It turns out that this
is indeed possible due to the sparse precision matrix formulation and
their efficient numerical
computations~\citep{rue2005gaussian,art451,art522,art632,art664,art375,art442}.
The \RINLA{}, \inlabru{}, and \rspde{}
packages~\citep{book126,art643,art527,bachl_inlabru_2019, BK2020rational} provide
accessible interfaces to many of the Gaussian SPDE-based models
discussed in this paper, whereas the \Rpkg{ngme}
package~\citep{Asar2020} implements the non-Gaussian models.

\subsection{Covariance or precision? Yes please!}

Some key insight to the good computational properties of the
SPDE-approach comes from considering the precision rather than the
covariance, which we will now discuss. More details will appear later
in Section~\ref{sec:INLA}.

A zero mean Gaussian random vector is traditionally represented by it
covariance matrix for which marginal properties can specified directly
or be immediately read off from the covariance matrix, while
conditional properties have to be computed. It can also be represented
by its precision (the inverse of the covariance) matrix, which is a
more ``modern'' representation related to graphical models
\citep{book24}. Here, the conditional properties can be specified
directly or be immediately read off, while its marginal properties
have to be computed.

Within the SPDE framework, we can associate precision to ``how the
model is generated'' and the covariance to (the derived) ``properties
of the model''. As a simple example, let us consider a standard
stationary first order auto-regressive process
$x_t = \phi x_{t-1} + \epsilon_t$, for $t=1, \ldots, T$. The precision
matrix is tridiagonal as only $x_{t-1}$ is required to generate
$x_{t}$, while the correlation matrix is dense with the $ij$'th
element $\phi^{|i-j|}$, since $x_{t}$ depends on $x_{t-1}$ that again
depends on $x_{t-2}$ and so on. The dense/sparse matrix properties
also hold if we move to continuous time \citep{art512} and the
Ornstein–Uhlenbeck process, $\mmd x_{t}=-\phi \,x_{t}\md t+\mmd B_{t}$
where $B_t$ denotes the Wiener process. The computational cost for
doing inference with a (general) tridiagonal precision matrix is
$\cO(T)$, while based on the (general) covariance matrix is
$\cO(T^{3})$. It follows that the sparse precision matrix formulation
is maintained when conditioning on conditional independent
observations, which is the key observation behind the
Kalman-recursions/updates and ensure that the computational cost
linear is still linear in $T$ \citep[Appendix]{art196}.

It would be beneficial if we could have it both ways, as the dense
covariance is useful for understanding marginal and bivariate
properties, while the sparse precision gives computational efficient
computations. However, this is complicated by the fact that Markov
properties in continuous space are more involved than Markov
properties in continuous time~\citep{art512}. The first main result in
\citet{lindgren11} say that we can have it both ways for Gaussian
fields with a Mat\'ern covariance function (for specific values of the
smoothness index), using finite dimensional Hilbert space of function
representations, and project the continuous domain functions onto this
space.

\subsection{Some recent applications}

As an initial illustration of the method's practical relevance, we
here present an incomplete list of recent applications. In the
time-period of May-Sep 2021, we find applications of the SPDE-approach
to Gaussian fields in \textbf{astronomy} \citep{bg:1}, \textbf{health}
\citep{bg:2,bg:3,bg:4,bg:5, bg:6,bg:7}, \textbf{engineering}
\citep{bg:8}, \textbf{theory}
\citep{bg:9,bg:10,bg:11,bolin2021efficient}, \textbf{environmetrics}
\citep{bg:12, bg:13, bg:14, bg:15, bg:16, bg:17, bg:18, bg:19, bg:20,
    bg:21, bg:22, bg:23}, \textbf{econometrics} \citep{bg:24,bg:25},
\textbf{agronomy} \citep{bg:26}, \textbf{ecology}
\citep{bg:27,bg:28,bg:29,bg:30,bg:31,bg:32,bg:33}, \textbf{urban
    planning} \citep{bg:34}, \textbf{imaging} \citep{bg:35},
\textbf{modelling of forest fires} \citep{bg:36,bg:37},
\textbf{fisheries}
\citep{bg:38,bg:39,bg:40,bg:41,bg:42,bg:43,bg:44,bg:45,bg:46,bg:47},
\textbf{dealing with barriers}
\citep{ddi.13392,bg:38,bg:27,EF2021,Cendoya2021.04.01.438042}, and so
on.

\subsection{Plan of this paper}

The plan for the rest of this paper is as follows. In
Section~\ref{sec:method} we provide an overview of the main ideas,
including the precision operator, Markov properties, intrinsic random
fields, random fields on manifolds, non-stationary fields, and finite
element methods (FEM).
Section~\ref{sec:INLA} introduces the main ideas for how to use the
SPDE approach for statistical inference.
Section~\ref{sec:extensions} discusses the extension to non-Gaussian
fields, and also extensions to non-Markovian fields with general
smoothness index, and extensions to separable and non-separable models
in space and time.
Theoretical properties of the SPDE-based models and the corresponding
computational methods are discussed in Section~\ref{sec:theory}, where
the main point that we want to convey is that these models and methods
by now are well understood from a theoretical point in very general
conditions.
Some key applications are presented in Section~\ref{sec:applications},
including Malaria modelling, the EUSTACE project, neuroimaging,
seismology and point process models in ecology. We end with a
discussion of related methods in Section~\ref{sec:related} and wrap it
all up with a general discussion in Section~\ref{sec:discussion}.


\section{Overview of the main ideas}\label{sec:method}

The initial motivation by \citet{lindgren11} was to address the
long-standing problem of how to construct precision matrices for
Gaussian Markov random fields (GMRFs) such that the resulting models
would be invariant to the geometry of the spatial neighbour defining
graph. The key to the solution was to construct a finite dimensional
Hilbert space of function representations, and project the continuous
domain functions onto this space. By choosing the finite dimensional
space to be spanned by local piecewise linear basis functions,
projections of Mat\'ern fields with Markov properties on the
continuous domain then lead to Markov properties for the basis
function weights. The triangulation graph for the basis functions
determine the Markov neighbourhood structure, with neighbourhood
diameter determined by the precision operator order of the model. This
solution combines classic results for the equivalence of Mat\'ern
covariance models and stochastic PDE models \citep{matern60,
    whittle54, whittle63} with Gaussian Markov random field theory
\citep{besag_spatial_1974,besag_conditional_1995,besag_first-order_2005,rue2005gaussian}
and Hilbert space projections via numerical finite element methods. We
here focus on the main results and connections between different
random field representations, and leave most of the technical detail
discussion and recent theory developments to Section~\ref{sec:theory}.

\subsection{Covariances and stochastic partial differential equations}

The classic stationary Mat\'ern covariance family is given by
\begin{equation}\label{eq:matern}
    \cov_M(\s,\s') = \frac{\sigma^2}{\Gamma(\nu)2^{\nu-1}}(\kappa\|\s-\s'\|)^\nu
    K_\nu(\kappa\|\s-\s'\|),
\end{equation}
where $K_\nu$ is the modified Bessel function of the second kind,
$\nu>0$ is the smoothness index, $\kappa>0$ controls the spatial
correlation range, and $\sigma^2$ is the marginal variance. For
ordinary pointwise evaluations of a field $u(\cdot)$ with Mat\'ern
covariance, $\pE\{u(\s)\}=0$ and
$\pCov\{u(\s),u(\s')\}=\cov_M(\s,\s')$.

A useful generalised characterisation of the dependence structure for
Gaussian random fields is that the covariance of linear functionals
$\scal{f}{u}$ and $\scal{g}{u}$ is given by
$$
\cR_u(f,g) = \pCov(\scal{f}{u},\scal{g}{u})=
\int_{\bbR^d} \int_{\bbR^d} f(\s) \cov(\s,\s') g(\s') \md\s \md\s'
$$
for any $f$ and $g$ such that $\cR_u(f,f)$ and $\cR_u(g,g)$ are
finite. The \emph{covariance function} or \emph{kernel} $\cov(\s,\s')$
is non-negative definite, so that $\cR_u(f,f)\geq 0$. Here
$\scal{f}{u} = \int_\cD f(\mv{s})u(\mv{s})\md \mv{s}$ is the inner
product on $L_2(\cD)$ for a domain $\cD$ (or in general a duality
pairing when $f$ and $u$ are in different function spaces). Pointwise
evaluation is obtained when $f$ and $g$ are Dirac delta functionals,
but this generalised covariance characterisation also extends to
generalised random fields that do not have pointwise meaning. Let
$\cW(\cdot)$ be a Gaussian white noise process on a general domain
$\cD$, characterised by $\pE(\scal{f}{\cW})=0$ and
$\cR_\cW(f,g)=\scal{f}{g}$. For $\cD=\bbR$, the $\cW(\cdot)$ process
is the formal derivative of a Brownian motion, but the $\cW$
definition here is valid on much more general manifolds. With this
definition, linear spatial stochastic partial differential equations
(SPDEs) of the form
\begin{align}\label{eq:generalspde}
  \cL u(\cdot) &= \cW(\cdot)
\end{align}
can be used to define random fields $u(\cdot)$, where the choice of
differential operator $\cL$ implicitly determines the covariance
structure of the solutions. For the choice
\begin{equation}\label{eq:whittle}
    \tau (\kappa^2 - \Delta)^{\alpha/2} u = \mathcal{W},
\end{equation}
on $\bbR^d$, \citet{whittle54} and \citet{whittle63} showed that the
stationary solutions have a Mat\'ern covariance of the form
\eqref{eq:matern}, with $\nu=\alpha-d/2$ and
$\sigma^2=\Gamma(\nu)\{\Gamma(\alpha)(4\pi)^{d/2}\kappa^{2\nu}\tau^2\}^{-1}$.
We use the term Whittle-Mat\'ern fields when referring to the more
general set of solutions of \eqref{eq:whittle}, that also includes
intrinsic stationary fields on $\bbR^d$ (discussed in
Section~\ref{sec:intrinsic}) , solutions on subdomains with boundary
conditions, and solutions on general manifold domains.

\subsection{Precision operators and reproducing kernel Hilbert spaces}
\label{sec:precisions}

The covariance characterisation of the solutions to
\eqref{eq:generalspde} is closely linked to the inner product, denoted
$\cQ_u(f,g)$, of an associated reproducing kernel Hilbert space
(RKHS). We will now sketch the main aspects of this connection, with
more details given in \ref{app:derivations}. Let $\adj{\cL}$ be the
adjoint of $\cL$, i.e.\ an operator such that
$\scal{\adj{\cL}f}{g}=\scal{f}{\cL g}$, and also assume that $\cL$ is
invertible. Then, by definition of the covariance product $\cR_\cW$
for the white noise process,
$\scal{f}{g}=\cR_\cW(f,g)=\cR_{\cL
    u}(f,g)=\cR_{u}(\adj{\cL}f,\adj{\cL}g)$, which shows that the
covariance function $\cov(\s,\s')$ fulfils
$\cL_\s \cL_{\s'}\cov(\s,\s')=\delta_{\s}(\s')$. This can be used to
show that $\cQ_u(f,g)=\scal{\cL f}{\cL g}$ fulfils
$\cQ_u\{\cov(\s,\cdot),g(\cdot)\}=g(\s)$ for all $\s\in\cD$ and any
suitable $g(\cdot)$, which means that this $\cQ_u(\cdot,\cdot)$ is the
inner product for the RKHS for $\cov(\cdot,\cdot)$. Furthermore,
$\scal{\adj{\cL}\cL\cov(\s,\cdot)}{g}=g(\s)$, which means that the
covariance function is a Green's function of what we call the
\emph{precision operator}, here $\adj{\cL}\cL$. Thus, for the
Whittle-Mat\'ern processes on $\bbR^d$,
\begin{equation}\label{eq:precisionproduct}
    \cQ_u(f,g) = \tau^2
    \scal{(\kappa^2 - \Delta)^{\alpha/2}f}{(\kappa^2 - \Delta)^{\alpha/2}g}
\end{equation}
and the precision operator is $\tau^2(\kappa^2 - \Delta)^{\alpha}$.

For Gaussian fields, high order Markov properties can be characterised
by the precision operator being defined by local differential
operators \citep{rozanov1977markov}. In the Mat\'ern cases with
integer $\alpha$ values, the precision operator expands into a sum of
integer powers of the negated Laplacian, showing directly that the
corresponding processes are Markov random fields. Furthermore, the
inner product \eqref{eq:precisionproduct} itself can be expanded into
a sum of inner products involving only integer and half-integer powers
of the Laplacian, and gradients,
\begin{equation}\label{eq:precisionproduct2}
    \cQ_u(f,g) =
    \sum_{k=0}^\alpha
    \binom{\alpha}{k} \kappa^{2(\alpha-k)}
    \scal{(-\Delta)^{k/2}f}{(-\Delta)^{k/2}g},
\end{equation}
where the half-integer Laplacian inner products can be converted to
equivalent inner products of gradient operators, from
$\scal{(-\Delta)^{1/2}f}{(-\Delta)^{1/2}g}=\scal{\nabla f}{\nabla g}$
\citep[][Appendix B]{lindgren11}. This shows that the Markov subset of
Mat\'ern fields only involve ordinary local differential operators in
the inner product integrals, despite the half-Laplacian being a
non-local operator. This is useful when constructing discrete
representations of the precision operator; the \emph{global Markov
    property} on the continuous domain, expressed via conditional
independence of subdomains with a \emph{separating set} in-between,
turn into a similar condition, transported to the topology of higher
order neighbourhoods of the triangulation graphs, leading to sparse
matrix representations of the precision operator
(Section~\ref{sec:FEM}).

The connection between RKHS constructions, splines, and random process
estimation has a long history, with
\citet{kimeldorf_correspondence_1970} showing how the conditional
expectations in Gaussian process regression can be formed as penalised
splines. An important distinction between the spline and Gaussian
random field aspects of RKHS theory is that while the splines are
members of the RKHS with finite precision norm on compact domains, a
random field realisation associated with the same RKHS do not have
finite norm. The reason for this is that the random field realisations
are less smooth, and only conditional expectations of the fields are
proper members of the RKHS.

\subsection{Uniqueness and intrinsic stationary random fields}
\label{sec:intrinsic}

The Green's functions of the precision operator are not necessarily
unique, such as when $\cQ_u(f,f)$ only generates a semi-norm. To avoid
the resulting extra solution space, the basic Mat\'ern case requires a
restriction to stationary solutions to eliminate functions in the
null-space of the operator, such as $\exp(\kappa\s\cdot \s_0)$ for any
$\s_0\in\bbR^d$ with $\|\s_0\|=1$. For compact domains these types of
fields are typically restricted by deterministic Neumann boundary
conditions instead, leading to non-stationary behaviour near the
domain boundary (further details in Section~\ref{sec:theory_constant})

When the null space of $\cQ_u(\cdot,\cdot)$ is not eliminated by
boundary or other conditions, but where stationarity is achieved when
applying some \emph{contrast filter} to the field, the result is a
family of \emph{intrinsic stationary} models where the probabilistic
structure is invariant to addition of functions in the null space of
the contrast operator. The classic grid-based intrinsic stationary
random fields \citep{besag_conditional_1995,besag_first-order_2005}
correspond to continuous domain models with $\kappa=0$, which gives
invariance to addition of constants (for $\alpha=1$) and planes (for
$\alpha=2$). However, more complex types of null spaces can appear
when the precision inner product is generalised to, e.g., oscillating
field models. This problem as well as opportunity is currently
underappreciated in the literature, that has focused on models
invariant to constants and planes. For example, on $\bbR^d$, the
undampened limiting case of oscillating fields \citep[Section
3.3]{lindgren11} gives invariance to sine and cosine functions in
arbitrary directions, with wave number $\kappa$.


\subsection{Spectral representations}\label{sec:spectral}

For theoretical treatment of stationary models, spectral
representations are essential. Classic linear filter theory for
Gaussian processes can be applied directly, and we refer to
\citet{Cramer1967}, which is a reprint of the original 1967 book, and
\citet{LindgrenStocProc} for the theoretical foundations. The
covariance of a stationary Gaussian process can be written as a
Fourier transform
$$
\cov(\s,\s') = \int_{\bbR^d} \exp(i(\s'-\s) \cdot \mv{k}) \md S(\mv{k})
$$
of a symmetric non-negative spectral measure $\md S(\mv{k})$,
$\mv{k}\in\bbR^d$. When the measure admits a density, we write
$\md S(\mv{k})=S(\mv{k})\md\mv{k}$, and the process itself can be
constructed as a stochastic Fourier integral
$$
u(\s) = \int_{\bbR^d} \exp(i\s \cdot \mv{k}) \sqrt{S(\mv{k})} \md Z(\mv{k}),
$$
where $\md Z(\mv{k})$ is a complex valued centred Gaussian white noise
measure with ${\mmd Z(-\mv{k}) = \ol{\mmd Z(\mv{k})}}$ and
$\pCov\{\mmd Z(\mv{k}),\mmd Z(\mv{k'})\} = \delta(\mv{k}-\mv{k}')
\md\mv{k}$.

The general connection between SPDEs and Mat\'ern covariances was
proven by \citet{whittle63}, using the spectral representation of the
differential operator to show that the Mat\'ern covariance
\eqref{eq:matern} is the Fourier transform of the spectral density
$S_M(\mv{k})=\{\tau^2(2\pi)^d(\kappa^2+\|\mv{k}\|^2)^\alpha\}^{-1}$,
$\mv{k}\in\bbR^d$, obtained from the reciprocal of the spectral
representation of the precision operator for the solutions of
\eqref{eq:whittle}. The factor $(2\pi)^d$ that appears in the spectral
density comes from the spectral density of the standardised white
noise definition where $\cR_\cW(f,g)=\scal{f}{g}$, and $\|\mv{k}\|^2$
are the eigenvalues of $-\Delta$ on $\bbR^d$. The local precision
operator characterisation of the Markov property
\citet{rozanov1977markov} implies that a stationary process on
$\bbR^d$ is Markov if and only if the reciprocal of its spectral
density is an even polynomial, which we see is fulfilled for integer
$\alpha$.

\subsection{Manifolds}\label{sec:manifolds}

A motivating example for \citet{lindgren11} was to construct
stationary MRF models on the sphere, to address geoscience problems
such as historical climate modelling. For a historical connection, see
\citet{wahba_spline_1981}, where spline penalties of similar form to
the inner product \eqref{eq:precisionproduct2} were used when
modelling data on the sphere. In contrast to explicit covariance
specification, the SPDE approach has the advantage that it is easy to
construct a wide range of valid models on any sufficiently smooth
manifold, and the Whittle representation provides a natural
generalisation of Mat\'ern field to such manifolds. The continuous
domain precision definition is directly applicable by letting $\Delta$
denote the Laplace-Beltrami operator on the sphere (which is the
restriction of the $\bbR^3$ Laplacian to the sphere), with
$\cD=\bbS^2$. The basic convergence proofs for the finite Hilbert
space representations (see Section~\ref{sec:FEM}) remain the same as
for compact flat subdomains of $\bbR^d$.

When dealing with non-Euclidean manifolds in particular, spectral
theory can be less intuitive than for $\bbR^d$, but for any smooth
compact manifold, the set of eigenfunctions of the Laplacian
generalised to manifolds, such as the Laplace-Beltrami operator on the
sphere, form a countable basis of eigenfunctions, that can be used to
construct Fourier-like representations. The spectral representation of
the precision operator, covariance function, and the process itself,
follow the same principles as for $\bbR^d$ but with a countable
harmonic basis. This technique was used to prove the generalised
Green's identity lemmas \citet[][Appendix D]{lindgren11}. On the
sphere, the spherical harmonic representation of a stationary Gaussian
field becomes
$$
u(\s) = \sum_{k=0}^{\infty} \sum_{m=-k}^k Y_{k,m}(\s) \sqrt{S(k)} Z_{k,m},\quad \s\in\bbS^2,
$$
where $\Delta Y_{k,m}(\cdot) = \lambda_k Y_{k,m}(\cdot)$, for
eigenvalues $\lambda_k=-k(k+1)$, $k=0,1,2,\dots$, with multiplicity
$2k+1$ across the modes $m=-k,\dots,k$, and independent $\pN(0,1)$
variables $Z_{k,m}$. Computing the covariance reveals the spectral
representation of the covariance function in terms of Legendre
polynomials $P_k(\cdot)$,
\begin{align}
  \label{eq:matern-cov-sphere}
  \cov(\s,\s') = \sum_{k=0}^{\infty} (2k+1) S(k) P_k(\s\cdot\s'),
  \quad \s,\s'\in\bbS^2,
\end{align}
where the $2k+1$ factor comes from the summation/product formula for
spherical harmonics,
$\sum_{m=-k}^k Y_{k,m}(\s)Y_{k,m}(\s')=(2k+1)P_k(\s\cdot\s')$. For
some further theoretical background, see
\citet{schoenberg_positive_1942} and \citet{wahba_spline_1981}.

Applying linear filter theory to the Whittle SPDE on the sphere leads
to the Whittle-Mat\'ern spectrum
$S_M(k) = (4\pi)^{-1}\tau^{-2}\{\kappa^2+k(k+1)\}^{-\alpha}$,
$k=0,1,2,\dots$. The covariance is not available in closed form, but
can be evaluated numerically via the infinite series
\eqref{eq:matern-cov-sphere}, which is convergent for $\alpha > 1$,
corresponding to smoothness $\nu >0$.

Note that the Markov characterisation from \citet{rozanov1977markov}
hinges on the precision operator being local, so integer $\alpha$
values will still generate Markov processes on the sphere, even though
the reciprocal of the functional form of the spectrum is not an even
polynomial in $k$.

Since the manifold curvature of the domain influences the Green's
functions of the precision operator, the notion of \emph{stationary}
fields is largely restricted to fields on $\bbR^d$ and $\bbS^d$. For
other manifolds, the differential geometry structure of the manifold
becomes important, in addition to the already present boundary effects
for compact domains with boundaries.

\subsection{Non-stationary models}

Once the connection between stationary Mat\'ern fields and stochastic
PDEs is made, the model family can be extended in many ways. Apart
from general manifold extensions, non-stationary models can be
constructed by modifying the differential operator. One immediate
non-stationary extension, that we refer to as a generalised
Whittle-Mat\'ern model, is to let $\kappa$ and $\tau$ depend on the
location, and to extend the Laplacian to non-stationary anisotropic
versions:
\begin{equation}\label{eq:nonstationary}
    \{\kappa(\s)^2 - \nabla\cdot \mv{H}(\s) \nabla\}^{\alpha/2}
    u(\s) = \frac{1}{\tau(\s)}\cW(\s) .
\end{equation}
With only mild regularity conditions on the parameter fields (see
Section~\ref{sec:generalcase}) this results in an implicitly defined
positive definite non-stationary covariance function, and the
precision operator is available in closed form. The temperature
application in \citet{lindgren11} used this generalised model with
$\alpha=2$, $\mv{H}(\s)\equiv\mv{I}$, and log-linear models for
$\kappa$ and $\tau$ expressed via 3 piecewise quadratic B-spline basis
functions in $\sin(\text{latitude})$. Spatial geographical covariates
were used by \citet{ingebrigtsen_spatial_2014}, \citet{art532}
discussed applications to adaptive spline models, and
\citet{fuglstad_exploring_2015} explored practical anisotropic
Laplacian representations.

The main challenge with more general non-stationary models is in
practical inference, as in many situations only a single noisy
realisation of the field is available, making practical
identifiability a challenge \citep[see][]{fuglstad2015non-stationary,
    bk-measure}. This is however not a unique trait of the SPDE
construction, since this is true of any sufficiently general
non-stationary model family. By exploiting the properties of the
operator, one possible approach is to estimate the operator locally,
avoiding global calculations. As long as basic regularity conditions
are enforced, the result will yield a valid global model, and more
effort can go into improving the local estimates rather than dealing
with cumbersome positive definiteness issues of non-stationary
covariance functions.

A special case of this approach was introduced by \cite{bakka2019} in
the form of a \emph{barrier model}, that disconnects the process
across physical barriers, blocking spurious dependence from travelling
between points that are near in the euclidean distance sense, but far
away in geodesic distance in the domain. The idea is to let $\kappa$
be close to $\infty$ in regions that form barriers, which has the same
effect as having spatial correlation range close to zero, and a
constant $\kappa$ value in the domain of interest. The resulting
fields have very little boundary effects compared with deterministic
Neumann conditions, making this an attractive alternative for many
practical situations, such as modelling fish in archipelagos, where
dependence should not travel across land.

Modifying the SPDE operator is equivalent to a change of metric on
Riemannian manifolds \citep{lindgren11}. This provides an illuminating
comparison to the classic \citet{sampson_nonparametric_1992}
deformation method for non-stationary random fields. The deformation
method works by warping the domain of the target field, optionally
into a manifold embedded in a higher dimension. Then, a stationary
covariance model is applied on the deformed manifold, with respect to
the Euclidean distance of the embedding domain. When mapped back to
the original space, a non-stationary model is obtained \citep[see][for
the general fractional case]{Hildeman2020}. If instead the
basic Whittle SPDE model \eqref{eq:whittle} is applied to the deformed
manifold domain, a different non-stationary model is obtained, that
relates to the geodesic distances within the manifold. By constructing
the metric for the manifold and converting the expression back to the
original manifold coordinates, a non-stationary SPDE model similar to
\eqref{eq:nonstationary} is obtained. This provides a different way to
parameterise certain types of non-stationarity. A major benefit of
this interpretation is that it provides geometric interpretability,
both for the SPDE models themselves and for how they differ from those
obtainable with the classic deformation method. For a given manifold,
the non-stationarity follows implicitly, but finding a manifold that
generates a specific pre-defined non-stationary behaviour is
difficult. An example of the latter is shown in Supplement~S7.1
of~\citet{fuglstad_constructing_2019}, where a piecewise linear change
in $\kappa(\s)$ corresponds to a cylindrically deformed manifold. It
is however challenging to design models of this type directly, as the
embedding space may need to be much larger than $\bbR^3$ to capture
desired properties. Instead, the main take-away is that non-stationary
SPDE operators and manifold metrics are closely connected.

\subsection{Locally supported Hilbert space basis and discretised
    precisions}
\label{sec:FEM}

In order to construct finite dimensional representations of the SPDE
solutions, \citet{lindgren11} used piecewise linear basis functions
with local support on spatial triangulations. This choice retains many
of the benefits of the continuous Markov properties, leading to sparse
matrices both also when conditioning on georeferenced observations, in
contrast to other non-local basis choices such as harmonic basis
functions and Karhunen-Lo\`eve expansions. The Hilbert space
projection theory works essentially the same for all these choices,
but we will focus on the Markov version for now and return to the
non-local basis choices in Section~\ref{sec:related}.

Let $\{\psi_j(\s),\,j=1,\dots,N\}$ denote a set of continuous piecewise linear
basis functions that sum to 1 for each spatial location $\s$, each with support on
triangles connected to a vertex. For planar triangulations, the
average number of triangles for each vertex is approximately 6. We
then seek a basis weight vector $\mv{u}=\{u_1,u_2,\dots,u_N\}$ such
that the distribution of the resulting function
$\wt{u}(\s)=\sum_{j=1}^N \psi_j(\s) u_j$ is close to that of the
continuously defined SPDE solutions. The solutions to the SPDE
\eqref{eq:generalspde} can be characterised by every finite
dimensional linear functional of the left and right hand sides of
\begin{align}\label{eq:leftright}
  \scal{f}{\cL u} &= \scal{f}{\cW}
\end{align}
having the same joint distributions, where the $f$ are denoted test
functions. For the finite representation $\wt{u}$ this cannot be
achieved for arbitrary collects of functionals, but by choosing
specific $N$-dimensional sets of functionals, the approximation properties can
be controlled. The approach used by \citet{lindgren11} for the Whittle
SPDE \eqref{eq:whittle} was to use the basis functions $\psi_j$ as
test functions for the case $\alpha=2$ (a Galerkin finite element
approach), and $\cL \psi_j$ as test functions for the case $\alpha=1$
(a least squares finite element approach), and then apply an iterated
Galerkin construction for higher order operators. Similarly to how the
covariance and precision products $\cR$ and $\cQ$ are connected for
the full SPDE solutions, these finite element constructions generate a
projection of the infinite dimensional solutions onto the finite
dimensional basis, such that the precision matrix $\mv{Q}$ for the
weight vector has a closed form expression in the model parameters.
For functions $f(\cdot)$ and $g(\cdot)$ in the finite dimensional
Hilbert space with weight vectors $\mv{f}$ and $\mv{g}$, the inner
product $\cQ_u(f,g)$ becomes $\mv{f}^{\top}\mv{Q}\mv{g}$, with a small
deviation depending on the details of the construction of $\mv{Q}$.
The inner products between the test functions and SPDE components can
be reduced to integrals over products of basis functions and over
products of gradients of the basis functions, which for piecewise
linear basis functions over triangles only involve straightforward
geometry. Let $\mv{C}$ and $\mv{G}$ be matrices with elements
$\mv{C}_{i,j}=\scal{\psi_i}{\psi_j}$ and
$\mv{G}_{i,j}=\scal{\nabla\psi_i}{\nabla\psi_j}$ respectively. To
illustrate the construction for $\alpha=2$ and $\tau=1$, we obtain
\begin{align*}
  \mat{\scal{\psi_i}{\cL \wt{u}}}_{i=1,\dots,N}
  &=
    \mat{\scal{\psi_i}{\sum_{j=1}^N
    (\kappa^2-\Delta)\psi_j(\cdot) u_j}}_{i=1,\dots,N}
  \\
  &=
    \mat{\sum_{j=1}^N \scal{\psi_i}{
    (\kappa^2-\Delta)\psi_j(\cdot)} u_j}_{i=1,\dots,N}
  \\
  &=
    (\kappa^2\mv{C} + \mv{G}) \mv{u}
\end{align*}
for the left hand side of \eqref{eq:leftright} and covariance
\begin{align*}
  \mat{
  \pCov\left(\scal{\psi_i}{\cW}\scal{\psi_j}{\cW}\right)
  }_{i,j=1,\dots,N} = \mv{C},
\end{align*}
for the right hand side of \eqref{eq:leftright}. This means that we
need $(\kappa^2\mv{C} + \mv{G}) \mv{u} \sim \pN(\mv{0},\mv{C})$, which
is achieved when the precision matrix for $\mv{u}$ is given by
$(\kappa^2\mv{C} + \mv{G})\mv{C}^{-1}(\kappa^2\mv{C} + \mv{G})$. As
discussed by \citet{lindgren11}, the inverse of $\mv{C}$ is
non-sparse, but this can be avoided by replacing $\mv{C}$ with a
diagonal version containing the row-sums of the original matrix,
giving $C_{i,i}=\scal{\psi_i}{1}$ due to the basis functions summing
to $1$. This \emph{mass-lumping} technique is common for finite
element methods with local basis functions, but is not applicable to
methods with globally supported basis functions.
\citet{bakka2019solve,art435} provide more mathematical details on
this construction.

The general precision matrix construction for general
$\alpha=1,2,3,\dots$ and $\tau$ is then given by
\begin{equation}\label{eq:precisionoper2}
    \mv{Q} = \tau^2 \mv{C}^{1/2} (\kappa^2\mv{I} +
    \mv{C}^{-1/2}\mv{G}\mv{C}^{-1/2})^{\alpha} \mv{C}^{1/2},
\end{equation}
where the diagonal version of $\mv{C}$ is used. It should be noted
that this construction works for any compact manifold that can be well
represented by a triangulation, and that Green's first identity that
is needed for integration by parts in
$\scal{f}{-\Delta g}=\scal{\nabla f}{\nabla g}$ (under suitable
boundary conditions) holds on polyhedral manifold surfaces and also
for some less than differentiable functions \citep[see][Appendix
B.3]{lindgren11}. The approximation properties of this Hilbert space
projection approach follow from common properties of finite element
methods, and will be discussed in more detail in
Section~\ref{sec:theory}.

As shown by \citet{bolin_comparison_2013}, the overall approximation
error can be reduced by using higher order B-spline basis functions or
wavelets for regularly gridded domains, and \citet{liu_efficient_2016}
implemented higher order bivariate splines on triangulations as basis
functions. In practice, increasing the resolution of the triangles
when needed is easier to implement, and avoids potential issues with
mass-lumping, which should be avoided for higher order basis
functions. One-dimensional domains are an important exception, where a
piecewise quadratic B-spline basis is easy to implement, and can lead
to clear improvements, especially for problems with irregularly spaced
observations within each spline knot interval. Where the piecewise
linear basis function can lead to a clear difference in marginal
variance between the nodes and the interval midpoints, similar to the
problems exhibited by kernel convolution methods
\citep{simpson_order_2012}, the higher order B-splines smooth out the
conditionally deterministic interval effects. This is useful even for
the case $\alpha=1$, where one might otherwise expect smoother basis
functions not to add value. Instead of mass-lumping, the complete
quint-diagonal $\mv{C}$ matrix should be used, and for $\alpha=2$, the
term $\mv{G}\mv{C}^{-1}\mv{G}$ is replaced with a second order matrix
$\mv{G}_2$ via elements $\scal{\Delta \psi_i}{\Delta \psi_j}$ that
represents a bi-harmonic operator.


\section{Practical spatial estimation and inference}\label{sec:INLA}

The plan of this section is to discuss why the precision matrix
representation of a Gaussian process plays nicely with conditioning on
observations, as well as when several components are joined into
larger statistical models.

\subsection{Conditional distribution under noisy observations}
\label{sec:INLA2}

The simplest hierarchical Gaussian process model with additive
observation noise and known Mat\'ern covariance can be written as
\begin{align*}
  u(\cdot) &\sim \pGRF\left(\mu_u(\cdot), \cov_M(\cdot,\cdot)\right), \\
  y_i|u(\cdot) & \sim \pN(u(\s_i), \tau_e^{-2}), \quad i=1,\dots,n.
\end{align*}
Replacing the full random field with the finite Hilbert space
representation from Section~\ref{sec:FEM} gives
\begin{align*}
  \mv{u} &\sim \pN(\mv{\mu}_u, \mv{Q}_u^{-1}) \\
  y_i|\mv{u} & \sim \pN\left(\sum_{j=1}^N \psi_j(\s_i) u_j,
               \tau_e^{-2}\right), \quad i=1,\dots,n .
\end{align*}
Introducing the basis function evaluation matrix $\mv{A}$ with
elements $A_{i,j}=\psi_j(\s_i)$, the full observation vector model
becomes $\mv{y}|\mv{u}\sim\pN(\mv{A}\mv{u},\mv{Q}_e^{-1})$, where
$\mv{Q}_e=\mv{I} \tau_e^2$ is the observation noise precision matrix.
By standard use of the precision matrix version of conditioning in
multivariate distributions \citep{rue2005gaussian}, the conditional
distribution of the basis weights for the field, given the
observations, becomes
\begin{align}
  \mv{u}|\mv{y} &\sim \pN(\mv{\mu}_{u|y}, \mv{Q}_{u|y}^{-1}), \\
  \mv{Q}_{u|y} &= \mv{Q}_u + \mv{A}^\top \mv{Q}_e \mv{A}, \label{eq:posterior-Q} \\
  \mv{\mu}_{u|y} &= \mv{\mu}_u + \mv{Q}_{u|y}^{-1} \mv{A}^\top
                   \mv{Q}_e (\mv{y} - \mv{A}\mv{\mu}_u). \label{eq:posterior-mu}
\end{align}
These equations provide the finite Hilbert space representation of the
kriging estimate of the field, as
$\sum_{j=1}^N \psi_j(\s) \mat{\mv{\mu}_{u|y}}_j$. For locally
supported basis functions, the conditional precision matrix is still
sparse and cheap to evaluate, and the conditional expectation only
involves a linear solve with that sparse matrix. By automatic
reordering of the Markov graph induced by the sparsity pattern of the
matrix, direct Cholesky factorisation can retain a high degree of
sparsity, making this the ideal direct solution method. By applying
the Takahashi recursions \citep{pro20} (see also \citet{art358},
\citet{art375} and \citet{col27} for easier access) to the Cholesky
factor of the posterior precision matrix, the posterior marginal
variances and neighbour covariances can be obtained as a by-product,
as implemented by the \verb!inla.qinv()! function in the \RINLA{}
package, without the need to compute a dense matrix inverse.

If we have unknown (hyper-)parameters $\mv{\theta}$ in this model, for
example the marginal variance or range, then we can compute the
posterior density $\pi(\mv{\theta}|\mv{y})$ \emph{directly}, like
\begin{displaymath}
    \pi(\mv{\theta} \mid \mv{y}) \propto
    \frac{\pi(\mv{\theta})\; \pi(\mv{u} \mid
        \mv{\theta}) \; \pi(\mv{y} \mid \mv{u},
        \mv{\theta})}{\pi(\mv{u}\mid\mv{y}, \mv{\theta})}
\end{displaymath}
since $\pi(\mv{u}|\mv{y},\mv{\theta})$ is Gaussian. The only new term
entering, beyond the conditional mean and precision that is already
computed, is $\log|\mv{Q}_{u|y}|$, which is directly available from
its Cholesky factorisation.

\subsection{Adding model components}

Our aim is to handle not only simpler model constructs as discussed in
Section~\ref{sec:INLA2}, but also cases where the linear predictor
$\mv{\eta}$ (in a GLM kind of model) is a sum of Gaussian model
components~\citep{art451}. As a simplified example, let us consider
$\mv{\eta} = \mv{A}_u\mv{u} + \mv{A}_v\mv{v} + \mv{A}_w\mv{w}$ where
$\mv{u} \sim \pN(\mv{0}, \mv{Q}_u^{-1})$,
$\mv{v} \sim \pN(\mv{0}, \mv{Q}_v^{-1})$ and
$\mv{w} \sim \pN(\mv{0}, \mv{Q}_w^{-1})$, and $\mv{A}_u$, $\mv{A}_v$,
and $\mv{A}_w$ are matrices that connect the latent variables in
$\mv{u}$, $\mv{v}$, and $\mv{w}$ to the linear predictor $\mv{\eta}$.
The latent model components may include both finite dimensional SPDE
representations, "fixed effect" coefficients, and other structured or
unstructured random effects. An important, if not \emph{the} most
important, property of adding model components via their precision
matrices, is that the structure of the joint precision matrix of
$(\mv{\eta}, \mv{u}, \mv{v},\mv{w})$ is directly available. This is
particularly important, as if we have covariance parameter
$\mv{\theta}$, we do not need to rebuild this entire joint precision
matrix if some elements of $\mv{\theta}$ change, but only re-evaluate
the elements that are directly affected. We can approach the
computations is various ways and we will discuss three of them.

A first strategy is to work directly with the joint precision for the
model components, and form the linear predictor $\mv{\eta}$
deterministically when conditioning the model on the observations. The
joint precision and predictor can be written as
\begin{align*}
    \pPrec
    \mat{\mv{u}\\
        \mv{v}\\
        \mv{w}}
        &=
    \mat{\mv{Q}_u & & \\
         & \mv{Q}_v & \\
         & & \mv{Q}_w
    },
    \quad\text{and}\quad
    \mv{\eta} = \mv{A} \mat{\mv{u} \\ \mv{v} \\ \mv{w}},
\end{align*}
using the combined matrix
$\mv{A} = \mat{ \mv{A}_u & \mv{A}_v & \mv{A}_w}$. With this
formulation, we then apply equations \eqref{eq:posterior-Q} and
\eqref{eq:posterior-mu} directly to the joint component model.

The second strategy is to build an approximate joint precision for the
linear predictor and the linear predictor, by adding a small noise
term with high precision $\tau$. With
$\mv{\eta} = \mv{u} + \mv{v} + \mv{w} + \tau^{-1/2} \mv{\epsilon}$
where $\mv{\epsilon} \sim \pN(\mv{0}, \mv{I})$, we get
\begin{displaymath}
    \pPrec
    \begin{bmatrix}
        \mv{\eta}\\
        \mv{u}\\
        \mv{v}\\
        \mv{w}
    \end{bmatrix} =
    \begin{bmatrix}
        \mv{0}& & &\\
        & \mv{Q}_u & & \\
        & & \mv{Q}_v & \\
        & & & \mv{Q}_w
    \end{bmatrix}
    + \tau
    \mat{\mv{I} \\ -\mv{A}_u^\top \\ -\mv{A}_v^\top \\ -\mv{A}_w^\top}
    \mat{\mv{I} & -\mv{A}_u & -\mv{A}_v & -\mv{A}_w}
    .
\end{displaymath}
Note that we need to keep all model components, as marginalisation
will destroy the Markov properties.

A third variant is to use cumulative sums. This approach can applied
to models where the combined effects can be written in a common
representations, such as for equal or nested triangulations at
different spatial resolution for SPDE models. Assume that
$\mv{B}_{uv}\mv{v}$ converts from coarse $v$-basis functions to finer
scale $u$-basis functions, and similarly for $\mv{B}_{vw}$. This
allows the linear predictor to be formulated as
$\mv{\eta}=\mv{A}_u\left\{\mv{u}+\mv{B}_{uv}(\mv{v}+\mv{B}_{vw}\mv{w})\right\}$.
We can then define $\wt{\mv{v}}=\mv{v}+\mv{B}_{vw}\mv{w}$ and
$\wt{\mv{u}}=\mv{u}+\mv{B}_{uv}\wt{\mv{v}}$, so that
$\mv{w}\sim\pN(\mv{0}, \mv{Q}_w^{-1})$,
$\wt{\mv{v}}|\mv{w}\sim\pN(\mv{B}_{vw}\mv{w}, \mv{Q}_v^{-1})$ and
$\wt{\mv{u}}|\wt{\mv{v}},\mv{w}\sim\pN(\mv{B}_{uv}\wt{\mv{v}},
\mv{Q}_u^{-1})$. The joint precision matrix is then
\begin{align*}
    \pPrec
    \mat{\wt{\mv{u}} \\
         \wt{\mv{v}} \\
         \mv{w}}
         &=
    \mat{\mv{0} \\ \mv{0} \\ \mv{I}}
    \mv{Q}_w
    \mat{\mv{0} & \mv{0} & \mv{I}}
    +
    \mat{\mv{0} \\ \mv{I} \\ -\mv{B}_{vw}^\top}
    \mv{Q}_v
    \mat{\mv{0} & \mv{I} & -\mv{B}_{vw}}
    +
    \mat{\mv{I} \\ -\mv{B}_{uv}^\top \\ \mv{0}}
    \mv{Q}_u
    \mat{\mv{I} & -\mv{B}_{uv} & \mv{0}}
    \\
         &=
    \mat{
        \mv{Q}_u & - \mv{Q}_u\mv{B}_{uv} & \\
        - \mv{B}_{uv}^\top\mv{Q}_u
        & \mv{Q}_v + \mv{B}_{uv}^\top\mv{Q}_u\mv{B}_{uv}
        & - \mv{Q}_v\mv{B}_{vw} \\
        &-\mv{B}_{vw}^\top\mv{Q}_v & \mv{Q}_w + \mv{B}_{vw}^\top\mv{Q}_v\mv{B}_{vw}
    }.
\end{align*}
This construction can then be used in combination with either the
first or second strategy, with
$\mv{\eta}=\mv{A}_u \wt{\mv{u}} + \mv{0} \wt{\mv{v}} + \mv{0} \mv{w}$,
so that the linear predictor is directly connected only to
$\wt{\mv{u}}$. The cumulative approach is very elegant and efficient,
and can also be helpful as a stepping stone towards multiscale
preconditioners for iterative solvers. A potential drawback is that
some of the original model components do not appear explicitly.
However, this is not an issue when the aim is to infer $\mv{\theta}$
or when a model component is not of direct interest.

As demonstrated, the joint precision matrices are directly available
and do not need to be rebuilt when the values of covariance parameters
$\mv{\theta}$ are changed. The \RINLA{} (\texttt{www.r-inla.org})
implementation currently uses a mix of all these three strategies for
various model components and the joint model.

\subsection{Bayesian inference and non-Gaussian observations}
\label{sec:INLA3}

When the SPDE model (or models) is used within a larger GLM kind of
model, for example with Poisson count data,
\begin{align*}
    y_i \mid \eta_i &\sim\pPo(E_i \exp(\eta_i)),
\end{align*}
for positive constants $E_i$, then the conditional distributions are
no longer available in closed form, like they are in the case of
Gaussian distributed observations. Deterministic inference
computations are still possible, but with some approximations. With a
well behaved Gaussian-like structure of the model we can use
\emph{Integrated Nested Laplace Approximations} (INLA), making the
impact of the approximation much smaller than the uncertainty in the
estimates themselves. INLA require, in short, to do the computations
outlined in Section~\ref{sec:INLA2} repeatedly in a nested way to
provide posterior marginal approximations to all model parameters, but
with using the second order Taylor approximation of the log-likelihood
instead of the $\mv{A}^\top\mv{Q}_e\mv{A}$ term from the Gaussian case
in equation \eqref{eq:posterior-Q}. The computational efficiency is
therefore crucial. \citet{art451} introduce the INLA approach,
\citet{art522} discuss some refinements, \citet{art632} review the
approach focusing on the underlying ideas, \citet{art664} describe
some recent extensions in the \RINLA{} package, while the book by
\citet{book126} provides a practical guide (with code) to using SPDE
models with the \RINLA{} package.

Priors for the (log-)range and (log-) marginal variance in the SPDE
model, are important since these parameters cannot both be estimated
consistently under infill asymptotics \citep{Zhang2004}.
\citet{fuglstad_constructing_2019} derive the joint penalised
complexity prior \citep{art631} for those, and also discuss how to
deal with non-stationary models. This family of priors is our
recommended one and works well in practice.




\section{Important extensions}\label{sec:extensions}

\subsection{Methods for general fractional powers}
\label{sec:fractional}

The original SPDE approach as developed in \cite{lindgren11} is only
applicable for integer values of $\alpha$. This is natural since it
constructs a GMRF approximation of the continuous process, which has
Markov properties only if $\alpha\in\bbN$, as discussed in
Section~\ref{sec:precisions}. For Markov processes on $\bbR^d$,
results from \cite{rozanov1977markov} imply that the spectral density
is the reciprocal of a polynomial,
$\widetilde{S}(\mv{k}) = (\sum_{j=0}^\alpha b_j
\|\mv{k}\|^{2j})^{-1}$.
However, by restricting $\alpha$ we are also restricting the available
values of the smoothness index, or the differentiability of the random
field. It would therefore be desirable to have a method that works for
general values of $\alpha>d/2$.

In the author's response in \cite{lindgren11}, a ``parsimonious''
Markov approximation, with spectral density
$\widetilde{S}(\mv{k}) = (\sum_{j=0}^m b_j \|\mv{k}\|^{2j})^{-1}$, was
presented for the stationary model \eqref{eq:matern}. This was based
on choosing the coefficients $\{b_j\}$ though the minimisation of a
weighted $L_2$-error
$\int w(\mv{k}) (S(\mv{k})-\widetilde{S}(\mv{k}))^2 \md\mv{k}$ for an
appropriately chosen weight function $w(\cdot)$. This approximation is
implemented in \RINLA{}, but it has a limited accuracy and is not
applicable for the more general non-stationary models such as
\eqref{eq:nonstationary}. Overcoming some of the shortcomings of the
parsimonious approach, \citet{roininen_sparse_2018} developed a more
general truncated Taylor series approximation for the reciprocal of
the spectrum.

To obtain a method that works for general, possibly non-stationary
models, the FEM approximation needs to be combined with an
approximation method for fractional powers of elliptic differential
operators such as $\cL = \kappa^2 - \nabla\cdot\mv{H}\nabla$. There
are a few different ways in which this can be done, and the first
method that was proposed for these more general SPDE models was the
quadrature approximation by \citet{BKK2020}. For $\alpha<2$, the
method first performs the same FEM approximation as in the $\alpha=2$
case, resulting in a discretised equation
$\cL_h^{\alpha/2} u_h = \cW_h$ on the space $V_h$ that is spanned by
the FEM basis functions $\{\psi_j(\s),\,j=1,\dots,N\}$. Here $\cL_h$
denotes the discrete version of $\cL$ which is defined on $V_h$. The
second step is to handle the fractional power of the discretised
operator, $\cL_h^{\alpha/2}$. In the deterministic case,
\cite{bonito2015} proposed the following quadrature approximation of
the application of the fractional inverse
\begin{equation*}
    \cL_h^{-\alpha/2}v
    =
    \frac{\sin(\pi\alpha/2)}{\pi}
    \int_0^{\infty} \lambda^{-\alpha/2} (\lambda \, \mathrm{Id} + \cL_h)^{-1} \, \rd \lambda v \approx
    \frac{2 k \sin(\pi\alpha/2)}{\pi} \sum_{\ell=-K^{-}}^{K^{+}} e^{\alpha \ell k} (\mathrm{Id} + e^{2 \ell k} \cL_h)^{-1}v,
\end{equation*}
where $k,K^{-},K^{+}>0$ are parameters which \cite{bonito2015} showed
can be chosen to obtain exponential convergence of the approximate
fractional inverse operator to $\cL_h^{-\alpha/2}$. \citet{BKK2020}
combined this strategy with the FEM approximation to a discretisation
method (the sinc-Galerkin method) that works for the non-stationary
generalised Whittle-Mat\'ern fields \eqref{eq:nonstationary} with
arbitrary smoothness parameter $\alpha>d/2$.

An alternative method that often is more accurate was later proposed
by \citet{BK2020rational}. This method replaces the quadrature
approximation by a rational approximation
\begin{equation*}
    \cL_h^{-\alpha/2}v \approx p_{\ell}(\cL_h)^{-1}p_r(\cL_h)v,
\end{equation*}
where $p_{\ell}(\cdot)$ and $p_r(\cdot)$ are polynomials with
coefficients that are obtained from a rational approximation of the
function $f(x) = x^{\alpha/2}$ on an interval that covers the spectrum
of $\cL_h^{-1}$. This method, as well as that in \citet{BKK2020},
produces an approximation
$u_h(\mv{s}) = \sum_{j=1}^N u_j \psi_j(\mv{s})$ where the weights no
longer form a GMRF, but instead has a distribution
$\mv{u} \sim \pN(\mv{0},\mv{P}\mv{Q}^{-1}\mv{P}^{\trsp})$ for two
sparse matrices $\mv{P}$ and $\mv{Q}$. Introducing an auxiliary GMRF
$\mv{x} \sim \pN(\mv{0},\mv{Q})$, we can thus write
$\mv{u} = \mv{P}\mv{x}$. We note that this approximation has the same
form as the nested SPDE models from \citet{bolin11}, which explains
why we can use all computational methods for GMRFs also for these
approximations, that are implemented in the R package \texttt{rSPDE}
\citep{BK2020rational} available on CRAN.

For Gaussian fields, these approaches can be improved further by
performing the rational approximation on the precision operator
$\cL_h^\alpha$ rather than on $\cL_h^{\alpha/2}$. This has the
additional benefit that it facilitates a representation of the weights
$\mv{u}$ as a sum of GMRFs, $\mv{u} = \sum_{i=1}^m \mv{u}_i$, where
$\mv{u}_i \sim \pN(\mv{0},\mv{Q}_i^{-1})$. Thus, this representation
fits into the framework of Section~\ref{sec:INLA}, so that the models
can be fitted in \RINLA{}. The \texttt{rSPDE} package contains an
interface to \RINLA{} that allows for the specification of SPDE-based
models of general smoothness \citep[see][for a recent tutorial on the
combination of the two packages]{rSPDEvignette}.

The idea of approximating fractional models with sums of Markov
processes was used by \citet{sorbye_approximate_2019} for long-memory
fractional Gaussian noise, and the \Rpkg{LatticeKrig} \RRR{} package
\citep{nychka_latticekrig_2016} uses a related technique to
approximate fractional Mat\'ern models. Another recent alternative
method for handling the fractional power, inspired by the methods
mentioned above, is the Galerkin--Chebyshev method proposed by
\citet{bg:11}. For an overview of other recent methods, we refer to
\citet[][Section 2]{BK2020rational}.

\subsection{Non-Gaussian models}

One of the important features of the SPDE approach is that it
facilitates extending the Gaussian Mat\'ern fields to flexible
non-Gaussian Mat\'ern fields that are still easy to work with in
applications. The main idea, as presented by \citet{bolin14}, is to
replace the driving Gaussian noise in \eqref{eq:whittle} by some
non-Gaussian noise.

To understand the approach, note that Gaussian white noise can be
viewed as an independently scattered measure, which for a Borel set
$B\in\cB(\cD)$ returns $\pN(0,\lambda(B))$, where $\lambda(B)$ is the
Lebesgue measure of the set $B$. We can now replace the normal
distribution with some other distribution, such as the Normal Inverse
Gaussian (NIG) or Generalised Asymmetric Laplace (GAL) distributions,
and obtain a model which still has a Mat\'ern covariance structure,
but more flexible sample path properties.

The reason for why these two particular distributions work well is
that they are closed under convolution and that they can be
represented as normal variance mean mixtures. Specifically, let
$\gamma,\mu \in \mathbb{R}$ be two parameters and $Z\sim \pN(0,1)$,
then $X = \gamma + \mu V + \sigma \sqrt{V} Z$ has a NIG distribution
if $V$ has an inverse Gaussian distribution, and $X$ has a GAL
distribution if $V$ has a Gamma distribution. Performing the same FEM
approximation as in the Gaussian case, we now instead obtain that the
stochastic weights $\mv{u}$ have a distribution
$$
\mv{u}|\mv{v} \sim \pN(\tau^{-1}\mv{K}^{-1}(\mv{v}-\mv{h}),
\tau^{-2}\mv{K}^{-1}\proper{diag}(\mv{v})\mv{K}^{-1}),
$$
where $\mv{v}$ is a vector of independent IG variables in the NIG
case, and Gamma variables in the GAL case. This conditional Gaussian
representation allows us to use the same computationally efficient
techniques for sparse matrices as in the Gaussian case. The difference
is that the conditioning on the ``hidden'' variances $\mv{v}$ needs to
be handled.

As an example, suppose that the model is included in the simple
hierarchical model in Section~\ref{sec:INLA2}, and let $\mv{\theta}$
again denote all parameter of the model, which now also includes the
parameters of $\pi(\mv{v})$.
Since the joint distribution $\pi(\mv{u},\mv{v}|\mv{\theta},\mv{y})$
has a closed form, an simple alternative for likelihood-based
inference for models like this is to use an EM algorithm
\citep{Wallin15}. However, a more computationally efficient
alternative is to use Fisher's identity to represent the gradient of
the log-likelihood as
$$
\nabla_{\theta} \log \pi(\mv{y}|\mv{\theta}) =
\pE_{\mv{v}}\{\nabla_{\theta}\log\pi(\mv{y},\mv{v}|\mv{\theta})|\mv{y}\}.
$$
Here $\log\pi(\mv{y},\mv{v}|\mv{\theta})$ is known in closed form and
the posterior expectation over $\mv{v}$ can be efficiently
approximated through Monte Carlo integration. This means that
stochastic gradient descent methods can be used to find maximum
likelihood or maximum aposteriori estimates of $\mv{\theta}$
\citep{bw20,Asar2020}.

This approach was recently used for multivariate random field models
by \citet{bw20} and for longitudinal data analysis in the RSS
discussion paper by \citet{Asar2020}. In particular, \citet{Asar2020}
also introduced the R package \texttt{ngme} that can be used to fit
all these different non-Gaussian SPDE-based models through stochastic
gradient descent methods.

As noted by \citet{Asar2020}, a limiting case of the NIG distribution
is the Cauchy distribution, the \texttt{ngme} package therefore also
includes methods for Cauchy random fields. Cauchy random fields were
also recently investigated by \cite{Chada2021} for Bayesian inverse
problems, an area where the SPDE approach previously has been used
extensively \citep[see, e.g.,][]{Roininen2014,Roininen2019}. Bayesian
methods for the non-Gaussian SPDE models were also recently
investigated by \citet{Walder2020}, who provided several examples of
their use.

\subsection{Spatio-temporal processes}
\label{sec:spacetime}

To extend the spatial models to space-time, a first step is to use a
Kronecker precision model, as introduced in this context by
\citet{cameletti_spatio-temporal_2013}. Using a unit variance AR(1)
process with tridiagonal precision matrix
\begin{align*}
  \mv{Q}_t &= \frac{1}{1-\phi^2}
             \mat{
             1 & -\phi & 0 & \dots \\
  -\phi & 1 + \phi^2 & -\phi & \ddots & \\
  0 & \ddots & \ddots & \ddots \\
  \vdots & 0 & -\phi & 1
                       },
\end{align*}
for some $\phi > 0$, the Kronecker product precision
$\mv{Q}_t\otimes\mv{Q}_s$ gives a space-time random field model
discretising an Ornstein-Uhlenbeck process where the driving temporal
white noise process has a Mat\'ern covariance structure in space, or
equivalently, stationary solutions to the space-time SPDE
\begin{align*}
  \left(a + \frac{\partial}{\partial t}\right)(\kappa^2 - \Delta_\s)^{\alpha/2} u(\s,t) &= \frac{\sqrt{2 a}}{\tau}\cW(\s,t),
\end{align*}
where $\cW(\s,t)$ is a space-time white noise process on
$\cD\times\bbR$, where $\cD$ is the spatial domain, such as a subset
of $\bbR^d$, $\bbS^2$, or a more general manifold. The normalisation
constant for the driving noise ensures that the marginal spatial
covariance is identical to the purely spatial SPDE model
\eqref{eq:whittle} on $\cD$ for any $a>0$. For discretisation time
step $h_t>0$, the parameter relation is given by $\phi=\exp(-h_t a)$.
These types of models are implemented via the \texttt{group} arguments
to model components in the \RINLA{} and \inlabru{} \RRR{} software
packages.

For realistic modelling, and in particular for temporal prediction
problems, the separable covariance construction is too simple. As part
of the EUSTACE project \citep{rayner_eustace_2020}, a family of
non-separable space-time models based on a generalised space-time
diffusion model
\begin{align*}
  \tau
  (\kappa^2-\Delta)^{\alpha_e/2}\left\{\gamma_t\frac{\partial}{\partial t} + (\kappa^2-\Delta)^{\alpha_s/2}\right\}^{\alpha_t} &= \cW(\s,t),\quad \s\in\cD,\,t\in\bbR,
\end{align*}
was developed. By varying the operator orders $\alpha_t$, $\alpha_s$,
and $\alpha_e$, a range of models from fully separable to fully
non-separable are obtained. All of these models have generalised
Whittle-Mat\'ern covariances for both the spatial marginal
distributions and for the temporal evolution of each component of the
harmonic spectral representation on $\cD$. A preliminary version of
this approach can be found in \citet{bakka_diffusion-based_2020}, that
shows how using a space-time Kronecker basis definition results in a
precision matrix that is expressed as a sum of basic space-time
Kronecker products, making implementation straightforward. Extensive
theory for a similar approach based on \citet{stein2005space} and
other SPDE models such as advection-diffusion models is provided by
\citet{vergara2022}. From a practical perspective,
\citet{sarkka_spatiotemporal_2013} developed Kalman filter
representations of spatio-temporal SPDE fields that are particularly
appealing when the spatial field is smooth enough to be represented by
a sufficiently small number of basis functions. The Kalman filter
approach effectively involves a direct construction of a temporal
factorisation of the spatio-temporal precision operator.


\section{Theoretical guarantees}\label{sec:theory}

Having introduced the main ideas of the SPDE approach and its
extensions, we are now ready to look at some of the more technical
theoretical properties of the SPDE-based models and relating
computational methods.
One can easily argue that the SPDE-based models is one of the most
well-understood classes of random field models, both from a
theoretical and practical point of view. To support this claim, we now
briefly summarise what we know abut the SPDE-based models and
corresponding computational methods. In
Subsection~\ref{sec:modelproperties} we present some of the most
important theoretical properties that are known about the SPDE-based
models, and in Subsection~\ref{sec:FEMproperties} we present the
current knowledge regarding the corresponding approximation methods.

\subsection{Properties of the SPDE models}\label{sec:modelproperties}
\subsubsection{The model with constant parameters}\label{sec:theory_constant}

It has been known already since the early 1960's (with partial results
from the 1950's) that stationary solutions to the SPDE
\eqref{eq:whittle} on $\cD= \mathbb{R}^d$ are centred Gaussian fields
with Mat\'ern covariance functions \citep{whittle54,whittle63}. Thus,
in this case, theoretical properties of the solutions can be obtained
from the standard theory for stationary Gaussian random fields
\citep[e.g.][]{Cramer1967}.

Properties of the solution to \eqref{eq:whittle} when considered on
manifolds such as the sphere are also well-understood since the
eigenvalues of the operator $\cL$ are explicitly defined in terms of
the eigenvalues of the Laplacian \citep[see,
e.g.,][]{Lang2015,borovitskiy2020mat}. In particular, as in the case
of $\cD= \mathbb{R}^d$, the exponent $\alpha$ controls H\"older
continuity and differentiability of the solution.

When considering the SPDE \eqref{eq:whittle} on a bounded domain
$\cD \subsetneq \mathbb{R}^d$, one has to add boundary conditions to
the operator. Typically, Neumann or Dirichlet boundary conditions are
used in practice. Because of this, the solution will be non-stationary
and no longer have a Mat\'ern covariance. However, \cite{lindgren11}
showed that for $d=1$ and Neumann boundary conditions, the solution
has a folded Mat\'ern covariance
$\widetilde{\cov}_{\kappa,\tau,\alpha}$ that will be similar to the
corresponding Mat\'ern covariance $\cov_{\kappa,\tau,\alpha}$ in the
interior of the domain. Because of this, it was argued that one should
use a domain $\cD$ that is extended by a distance $\delta$ which is at
least two times the practical correlation range
$\rho = \sqrt{8\nu}/\kappa$ outside the domain of interest $\cD_0$.
This procedure was further validated by \citet{Ullmann2019} who
extended the result to $d>1$ as well as Neumann and periodic boundary
conditions when $\cD$ is a box in $\bbR^d$. They in particular showed
that the supremum norm
$\|\widetilde{\cov}_{\kappa,\tau,\alpha}-\cov_{\kappa,\tau,\alpha}\|_{L_{\infty}(\cD_0)}$
can be bounded in terms of $\delta/\rho$ and that the error
asymptotically decreases exponentially as this term increases. Thus,
the SPDE with stationary parameters, as well as the effects of the
boundary conditions in this case are well understood.

\subsubsection{The non-stationary Whittle-Mat\'ern
    generalisation}\label{sec:generalcase}

The theory in the non-stationary case is more involved; however, the
properties of the processes are well-understood also in this case.
Considering the generalised Whittle--Mat\'ern fields
\eqref{eq:nonstationary} for a convex polytope
$\cD\subset\bbR^d, d\in\{1,2,3\}$, we know from
\citet{BKK2020,BK2020rational} that there exists a unique solution to
the SPDE given that $\alpha>d/2$ (which corresponds to $\nu>0$ in the
stationary case), $\kappa$ is an essentially bounded function,
$\kappa\in L_{\infty}(\cD)$, and $\mv{H}$ is a sufficiently nice (Lipschitz
continuous on $\bar{\cD}$ and uniformly positive definite) matrix
valued function. \citet{Herrmann2020} extended this existence result
by showing that it holds also when considering the SPDE
\eqref{eq:nonstationary} on a closed, connected, orientable, smooth,
compact 2-surface in $\bbR^3$, under the assumption that $\kappa,\mv{H}$
are smooth, and \citet{Harbrecht2021} derived similar results for the
model on more general manifolds without boundaries.

\citet{cox2020} generalised the case $\cD\subset\bbR^d$ further by
only requiring that the domain $\cD$ has a Lipschitz boundary, and by
relaxing the requirement on $\mv{H}$ to only assume essential boundedness
and uniformly positive definiteness. More importantly, they also
characterised the Sobolev and H\"older regularity of the solution $u$
and its covariance function. Thus, the regularity of the SPDE-based
model is known also in the non-stationary case.

\subsubsection{Induced Gaussian measures and kriging}

One of the key tools in the theory of Gaussian fields is equivalence
and orthogonality of Gaussian measures. This is, for example, often
used to derive consistency of maximum likelihood estimators
\citep{Zhang2004}. The question of when two different SPDE models
\eqref{eq:whittle} with constant parameters generate equivalent
measures was shown by \citet{BK2020rational}. That also showed that
for a fixed value of $\alpha$, one can estimate $\tau$ consistently
under infill asymptotics, but not $\kappa$, which is in accordance
with the results for Gaussian Mat\'ern fields \citep{Zhang2004}.

For statistical applications, it is also important to understand the
effects of misspecifying the parameters in the model. This has for
example been investigated thoroughly for stationary Gaussian random
fields by \citet{stein99} in the context of kriging. Similar results
are known in great generality also for the non-stationary generalised
Whittle--Mat\'ern fields.
For example, \citet{kb-kriging} derived conditions for uniform
asymptotic optimality of linear prediction for isotropic
Whittle--Mat\'ern fields on the sphere based on misspecified
parameters $\alpha,\kappa,\tau$. Further, \citet{bk-measure} derived
conditions for uniform asymptotic optimality of linear prediction for
generalised Whittle--Mat\'ern fields on bounded domains in $\bbR^d$
based on misspecified parameters $\alpha,\kappa,\mv{H}$. They further
generalised the results in \citet{BK2020rational} by deriving explicit
conditions for when two generalised Whittle--Mat\'ern fields induce
equivalent Gaussian measures. As far as we know, the generalised
Whittle--Mat\'ern is the only class of non-stationary models where
theoretical results like these are known.

\subsection{Properties of the approximations}\label{sec:FEMproperties}

Also the computational methods for the SPDE models are well
understood. Already by \citet{lindgren11} it was shown that the
distribution of the finite element approximation in the case
$\alpha\in\bbN$ converges to that of the exact solution as the mesh
becomes finer. These results were extended by \citet{Simpson2016} who
considered log-Gaussian Cox processes based on the SPDE-model
\eqref{eq:whittle}.

Later, the general fractional case has been thoroughly investigated.
This started with the results by \citet{BKK2020} who derived explicit
convergence rates of the strong error $\pE(\|u-u_h\|_{L_2(\cD)})$ of
the sinc-Galerkin approximations $u_h$ introduced in
Section~\ref{sec:fractional}. \citet{BKK2018} extended these results
by deriving explicit convergence rates also for weak errors
$|\pE(g(u)) - \pE(g(u_h))|$ for sufficiently smooth functionals
$g(\cdot)$. \citet{cox2020} extended the results further by also
providing explicit convergence rates for the error of the covariance
function of the approximation. All these results also hold for the
rational SPDE approach by \citet{BK2020rational} and for models on
surfaces \citep{Herrmann2020}.

Recently, posterior contraction rates of the FEM approximation when
included in the simple hierarchical model in Section~\ref{sec:INLA2}
as well as in a binary classification model was derived by
\citet{bg:10}. This provides theoretical justifications for how to
choose the number of FEM basis functions relative to the size of the
dataset that is considered.


\section{Applications}\label{sec:applications}

Since the publication of the \citet{lindgren11} paper, a wide variety
of applications have taken advantage of the available software
implementation in the \RINLA{} package, as well as used specialised
implementations of the SPDE constructions. We will highlight some key
applications that demonstrate the utility of these models in applied
problems with realistically complex observation models and large
hierarchical random field structure.

\subsection{Malaria modelling with spatio-temporal SPDEs}

One of the first large scale applications of the GMRF/SPDE models was
\citet{bhatt_effect_2015,bg:5}, that modelled the effect of malaria
control over time. The results showed that infection prevalence in
Africa had halved between 2000 and 2015, with an estimated 542--763
million (95\% credible interval) of averted cases attributable to
preventative interventions such as insecticide-treated nets.

As is common with medical data, the complex measurement structure had
to be considered carefully, with a spatio-temporal SPDE model used to
capture spatially structured effects that the rest of the model
components could not handle by themselves. Since Africa is large
enough that any map projection would introduce deformation, the model
was built directly on a subset of a spherical manifold. Although a
spatially stationary model was used, by eliminating spurious
non-stationarities due to arbitrary map projections, the results are
interpretable with respect to geodesic distances on the globe. The
implementation used a triangulated spherical mesh covering Africa,
with the space-time Kronecker precision model from
Section~\ref{sec:spacetime}.

\subsection{The EUSTACE project}\label{sec:EUSTACE}

When analysing past weather and climate, one challenge is to merge
information from multiple data sources and types of measurements.
Satellites provide large quantities of data from recent years, but
with complicated relationships between what is measured and the
quantities of interest, including spatially and temporally dependent
noise and satellite specific biases. Weather station data goes much
further back in time in some locations on the globe, but have
temporally persistent and changing biases, due to both local weather
variation and changes in instrumentation. Similar challenges apply to
air temperature measurements for ships. The large collaborative
EUSTACE project \citep{rayner_eustace_2020} aimed to construct a
reconstruction of weather and climate on a daily timescale and
$1/4 \times 1/4$ degree spatial resolution for the entire globe. The
full problem, including modelling both daily maximum and minimum
temperature for all $\sim 60,000$ days since 1850, has on the order
$10^{11}$ values to be estimated.

In basic applications of Gaussian processes and kriging, the typical
model combines a single random field with a few covariates, and links
those to georeferenced observations with Gaussian additive noise. The
global weather field has a dependence structure that operates on a
wide range of spatial and temporal scales, and explicitly designing a
covariance model for the weather would be unrealistic. The EUSTACE
project instead constructed a hierarchical model, where each node
contributes to just one aspect of the full behaviour, such as a slowly
varying climatological mean temperature field, a systematic latitude
effect, and daily weather residuals. Jointly, this defines a Markov
random field with respect to a graph connecting several
spatio-temporal graphs. Just as in the fractional SPDE constructions
discussed in Section~\ref{sec:fractional}, the resulting sum of fields
is not a Markov random field, but the computational benefits of Markov
properties are still present.

The project explored methods for handling the non-Gaussian behaviour
of daily maximum and minimum temperatures, via the same approximation
techniques for non-Gaussian observations as in the \RINLA{} software.
In order to keep the implementation and computational time manageable,
the final implemented method did not include this, but aspects of that
work can be found in \citet{vandeskog_quantile_2021}. Instead, a fully
Gaussian method was implemented, and an iterative linear solver used
to compute the conditional distributions for all the latent variables,
reduced to $\sim 1.5\cdot 10^{10}$ by only estimating the daily mean
temperatures, and reducing the spatial resolution to $0.5$ degrees.
The components were grouped into three categories:
\begin{itemize}
\item Climatological variation ($\sim 3.5\cdot 10^5$ nodes): 2-monthly
    1-degree seasonal patterns, 5-year 5-degree scale climate
    variation, non-linear latitude effect, altitude effect, coastal
    effect, and an overall mean
\item Large-scale variation ($\sim 1.8 \cdot 10^6$ nodes): 3-monthly
    5-degree field and weather station bias random effects
\item Daily fields
    ($\sim 6\cdot 10^4 \times 2.5 \cdot 10^5 \approx 1.5 \cdot
    10^{10}$ nodes): local 0.5 degree weather, satellite bias fields
    at 2.5 degree resolution
\end{itemize}
To compute the conditional expectations, an iterated conditional
approach was used, where each category was solved conditionally on the
others, rotating through until convergence. The daily category treated
each day as conditionally independent, which allowed massively
parallel computations, with $60\,000$ individual server tasks. Thus, the
largest individual solves were for the large-scale variation category,
with a computational graph size of around 2 million, using direct
Cholesky factorisation of the precision matrix.

\subsection{Neuroimaging}

\begin{figure}[t]
    \centerline{\includegraphics[width=0.6\linewidth]{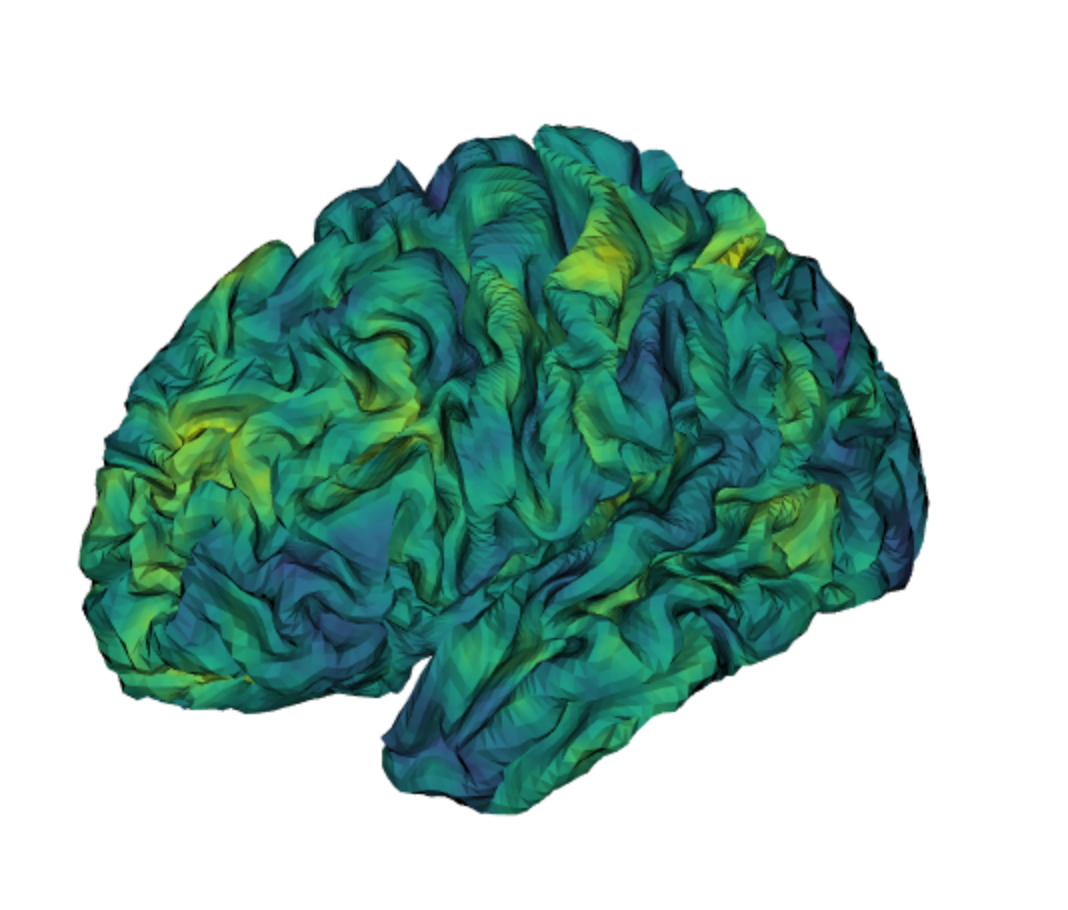}}
    \caption{Simulation of a Generalised Whittle--Mat\'ern field on
        the cortical surface.}
    \label{fig:brain}
\end{figure}
The ability to formulate Mat\'ern-like random fields on manifolds
opens up for applications to areas that are very difficult to approach
with covariance-based models. One such application is to neuroimaging.
Specifically, functional magnetic resonance imaging (fMRI) is a
popular neuroimaging technique for localising regions of the brain
which are activated by a task or stimulus. Traditional volumetric fMRI
data consist of observations of time series measured at thousands of
three-dimensional volumes in the brain. A common approach to analyse
fMRI data is through the classical general linear modelling (GLM)
approach \citep{friston1994statistical}.

The GLM approach does not have any explicit statistical model for the
spatial dependence, but it is implicitly taken into account through
pre-smoothing of data and post-correction of multiple hypothesis tests
for finding areas of activation, which is known to be problematic
\citep[see, e.g.][]{eklund2016cluster}. However, it is rare to
explicitly model the spatial dependence in fMRI, mainly because of the
high computational costs of spatial models.
This problem was recently overcome through applying the SPDE approach
to define a method for whole-brain fMRI analysis with spatial priors
\citep{Siden2021}.

Despite its popularity, traditional fMRI analysis has some
limitations, such as the fact that the data includes observations from
many different tissue types, whereas it is known that neuronal
activity only occurs in grey matter. Cortical surface fMRI (cs-fMRI)
solves this problem by representing the cortical grey matter as a
two-dimensional manifold surface \citep{fischl2012freesurfer}. The
approach has recently grown in popularity because it also offers
better visualisation, dimension reduction, and improved alignment of
cortical areas across subjects. Furthermore, the manifold
representation allows for more neurobiologically meaningful measures
of distance between locations, which is of great importance for
analysis. The main difficulty with analysing such data from a
statistical point of view is that the spatial dependence across the
cortical surface needs to be modelled. This difficulty was recently
overcome by using the SPDE approach to define the first Bayesian GLM
approach to cs-fMRI data \citep{Mejia2020fmri} which has shown to be
highly accurate compared to the standard GLM approach that ignores
spatial dependence \citep{spencer2021spatial}. In
Figure~\ref{fig:brain}, we show a simulation of a Generalised
Whittle--Mat\'ern field on the cortical surface of a patient.

The models mentioned above are effective in finding which areas of the
brain that are active when a particular task is being performed.
Another way to study the brain, which recently has gained much
attention, is to take a more holistic approach and to look at the
functional organisation of the brain in the absence of a particular
stimulus. A common method for this task is independent component
analysis (ICA), which works reliably when looking at groups of
patients \citep{calhoun2001method}. Performing estimation at a subject
level is more challenging, and taking spatial dependence into account
becomes more important at the subject level to reduce noise. This is
again difficult to do for both fMRI and cs-fMRI data, but the SPDE
approach has recently also facilitated the development of a spatial
ICA method for the estimation of functional brain networks
\citep{mejia2020spatial}.

Similar techniques have also been applied to other anatomical
manifolds, such as hearts. In order to model local activation times
from electrograms, \citet{coveney_probabilistic_2020} estimated
probabilistic activation times on a manifold of human atria. The
triangulated manifold mesh was estimated from the measured individual,
and then smoothed. The electrode positions were then projected to the
nearest mesh point, and a the Kronecker product precision model from
Section~\ref{sec:spacetime} used for the spatio-temporal activation
process on the manifold, with the \RINLA{} software.

\subsection{Seismology and material science}

While the most common application of SPDE models in spatial statistics
is to 2-dimensional space, sometimes with added time, in seismology
the data and problems of interest are often 3-dimensional. The finite
element methods used in the Hilbert space construction works in the
same way for tetrahedralisations as it does for triangulations. This
was exploited by \citet{zhang_bayesian_2016}, to estimate seismic
velocity under the western USA down to 700~km depth. In this complex
inverse problem, they applied the SPDE models to both the sub-surface
velocity field, and to the seismic source and receiver fields on the
surface. They also provide a geometrical derivation of the finite
element matrices $\mv{C}$ and $\mv{G}$ for tetrahedral meshes, that
have since been implemented in an experimental \RRR{} package at
\verb!https://github.com/finnlindgren/inlamesh3d!.

Statistical modelling of porous materials is another application, on a
very different spatial scale, that also requires models in
3-dimensional space. \citet{Barman2018} used the SPDE approach to
design a model for the study of a porous
ethylcellulose/hydroxypropylcellulose polymer blend that is used as a
coating to control drug release from pharmaceutical tablets. 
This model has later been used to study how to pore geometry affects the
diffusive transport through polymer films \citep{Barman2019}. 

\subsection{Point processes in ecology}
\label{sec:poissonprocess}

One of the benefits of using local basis functions in SPDE
constructions is the close resemblance to graph based Markov random
field models that are common in ecology and epidemiology. This makes
the step from aggregated counts on regions to continuous domain point
process models very small. Direct approximation of inhomogeneous
Poisson process likelihoods was analysed in detail by
\citet{Simpson2016}, and has since been automated as part of the
\texttt{inlabru} software \citep{bachl_inlabru_2019,yuan_point_2017}.

When $\eta(\s)$ is a linear predictor expression evaluated spatially,
the inhomogeneous Poisson point process
$\cY=\{\mv{y}_1,\dots,\mv{y}_M\}$, $\mv{y}_i\in\cD$, with intensity
$\lambda(\s)=\exp\{\eta(\s)\}$ on a domain $\cD$ has log-likelihood
$$
l(\eta; \cY) = - \int_\cD \exp\{\eta(\s)\} \md\cD(\s) + \sum_{i=1}^M \eta(\mv{y}_i)
$$
where the integral is a surface integral for manifold domains $\cD$.
The integral is in general intractable, but when $\eta(\s)$ is built
from local basis functions as in Section~\ref{sec:FEM}, several
options are available. For example, the discretisation
$\sum_{j=1}^N w_j \exp\{\eta(\s_j)\}$, where $\s_j$ are the mesh nodes
and $w_j=\scal{\psi_j}{1}$, provides a good and stable approximation.
The resulting likelihood expression is not a Poisson model likelihood
for $\eta$ since the two sums involve different spatial locations.
However, it is similar enough that it is easy to implement in the
\RINLA{} software, and to automate it, as in \inlabru{}.

A key motivation for the \inlabru{} software was to allow easier
specification of models of this type, as well as more complex models,
such as distance sampling from transect surveys, where the point
process is only observed along lines, e.g.\ from a ship traversing the
ocean. The surface integral is then effectively replaced by a line
integral, and the probability of detecting a point (usually an animal
or a group of animals) can also be incorporated. Such models however
do not necessarily result in a log-linear expression with respect to
the parameters of the intensity of the resulting thinned Poisson point
process, $\lambda(\s)\pP(\text{point detected at $\s$}\mid\text{point
    exists at $\s$})$. To deal with this, the \inlabru{} method
iteratively linearises the given predictor expression, and applies the
INLA method for the linearised versions, until the posterior mode has
been found.

Frequentist methods for distance sampling, that use penalised spline
smoothers closely related to intrinsic stationary random field models,
are available in the \RRR{} package \texttt{dsm}
\citep{miller_dsm_2021}.


\section{Related methods}\label{sec:related}

Here we highlight a few related or contrasting techniques and methods.
Further connections, related and contrasting methods can be found in
\citet{cressie_statistics_2011} and
\citet{wikle_spatio-temporal-with-R_2019}, that cover space-time
models in the hierarchical model setting, and a variety of
computational methods, including model assessment. The latter is
becoming an increasingly important topic, since traditional basic
measures such as the mean square error between the posterior mean (or
frequentistic point estimate) only provide limited insights. In order
to meaningfully compare complex spatial and spatio-temporal estimation
techniques, it is essential to use comparison scores that take the
estimated uncertainty of the predictions into account, such as
log-density and log-probability scores (for continuous and discrete
outcomes, respectively), CRPS (continuous ranked probability score),
or Dawid-Sebastiani (the log-density score from the Gaussian
distribution, which is a proper score with respect to the expectation
and variance also for other distributions), see
\citet{gneiting_strictly_2007}. For the sparse precision matrices
generated by the SPDE approach, leave-one-out cross validation
\citep{vehtari_practical_2017} with such scores can also be applied
\citep{art638}.

\subsection{Process priors versus smoothing penalties}

The theory of reproducing kernel Hilbert spaces provide a close
connection between frequentist penalised spline estimators and
Bayesian Gaussian process priors, as was recently discussed by
\citet{miller_understanding_2020}. In parallel with the development of
the stochastic PDE approaches, related development has been ongoing
for PDE based penalties, addressing similar problems, such as complex
shaped domains and non-Euclidean manifolds
\citep{sangalli_spatial_2013, sangalli_spatial_2021}.

As alluded to in Section~\ref{sec:precisions}, a main difference
between penalty minimisers (here usually the same as or similar to
conditional expectations) and full stochastic processes, is that the
penalty minimisers, associated with the same RKHS as the stochastic
process, are fundamentally smoother than the process realisations.
This also manifests in that for high dimensional Gaussian
distributions (where random fields are at the extreme, infinite
limit), the bulk of the probability mass lies far away from the
expectation of the distribution, and this deviation is essential when
quantifying prediction uncertainty. The variance of the point
estimator or conditional expectation provided by a penalty method
should not be confused with the prediction uncertainty provided by a
full posterior distribution of the process.

\subsection{Spectral model constructions and generalised
    Whittle-Mat\'ern fields}

An important aspect of using the Whittle SPDE is to generalise the
Mat\'ern covariance family to processes on manifolds, while keeping
the local geometric interpretations. The SPDE generalisation of the
Mat\'ern covariance models to smooth manifolds retains all the
differentiability and Markov subset properties of the original models,
and are asymptotically equivalent for short range. The spectral
representations are linked to the eigenfunctions and eigenvalues of
the Laplacian and its manifold versions. The methods for fractional
operators discussed in Section~\ref{sec:fractional} have recently been
extended using high order numerical methods for PDEs on Riemannian
manifolds by \citet{bg:11} and \citet{Harbrecht2021}, involving
polynomial and wavelet basis expansions. Other approaches to
constructing valid models on spheres can be found in
\citet{porcu_spatio-temporal_2016}.

On $\bbR^d$, the eigenvalues of the Laplacian are $-\|\mv{k}\|^2$ for
harmonic $\exp(i\mv{s}\cdot\mv{k})$, $\mv{k}\in\bbR^d$, and on the
sphere, $\lambda_k=-k(k+1)$ is the eigenvalue for the spherical
harmonic $Y_{k,m}$ of order $k\in\{0,1,2,\dots\}$,
$m\in\{-k,\dots,k\}$. In the literature, spectral constructions are
often used to define new families of models, e.g.\ for space-time
models \citep{stein2005space}. When using the Whittle SPDE
representation to generalise the Mat\'ern models to the sphere and
other non-Euclidean manifolds, the spectral representation of the
Laplacian plays a key role. In contrast, the spherical covariance
introduced by \citet{guinness_isotropic_2016} under the name
Legendre-Mat\'ern used $k^2$ in the spectrum definition instead of the
$k(k+1)$ that appears in the Whittle SPDE generalisation. In addition,
the construction did not take into account the implications on the
spectral representation of the model itself, leading to a different
operator power, and lacking the $2k+1$ factor from
Section~\ref{sec:manifolds} in the spectrum-to-covariance
transformation. Thus, where the Whittle-Mat\'ern generalisation has
$\{\kappa^2+k(k+1)\}^\alpha$, the Legendre-Mat\'ern version has
$(2k+1)(\kappa^2+k^2)^{\alpha-1/2}$, which can not easily be
reformulated using the eigenvalues of the Laplacian. These issues make
that alternative generalisation of Mat\'ern models less natural, as it
loses all the Markov connections of the generalised Whittle
representation, as well as the simple form of the precision operator,
that can not be easily expressed via powers of the Laplacian. When
only considering the theoretically valid expressions for positive
definiteness, these kinds of effects are easily missed, and grounding
more complex model construction in geometrical locally interpretable
differential operators may provide better intuitive insight into the
theory.

\subsection{Global basis functions}

In Section~\ref{sec:FEM}, locally supported basis functions were used
in order to produce sparse precision matrices as well as retain this
sparsity when conditioning on measurements. In the other extreme,
global basis functions can be used, chosen so that the model precision
for the basis weights are diagonal, but the posterior precisions
generally dense. The Karhunen-Lo\`eve (K-L) expansion uses
eigenfunctions of the covariance operator, or equivalently of the
precision operator. The benefit of this type of finite dimensional
spectral representation is that it results in a close approximation to
the true covariance for a smaller number of basis functions. The main
downside is the computational cost of evaluating the eigenfunctions,
as these depend on the model parameters. Another choice is to use the
eigenfunctions of the Laplacian that are involved in the Fourier-like
spectral representations. These only depend on the shape of the
domain, and can be calculated numerically at the start of the analysis
\citep[see][for one such approach]{solin_hilbert_2020}. However, since
high frequencies may be needed to provide a good approximation of the
true model, this approach is still expensive for general domains. In
addition, the typical numerical solution is to use the same finite
element methods to solve for the eigenvalues as is used to compute the
entire conditional expectation. Hence, computing the eigenfunctions
numerically is only helpful if they can then be used more effectively
than the sparse precision matrices themselves. For both K-L and
harmonic basis functions, the Hilbert space inner product yields
diagonal prior precision matrices. The main problem in practical
applications appear when the data structure does not admit efficient
posterior distribution calculations due to the resulting dense
precision matrices, due to the structure of the
$\mv{A}^\top\mv{Q}_e\mv{A}$ term in \eqref{eq:posterior-Q}. This means
that harmonic basis functions are mostly useful for very smooth fields
that only need a few frequencies, or where fast Fourier transforms are
possible.

When using spherical harmonics on the sphere, care must be taken when
choosing a Legendre polynomial implementation, as some implementations
are numerically unstable for orders above $\sim 40$. For example,
methods that first construct the polynomial coefficients explicitly,
and then evaluate them, break down for orders above $\sim 40$,
including the implementations in the \texttt{pracma} and
\texttt{orthopolynom} packages
\citep{borchers_pracma_2021,novomestky_orthopolynom_2013}. With these
unstable implementations, the spatial resolution on the globe is
limited to wavelengths of about $360/40=9$ degrees, or
$\sim 1,000\,\text{km}$, making it unsuited for fine resolution
problems such as the EUSTACE project discussed in
Section~\ref{sec:EUSTACE}. The GSL library implementation
\citep{galassi2018scientific,hankin_r-gsl_2006} however appears to be
stable for much higher orders.

A numerical alternative is to compute the Laplacian eigenfunctions
on a triangulation, via the generalised eigenvalue problem
$\mv{G}\mv{V}=\mv{C}\mv{V}\mv{\Lambda}$. This closely approximates the
eigenfunctions of the continuous domain Laplacian, and works also on
other manifolds than the sphere. This would be a slight modification
of the directly graph based eigenfunctions used by
\citet{lee_picar_2021}.

\subsection{Other precision approximation methods}

Every computational approach to spatial processes can be viewed in (at
least) to ways; as an approximation of a given model, or as a
specially constructed model in itself. In the literature, both
viewpoints are used, but since the structure of the approximations can
be very specific to the details of the constructions, the second
approach does not provide useful insights into the difference and
similarities between the approaches. Instead, we find it more useful
to take the first approach, and consider the continuous domain
interpretability of the methods.

The Markov models resulting from the finite element Hilbert space
approach to representing SPDE solutions has close links to several
other methods for constructing computationally efficient
approximations to given continuous domain models.

Just as one of the motivations for developing the SPDE/GMRF link in
\citet{lindgren11} was to bridge the gap between discrete Markov
models and continuous covariance models, there are further links
between triangulation graphs for finite elements and other graph
methods. Two recent examples are \citet{Alonso2021} and
\citet{dunson_graph_2021}, that construct different local Laplacian
approximations on the same kind of graphs. The first of these two
papers also show the extensive connections between the approaches, and
how the finite element constructions provide better approximations to
the continuous domain models, in the settings where they are
available.

Another approach is the nearest neighbour Gaussian process (NNGP)
construction from \citet{datta_hierarchical_2016}, that takes a given
covariance function and, in essence, constructs an incomplete Cholesky
factorisation of the precision matrix for a given, ordered, sequence
of spatial locations. By computing the exact conditional distribution
given the previously included points, a discretised model is obtained.
The benefit is that, instead of including the Cholesky \emph{infill}
that appears in the Cholesky factorisation of a Markovian precision
matrix, only the previous neighbours are included. The downside is
that the resulting representation depends on the order in which the
spatial locations are included, which can lead to large discrepancies
between the source covariance and the NNGP covariance. In contrast,
for full Cholesky factorisations, the ordering of the nodes only
affects the sparsity of the computed factorisation
\citep{rue2005gaussian}. However, due to the relative speed of the
NNGP construction, a potential extension could be to use the NNGP
model as a fast preconditioner in large scale iterative solver
methods. This would also be applicable to block-wise constructions,
such as
\citet{katzfuss_multi-resolution_2017,art681,peruzzi_highly_2020}.


\section{Discussion}\label{sec:discussion}

After 10 years since the publication of \citet{lindgren11}, the
stochastic PDE method for constructing computationally efficient
representations of spatial and spatio temporal Gaussian random fields
has proven its value in a wide range of practical applications. There
is now a wide variety in computational methods for spatial statistics,
but many lack strong theoretical guarantees of fidelity in relation to
desired continuous domain properties, or are limited to specific
applications. In contrast, the SPDE approach exploits the close
connections between different representations of Gaussian processes
and their dependence structure, and the relative separation of the
model construction itself from the computational methods, providing
strong theoretical guarantees as well as efficient implementations.
This has allowed Gaussian random field models to be incorporated in a
variety of hierarchical models, and the method continues to be
expanded to new application domains and more general dependency
structures.

There is always a trade-off between generality and computational
efficiency. The SPDE methods have the advantage that methods can be
adapted from well-studied methods for deterministic PDE solvers, that
are already capable of solving large systems. Part of the challenge is
that the typical space-time precision operators for random fields are
of much higher order than the typical ordinary Laplacian that is
common in purely physics-motivated PDEs, which makes the standard
preconditioning methods for iterative solvers inefficient. However, by
exploiting local block structures, as well as multigrid methods and
parallel solvers, these methods show great promise for even larger
problems than can be solved with the current direct Cholesky methods,
that are useable up to a few million elements; the largest single
field solve in the EUSTACE project had nearly 2 million nodes.

Among the currently active areas of research are more general
space-time extensions, non-stationarity and anisotropic models, and
combinations thereof. We anticipate that this will serve to broaden
the practical usefulness of SPDE models to realistic models for
complex spatio-temporal data.

While it is still true that there is a practical implementation and
preprocessing cost associated with the methods, and that this is true
of all advanced spatial and spatio-temporal modelling techniques, the
closing remark from \citet{lindgren11} that ``such costs are
unavoidable when efficient computations are required'' is no longer as
true now as it was in 2011, from the software users' perspective.
Interfaces such as \texttt{inlabru} are able to encapsulate all the
internal bookkeeping and linear algebra in the internal code, allowing
the software users to focus more on building structured and
geometrically interpretable models.


\section*{Acknowledgements}
As part of the EUSTACE project, Finn Lindgren received funding from the European Union’s
“Horizon 2020 Programme for Research and Innovation”, under Grant Agreement no 640171.

\bibliographystyle{elsarticle-harv}
\bibliography{spde10years-combined}

\begin{thebibliography}{168}
\expandafter\ifx\csname natexlab\endcsname\relax\def\natexlab#1{#1}\fi
\providecommand{\url}[1]{\texttt{#1}}
\providecommand{\href}[2]{#2}
\providecommand{\path}[1]{#1}
\providecommand{\DOIprefix}{doi:}
\providecommand{\ArXivprefix}{arXiv:}
\providecommand{\URLprefix}{URL: }
\providecommand{\Pubmedprefix}{pmid:}
\providecommand{\doi}[1]{\href{http://dx.doi.org/#1}{\path{#1}}}
\providecommand{\Pubmed}[1]{\href{pmid:#1}{\path{#1}}}
\providecommand{\bibinfo}[2]{#2}
\ifx\xfnm\relax \def\xfnm[#1]{\unskip,\space#1}\fi
\bibitem[{Aquino et~al.(2021)Aquino, Castruccio, Gupta and Howard}]{bg:35}
\bibinfo{author}{Aquino, B.}, \bibinfo{author}{Castruccio, S.},
  \bibinfo{author}{Gupta, V.}, \bibinfo{author}{Howard, S.},
  \bibinfo{year}{2021}.
\newblock \bibinfo{title}{Spatial modeling of mid-infrared spectral data with
  thermal compensation using integrated nested {L}aplace approximation}.
\newblock \bibinfo{journal}{Appl. Opt.} \bibinfo{volume}{60},
  \bibinfo{pages}{8609--8615}.
\newblock \DOIprefix\doi{10.1364/AO.435918}.
\bibitem[{Asar et~al.(2020)Asar, Bolin, Diggle and Wallin}]{Asar2020}
\bibinfo{author}{Asar, O.}, \bibinfo{author}{Bolin, D.},
  \bibinfo{author}{Diggle, P.J.}, \bibinfo{author}{Wallin, J.},
  \bibinfo{year}{2020}.
\newblock \bibinfo{title}{Linear mixed effects models for non-{G}aussian
  continuous repeated measurement data}.
\newblock \bibinfo{journal}{J. R. Stat. Soc. Ser. C. Appl. Stat.}
  \bibinfo{volume}{69}, \bibinfo{pages}{1015--1065}.
\newblock \DOIprefix\doi{10.1111/rssc.12405}. \bibinfo{note}{with discussion
  and a reply by the authors}.
\bibitem[{Asri and Benamirouche(2021)}]{bg:7}
\bibinfo{author}{Asri, A.}, \bibinfo{author}{Benamirouche, R.},
  \bibinfo{year}{2021}.
\newblock \bibinfo{title}{Using {INLA}/{SPDE} approach for estimating a spatial
  model for lung cancer mortality in {Algeria} 2016}.
\newblock \bibinfo{journal}{Revue d{\'E}conomie et de Statistique
  Appliqu{\'e}e} \bibinfo{volume}{18}, \bibinfo{pages}{261--277}.
\bibitem[{Babyn et~al.(2021)Babyn, Varkey, Regular, Ings and {Mills
  Flemming}}]{bg:38}
\bibinfo{author}{Babyn, J.}, \bibinfo{author}{Varkey, D.},
  \bibinfo{author}{Regular, P.}, \bibinfo{author}{Ings, D.},
  \bibinfo{author}{{Mills Flemming}, J.}, \bibinfo{year}{2021}.
\newblock \bibinfo{title}{A {G}aussian field approach to generating spatial age
  length keys}.
\newblock \bibinfo{journal}{Fisheries Research} \bibinfo{volume}{240},
  \bibinfo{pages}{105956}.
\newblock \DOIprefix\doi{10.1016/j.fishres.2021.105956}.
\bibitem[{Bachl et~al.(2019)Bachl, Lindgren, Borchers and
  Illian}]{bachl_inlabru_2019}
\bibinfo{author}{Bachl, F.E.}, \bibinfo{author}{Lindgren, F.},
  \bibinfo{author}{Borchers, D.L.}, \bibinfo{author}{Illian, J.B.},
  \bibinfo{year}{2019}.
\newblock \bibinfo{title}{inlabru: an {R} package for {Bayesian} spatial
  modelling from ecological survey data}.
\newblock \bibinfo{journal}{Methods Ecol. Evol} \bibinfo{volume}{10},
  \bibinfo{pages}{760--766}.
\newblock \DOIprefix\doi{10.1111/2041-210X.13168}.
\bibitem[{Bakka(2019)}]{bakka2019solve}
\bibinfo{author}{Bakka, H.}, \bibinfo{year}{2019}.
\newblock \bibinfo{title}{How to solve the stochastic partial differential
  equation that gives a {M}atérn random field using the finite element
  method}.
\newblock \href{http://arxiv.org/abs/1803.03765}{{\tt arXiv:1803.03765}}.
\bibitem[{Bakka et~al.(2020)Bakka, Krainski, Bolin, Rue and
  Lindgren}]{bakka_diffusion-based_2020}
\bibinfo{author}{Bakka, H.}, \bibinfo{author}{Krainski, E.},
  \bibinfo{author}{Bolin, D.}, \bibinfo{author}{Rue, H.},
  \bibinfo{author}{Lindgren, F.}, \bibinfo{year}{2020}.
\newblock \bibinfo{title}{The diffusion-based extension of the {Mat}\'{e}rn
  field to space-time}.
\newblock \href{http://arxiv.org/abs/2006.04917}{{\tt arXiv:2006.04917}}.
\bibitem[{Bakka et~al.(2018)Bakka, Rue, Fuglstad, Riebler, Bolin, Illian,
  Krainski, Simpson and Lindgren}]{art643}
\bibinfo{author}{Bakka, H.}, \bibinfo{author}{Rue, H.},
  \bibinfo{author}{Fuglstad, G.A.}, \bibinfo{author}{Riebler, A.},
  \bibinfo{author}{Bolin, D.}, \bibinfo{author}{Illian, J.},
  \bibinfo{author}{Krainski, E.}, \bibinfo{author}{Simpson, D.},
  \bibinfo{author}{Lindgren, F.}, \bibinfo{year}{2018}.
\newblock \bibinfo{title}{Spatial modelling with {R-INLA}: {A} review}.
\newblock \bibinfo{journal}{Wiley Interdiscip. Rev. Comput. Stat.}
  \bibinfo{volume}{10:e1443}.
\newblock \DOIprefix\doi{10.1002/wics.1443}.
\bibitem[{Bakka et~al.(2019)Bakka, Vanhatalo, Illian, Simpson and
  Rue}]{bakka2019}
\bibinfo{author}{Bakka, H.}, \bibinfo{author}{Vanhatalo, J.},
  \bibinfo{author}{Illian, J.B.}, \bibinfo{author}{Simpson, D.},
  \bibinfo{author}{Rue, H.}, \bibinfo{year}{2019}.
\newblock \bibinfo{title}{Non-stationary {G}aussian models with physical
  barriers}.
\newblock \bibinfo{journal}{Spat.\ Stat.} \bibinfo{volume}{29},
  \bibinfo{pages}{268--288}.
\bibitem[{Barman and Bolin(2018)}]{Barman2018}
\bibinfo{author}{Barman, S.}, \bibinfo{author}{Bolin, D.},
  \bibinfo{year}{2018}.
\newblock \bibinfo{title}{A three-dimensional statistical model for imaged
  microstructures of porous polymer films}.
\newblock \bibinfo{journal}{J. Microsc.} \bibinfo{volume}{269},
  \bibinfo{pages}{247--258}.
\newblock \DOIprefix\doi{10.1111/jmi.12623}.
\bibitem[{Barman et~al.(2019)Barman, Rootz{\'e}n and Bolin}]{Barman2019}
\bibinfo{author}{Barman, S.}, \bibinfo{author}{Rootz{\'e}n, H.},
  \bibinfo{author}{Bolin, D.}, \bibinfo{year}{2019}.
\newblock \bibinfo{title}{Prediction of diffusive transport through polymer
  films from characteristics of the pore geometry}.
\newblock \bibinfo{journal}{AIChE Journal} \bibinfo{volume}{65},
  \bibinfo{pages}{446--457}.
\bibitem[{Bell et~al.(2021)Bell, Jones, Cunningham, {Ruiz-Aravena}, Hamilton,
  Comte, Hamede, Bearhop and McDonald}]{bg:30}
\bibinfo{author}{Bell, O.}, \bibinfo{author}{Jones, M.E.},
  \bibinfo{author}{Cunningham, C.X.}, \bibinfo{author}{{Ruiz-Aravena}, M.},
  \bibinfo{author}{Hamilton, D.G.}, \bibinfo{author}{Comte, S.},
  \bibinfo{author}{Hamede, R.K.}, \bibinfo{author}{Bearhop, S.},
  \bibinfo{author}{McDonald, R.A.}, \bibinfo{year}{2021}.
\newblock \bibinfo{title}{{I}sotopic niche variation in {T}asmanian devils
  {S}arcophilus harrisii with progression of devil facial tumor disease}.
\newblock \bibinfo{journal}{Ecology and Evolution} \bibinfo{volume}{11},
  \bibinfo{pages}{8038--8053}.
\newblock \DOIprefix\doi{10.1002/ece3.7636}.
\bibitem[{Beloconi et~al.(2021)Beloconi, {Probst-Hensch} and Vounatsou}]{bg:14}
\bibinfo{author}{Beloconi, A.}, \bibinfo{author}{{Probst-Hensch}, N.M.},
  \bibinfo{author}{Vounatsou, P.}, \bibinfo{year}{2021}.
\newblock \bibinfo{title}{Spatio-temporal modelling of changes in air pollution
  exposure associated to the {COVID}-19 lockdown measures across {Europe}}.
\newblock \bibinfo{journal}{Sci. Total Environ.} \bibinfo{volume}{787},
  \bibinfo{pages}{147607}.
\newblock \DOIprefix\doi{10.1016/j.scitotenv.2021.147607}.
\bibitem[{Berg et~al.(2021)Berg, Shirajee, Folkvord, Godiksen, Skaret and
  Slotte}]{bg:43}
\bibinfo{author}{Berg, F.}, \bibinfo{author}{Shirajee, S.},
  \bibinfo{author}{Folkvord, A.}, \bibinfo{author}{Godiksen, J.A.},
  \bibinfo{author}{Skaret, G.}, \bibinfo{author}{Slotte, A.},
  \bibinfo{year}{2021}.
\newblock \bibinfo{title}{Early life growth is affecting timing of spawning in
  the semelparous {B}arents {S}ea capelin ({M}allotus villosus)}.
\newblock \bibinfo{journal}{Prog. Oceanogr.} \bibinfo{volume}{196},
  \bibinfo{pages}{102614}.
\newblock \DOIprefix\doi{10.1016/j.pocean.2021.102614}.
\bibitem[{Bertozzi-Villa et~al.(2021)Bertozzi-Villa, Bever, Koenker, Weiss,
  Vargas-Ruiz, Nandi, Gibson, Harris, Battle, Rumisha, Keddie, Amratia,
  Arambepola, Cameron, Chestnutt, Collins, Millar, Mishra, Rozier, Symons,
  Twohig, Hollingsworth, Gething and Bhatt}]{bg:5}
\bibinfo{author}{Bertozzi-Villa, A.}, \bibinfo{author}{Bever, C.A.},
  \bibinfo{author}{Koenker, H.}, \bibinfo{author}{Weiss, D.J.},
  \bibinfo{author}{Vargas-Ruiz, C.}, \bibinfo{author}{Nandi, A.K.},
  \bibinfo{author}{Gibson, H.S.}, \bibinfo{author}{Harris, J.},
  \bibinfo{author}{Battle, K.E.}, \bibinfo{author}{Rumisha, S.F.},
  \bibinfo{author}{Keddie, S.}, \bibinfo{author}{Amratia, P.},
  \bibinfo{author}{Arambepola, R.}, \bibinfo{author}{Cameron, E.},
  \bibinfo{author}{Chestnutt, E.G.}, \bibinfo{author}{Collins, E.L.},
  \bibinfo{author}{Millar, J.}, \bibinfo{author}{Mishra, S.},
  \bibinfo{author}{Rozier, J.}, \bibinfo{author}{Symons, T.},
  \bibinfo{author}{Twohig, K.A.}, \bibinfo{author}{Hollingsworth, T.D.},
  \bibinfo{author}{Gething, P.W.}, \bibinfo{author}{Bhatt, S.},
  \bibinfo{year}{2021}.
\newblock \bibinfo{title}{Maps and metrics of insecticide-treated net access,
  use, and nets-per-capita in {Africa} from 2000-2020}.
\newblock \bibinfo{journal}{Nat. Commun.} \bibinfo{volume}{12},
  \bibinfo{pages}{3589}.
\newblock \DOIprefix\doi{10.1038/s41467-021-23707-7}.
\bibitem[{Besag(1974)}]{besag_spatial_1974}
\bibinfo{author}{Besag, J.}, \bibinfo{year}{1974}.
\newblock \bibinfo{title}{Spatial {Interaction} and the {Statistical}
  {Analysis} of {Lattice} {Systems}}.
\newblock \bibinfo{journal}{J.\ R.\ Stat.\ Soc.\ Ser.\ B Stat.\ Methodol.}
  \bibinfo{volume}{36}, \bibinfo{pages}{192--225}.
\newblock \DOIprefix\doi{10.1111/j.2517-6161.1974.tb00999.x}.
\bibitem[{Besag and Kooperberg(1995)}]{besag_conditional_1995}
\bibinfo{author}{Besag, J.}, \bibinfo{author}{Kooperberg, C.},
  \bibinfo{year}{1995}.
\newblock \bibinfo{title}{On {Conditional} and {Intrinsic} {Autoregression}}.
\newblock \bibinfo{journal}{Biometrika} \bibinfo{volume}{82},
  \bibinfo{pages}{733--746}.
\newblock \DOIprefix\doi{10.2307/2337341}.
\bibitem[{Besag and Mondal(2005)}]{besag_first-order_2005}
\bibinfo{author}{Besag, J.}, \bibinfo{author}{Mondal, D.},
  \bibinfo{year}{2005}.
\newblock \bibinfo{title}{First-{Order} {Intrinsic} {Autoregressions} and the
  {De} {Wijs} {Process}}.
\newblock \bibinfo{journal}{Biometrika} \bibinfo{volume}{92},
  \bibinfo{pages}{909--920}.
\newblock \URLprefix \url{https://www.jstor.org/stable/20441244}.
\bibitem[{Bhatt et~al.(2015)Bhatt, Weiss, Cameron, Bisanzio, Mappin, Dalrymple,
  Battle, Moyes, Henry, Eckhoff, Wenger, Briët, Penny, Smith, Bennett, Yukich,
  Eisele, Griffin, Fergus, Lynch, Lindgren, Cohen, Murray, Smith, Hay,
  Cibulskis and Gething}]{bhatt_effect_2015}
\bibinfo{author}{Bhatt, S.}, \bibinfo{author}{Weiss, D.J.},
  \bibinfo{author}{Cameron, E.}, \bibinfo{author}{Bisanzio, D.},
  \bibinfo{author}{Mappin, B.}, \bibinfo{author}{Dalrymple, U.},
  \bibinfo{author}{Battle, K.E.}, \bibinfo{author}{Moyes, C.L.},
  \bibinfo{author}{Henry, A.}, \bibinfo{author}{Eckhoff, P.A.},
  \bibinfo{author}{Wenger, E.A.}, \bibinfo{author}{Briët, O.},
  \bibinfo{author}{Penny, M.A.}, \bibinfo{author}{Smith, T.A.},
  \bibinfo{author}{Bennett, A.}, \bibinfo{author}{Yukich, J.},
  \bibinfo{author}{Eisele, T.P.}, \bibinfo{author}{Griffin, J.T.},
  \bibinfo{author}{Fergus, C.A.}, \bibinfo{author}{Lynch, M.},
  \bibinfo{author}{Lindgren, F.}, \bibinfo{author}{Cohen, J.M.},
  \bibinfo{author}{Murray, C.L.J.}, \bibinfo{author}{Smith, D.L.},
  \bibinfo{author}{Hay, S.I.}, \bibinfo{author}{Cibulskis, R.E.},
  \bibinfo{author}{Gething, P.W.}, \bibinfo{year}{2015}.
\newblock \bibinfo{title}{The effect of malaria control on {Plasmodium}
  falciparum in {Africa} between 2000 and 2015}.
\newblock \bibinfo{journal}{Nature} \bibinfo{volume}{526},
  \bibinfo{pages}{207--211}.
\newblock \DOIprefix\doi{10.1038/nature15535}.
\bibitem[{Bleuel et~al.(2021)Bleuel, Pennino and Longo}]{bg:20}
\bibinfo{author}{Bleuel, J.}, \bibinfo{author}{Pennino, M.},
  \bibinfo{author}{Longo, G.}, \bibinfo{year}{2021}.
\newblock \bibinfo{title}{Coral distribution and bleaching vulnerability areas
  in {Southwestern} {Atlantic} under ocean warming}.
\newblock \bibinfo{journal}{Scientific Reports} \bibinfo{volume}{11},
  \bibinfo{pages}{12833}.
\newblock \DOIprefix\doi{10.1038/s41598-021-92202-2}.
\bibitem[{Bolin(2012)}]{phd11}
\bibinfo{author}{Bolin, D.}, \bibinfo{year}{2012}.
\newblock \bibinfo{title}{Models and Methods for Random Fields in Spatial
  Statistics with Computational Efficiency from {M}arkov Properties}.
\newblock Ph.D. thesis. Faculty of Engineering, Centre for Mathematical
  Statistics, Lund University, Sweden.
\bibitem[{Bolin(2014)}]{bolin14}
\bibinfo{author}{Bolin, D.}, \bibinfo{year}{2014}.
\newblock \bibinfo{title}{Spatial {M}at\'{e}rn fields driven by non-{G}aussian
  noise}.
\newblock \bibinfo{journal}{Scand. J. Stat.} \bibinfo{volume}{41},
  \bibinfo{pages}{557--579}.
\newblock \DOIprefix\doi{10.1111/sjos.12046}.
\bibitem[{Bolin and Kirchner(2020)}]{BK2020rational}
\bibinfo{author}{Bolin, D.}, \bibinfo{author}{Kirchner, K.},
  \bibinfo{year}{2020}.
\newblock \bibinfo{title}{The rational {SPDE} approach for {G}aussian random
  fields with general smoothness}.
\newblock \bibinfo{journal}{J. Comp. Graph. Stat.} \bibinfo{volume}{29},
  \bibinfo{pages}{274--285}.
\bibitem[{Bolin and Kirchner(2021)}]{bk-measure}
\bibinfo{author}{Bolin, D.}, \bibinfo{author}{Kirchner, K.},
  \bibinfo{year}{2021}.
\newblock \bibinfo{title}{Equivalence of measures and asymptotically optimal
  linear prediction for gaussian random fields with fractional-order covariance
  operators}.
\newblock \bibinfo{note}{Preprint, arXiv:2101.07860}.
\bibitem[{Bolin et~al.(2018)Bolin, Kirchner and Kov\'{a}cs}]{BKK2018}
\bibinfo{author}{Bolin, D.}, \bibinfo{author}{Kirchner, K.},
  \bibinfo{author}{Kov\'{a}cs, M.}, \bibinfo{year}{2018}.
\newblock \bibinfo{title}{Weak convergence of {G}alerkin approximations for
  fractional elliptic stochastic {PDE}s with spatial white noise}.
\newblock \bibinfo{journal}{BIT} \bibinfo{volume}{58},
  \bibinfo{pages}{881--906}.
\bibitem[{Bolin et~al.(2020)Bolin, Kirchner and Kov\'{a}cs}]{BKK2020}
\bibinfo{author}{Bolin, D.}, \bibinfo{author}{Kirchner, K.},
  \bibinfo{author}{Kov\'{a}cs, M.}, \bibinfo{year}{2020}.
\newblock \bibinfo{title}{Numerical solution of fractional elliptic stochastic
  {PDE}s with spatial white noise}.
\newblock \bibinfo{journal}{IMA J. Numer. Anal.} \bibinfo{volume}{40},
  \bibinfo{pages}{1051--1073}.
\bibitem[{Bolin and Lindgren(2011)}]{bolin11}
\bibinfo{author}{Bolin, D.}, \bibinfo{author}{Lindgren, F.},
  \bibinfo{year}{2011}.
\newblock \bibinfo{title}{Spatial models generated by nested stochastic partial
  differential equations, with an application to global ozone mapping}.
\newblock \bibinfo{journal}{Ann.\ Appl.\ Stat.} \bibinfo{volume}{5},
  \bibinfo{pages}{523--550}.
\bibitem[{Bolin and Lindgren(2013)}]{bolin_comparison_2013}
\bibinfo{author}{Bolin, D.}, \bibinfo{author}{Lindgren, F.},
  \bibinfo{year}{2013}.
\newblock \bibinfo{title}{A comparison between {Markov} approximations and
  other methods for large spatial data sets}.
\newblock \bibinfo{journal}{Comput. Statist. Data Anal.} \bibinfo{volume}{61},
  \bibinfo{pages}{7--21}.
\newblock \DOIprefix\doi{10.1016/j.csda.2012.11.011}.
\bibitem[{Bolin and Simas(2021)}]{rSPDEvignette}
\bibinfo{author}{Bolin, D.}, \bibinfo{author}{Simas, A.B.},
  \bibinfo{year}{2021}.
\newblock \bibinfo{title}{{R-INLA} implementation of the covariance-based
  rational approximation}.
\newblock \URLprefix
  \url{https://davidbolin.github.io/rSPDE/articles/rspde_inla.html}.
\bibitem[{Bolin and Wallin(2020)}]{bw20}
\bibinfo{author}{Bolin, D.}, \bibinfo{author}{Wallin, J.},
  \bibinfo{year}{2020}.
\newblock \bibinfo{title}{Multivariate type {G} {M}at\'{e}rn stochastic partial
  differential equation random fields}.
\newblock \bibinfo{journal}{J. R. Stat. Soc. Ser. B. Stat. Methodol.}
  \bibinfo{volume}{82}, \bibinfo{pages}{215--239}.
\bibitem[{Bolin and Wallin(2021)}]{bolin2021efficient}
\bibinfo{author}{Bolin, D.}, \bibinfo{author}{Wallin, J.},
  \bibinfo{year}{2021}.
\newblock \bibinfo{title}{Efficient methods for {G}aussian {M}arkov random
  fields under constraints}, in: \bibinfo{booktitle}{Advances in Neural
  Information Processing Systems 34}, \bibinfo{publisher}{Curran Associates,
  Inc.}. pp. \bibinfo{pages}{1--13}.
\bibitem[{Boman et~al.(2021)Boman, de~Graaf, Kough, Izioka-Kuramae, Zuur, Smaal
  and Nagelkerke}]{ddi.13392}
\bibinfo{author}{Boman, E.M.}, \bibinfo{author}{de~Graaf, M.},
  \bibinfo{author}{Kough, A.S.}, \bibinfo{author}{Izioka-Kuramae, A.},
  \bibinfo{author}{Zuur, A.F.}, \bibinfo{author}{Smaal, A.},
  \bibinfo{author}{Nagelkerke, L.}, \bibinfo{year}{2021}.
\newblock \bibinfo{title}{Spatial dependency in abundance of {Q}ueen conch,
  {A}liger gigas, in the {C}aribbean, indicates the importance of surveying
  deep-water distributions}.
\newblock \bibinfo{journal}{Diversity and Distributions}
  \DOIprefix\doi{10.1111/ddi.13392}.
\bibitem[{Bonito and Pasciak(2015)}]{bonito2015}
\bibinfo{author}{Bonito, A.}, \bibinfo{author}{Pasciak, J.E.},
  \bibinfo{year}{2015}.
\newblock \bibinfo{title}{Numerical approximation of fractional powers of
  elliptic operators}.
\newblock \bibinfo{journal}{Math.\ Comp.} \bibinfo{volume}{84},
  \bibinfo{pages}{2083--2110}.
\bibitem[{Borchers(2021)}]{borchers_pracma_2021}
\bibinfo{author}{Borchers, H.W.}, \bibinfo{year}{2021}.
\newblock \bibinfo{title}{pracma: Practical Numerical Math Functions}.
\newblock \URLprefix \url{https://CRAN.R-project.org/package=pracma}.
  \bibinfo{note}{{R} package version 2.3.3}.
\bibitem[{{Borges da Silva} et~al.(2021){Borges da Silva}, Xavier and
  Faria}]{bg:26}
\bibinfo{author}{{Borges da Silva}, E.D.}, \bibinfo{author}{Xavier, A.},
  \bibinfo{author}{Faria, M.V.}, \bibinfo{year}{2021}.
\newblock \bibinfo{title}{Joint modeling of genetics and field variation in
  plant breeding trials using relationship and different spatial methods: A
  simulation study of accuracy and bias}.
\newblock \bibinfo{journal}{Agronomy} \bibinfo{volume}{11}.
\newblock \DOIprefix\doi{10.3390/agronomy11071397}.
\bibitem[{Borovitskiy et~al.(2020)Borovitskiy, Terenin, Mostowsky and
  Deisenroth}]{borovitskiy2020mat}
\bibinfo{author}{Borovitskiy, V.}, \bibinfo{author}{Terenin, A.},
  \bibinfo{author}{Mostowsky, P.}, \bibinfo{author}{Deisenroth, M.P.},
  \bibinfo{year}{2020}.
\newblock \bibinfo{title}{Mat\'{e}rn {Gaussian} processes on {Riemannian}
  manifolds}.
\newblock \href{http://arxiv.org/abs/2006.10160}{{\tt arXiv:2006.10160}}.
\bibitem[{Breivik et~al.(2021)Breivik, Aanes, S{\o}vik, Aglen, Mehl and
  Johnsen}]{bg:44}
\bibinfo{author}{Breivik, O.N.}, \bibinfo{author}{Aanes, F.},
  \bibinfo{author}{S{\o}vik, G.}, \bibinfo{author}{Aglen, A.},
  \bibinfo{author}{Mehl, S.}, \bibinfo{author}{Johnsen, E.},
  \bibinfo{year}{2021}.
\newblock \bibinfo{title}{{Predicting abundance indices in areas without
  coverage with a latent spatio-temporal {Gaussian} model}}.
\newblock \bibinfo{journal}{ICES J. Mar. Sci.}
  \DOIprefix\doi{10.1093/icesjms/fsab073}.
\bibitem[{Calhoun et~al.(2001)Calhoun, Adali, Pearlson and
  Pekar}]{calhoun2001method}
\bibinfo{author}{Calhoun, V.D.}, \bibinfo{author}{Adali, T.},
  \bibinfo{author}{Pearlson, G.D.}, \bibinfo{author}{Pekar, J.J.},
  \bibinfo{year}{2001}.
\newblock \bibinfo{title}{A method for making group inferences from functional
  {MRI} data using independent component analysis}.
\newblock \bibinfo{journal}{Human brain mapping} \bibinfo{volume}{14},
  \bibinfo{pages}{140--151}.
\bibitem[{Cameletti et~al.(2013)Cameletti, Lindgren, Simpson and
  Rue}]{cameletti_spatio-temporal_2013}
\bibinfo{author}{Cameletti, M.}, \bibinfo{author}{Lindgren, F.},
  \bibinfo{author}{Simpson, D.}, \bibinfo{author}{Rue, H.},
  \bibinfo{year}{2013}.
\newblock \bibinfo{title}{Spatio-temporal modeling of particulate matter
  concentration through the {SPDE} approach}.
\newblock \bibinfo{journal}{Adv. Stat. Anal.} \bibinfo{volume}{97},
  \bibinfo{pages}{109--131}.
\newblock \DOIprefix\doi{10.1007/s10182-012-0196-3}.
\bibitem[{Carrizo~Vergara et~al.(2022)Carrizo~Vergara, Allard and
  Dessasis}]{vergara2022}
\bibinfo{author}{Carrizo~Vergara, R.}, \bibinfo{author}{Allard, D.},
  \bibinfo{author}{Dessasis, N.}, \bibinfo{year}{2022}.
\newblock \bibinfo{title}{A general framework for {SPDE}-based stationary
  random fields}.
\newblock \bibinfo{journal}{Bernoulli} \bibinfo{volume}{28},
  \bibinfo{pages}{1--32}.
\newblock \DOIprefix\doi{10.3150/20-BEJ1317}.
\bibitem[{Cavieres et~al.(2021)Cavieres, Monnahan and Vehtari}]{bg:41}
\bibinfo{author}{Cavieres, J.}, \bibinfo{author}{Monnahan, C.C.},
  \bibinfo{author}{Vehtari, A.}, \bibinfo{year}{2021}.
\newblock \bibinfo{title}{Accounting for spatial dependence improves relative
  abundance estimates in a benthic marine species structured as a
  metapopulation}.
\newblock \bibinfo{journal}{Fisheries Research} \bibinfo{volume}{240},
  \bibinfo{pages}{105960}.
\newblock \DOIprefix\doi{10.1016/j.fishres.2021.105960}.
\bibitem[{Cendoya et~al.(2021)Cendoya, Hubel, Conesa and
  Vicent}]{Cendoya2021.04.01.438042}
\bibinfo{author}{Cendoya, M.}, \bibinfo{author}{Hubel, A.},
  \bibinfo{author}{Conesa, D.}, \bibinfo{author}{Vicent, A.},
  \bibinfo{year}{2021}.
\newblock \bibinfo{title}{Barrier effects on the spatial distribution of
  {X}ylella fastidiosa in {A}licante, {S}pain}.
\newblock \bibinfo{journal}{bioRxiv} \DOIprefix\doi{10.1101/2021.04.01.438042}.
\bibitem[{Chada et~al.(2021)Chada, Roininen and Suuronen}]{Chada2021}
\bibinfo{author}{Chada, N.K.}, \bibinfo{author}{Roininen, L.},
  \bibinfo{author}{Suuronen, J.}, \bibinfo{year}{2021}.
\newblock \bibinfo{title}{{Cauchy} {Markov} random field priors for {Bayesian}
  inversion}.
\newblock \bibinfo{note}{Preprint, arXiv:2105.12488}.
\bibitem[{Coveney et~al.(2020)Coveney, Corrado, Roney, Wilkinson, Oakley,
  Lindgren, Williams, O’Neill, Niederer and
  Clayton}]{coveney_probabilistic_2020}
\bibinfo{author}{Coveney, S.}, \bibinfo{author}{Corrado, C.},
  \bibinfo{author}{Roney, C.H.}, \bibinfo{author}{Wilkinson, R.D.},
  \bibinfo{author}{Oakley, J.E.}, \bibinfo{author}{Lindgren, F.},
  \bibinfo{author}{Williams, S.E.}, \bibinfo{author}{O’Neill, M.D.},
  \bibinfo{author}{Niederer, S.A.}, \bibinfo{author}{Clayton, R.H.},
  \bibinfo{year}{2020}.
\newblock \bibinfo{title}{Probabilistic {Interpolation} of {Uncertain} {Local}
  {Activation} {Times} on {Human} {Atrial} {Manifolds}}.
\newblock \bibinfo{journal}{IEEE Trans. Biomed. Eng.} \bibinfo{volume}{67},
  \bibinfo{pages}{99--109}.
\newblock \DOIprefix\doi{10.1109/TBME.2019.2908486}.
\bibitem[{Cox and Kirchner(2020)}]{cox2020}
\bibinfo{author}{Cox, S.G.}, \bibinfo{author}{Kirchner, K.},
  \bibinfo{year}{2020}.
\newblock \bibinfo{title}{Regularity and convergence analysis in {S}obolev and
  {H}\"older spaces for generalized {W}hittle--{M}atérn fields}.
\newblock \bibinfo{journal}{Numer.\ Math.} \bibinfo{volume}{146},
  \bibinfo{pages}{819--873}.
\bibitem[{Cram\'{e}r and Leadbetter(2004)}]{Cramer1967}
\bibinfo{author}{Cram\'{e}r, H.}, \bibinfo{author}{Leadbetter, M.R.},
  \bibinfo{year}{2004}.
\newblock \bibinfo{title}{Stationary and related stochastic processes}.
\newblock \bibinfo{publisher}{Dover Publications, Inc., Mineola, NY}.
\newblock \bibinfo{note}{Sample function properties and their applications,
  Reprint of the 1967 original}.
\bibitem[{Cressie and Wikle(2011)}]{cressie_statistics_2011}
\bibinfo{author}{Cressie, N.}, \bibinfo{author}{Wikle, C.K.},
  \bibinfo{year}{2011}.
\newblock \bibinfo{title}{Statistics for Spatio-Temporal Data}.
\newblock \bibinfo{publisher}{Wiley}.
\bibitem[{Datta et~al.(2016)Datta, Banerjee, Finley and
  Gelfand}]{datta_hierarchical_2016}
\bibinfo{author}{Datta, A.}, \bibinfo{author}{Banerjee, S.},
  \bibinfo{author}{Finley, A.O.}, \bibinfo{author}{Gelfand, A.E.},
  \bibinfo{year}{2016}.
\newblock \bibinfo{title}{Hierarchical {Nearest}-{Neighbor} {Gaussian}
  {Process} {Models} for {Large} {Geostatistical} {Datasets}}.
\newblock \bibinfo{journal}{J. Amer. Statist. Assoc.} \bibinfo{volume}{111},
  \bibinfo{pages}{800--812}.
\newblock \DOIprefix\doi{10.1080/01621459.2015.1044091}.
\bibitem[{Dunson et~al.(2021)Dunson, Wu and Wu}]{dunson_graph_2021}
\bibinfo{author}{Dunson, D.B.}, \bibinfo{author}{Wu, H.T.},
  \bibinfo{author}{Wu, N.}, \bibinfo{year}{2021}.
\newblock \bibinfo{title}{Graph based {Gaussian} processes on restricted
  domains}.
\newblock \href{http://arxiv.org/abs/2010.07242}{{\tt arXiv:2010.07242}}.
  \bibinfo{note}{accepted}.
\bibitem[{Eidsvik et~al.(2009)Eidsvik, Martino and Rue}]{art442}
\bibinfo{author}{Eidsvik, J.}, \bibinfo{author}{Martino, S.},
  \bibinfo{author}{Rue, H.}, \bibinfo{year}{2009}.
\newblock \bibinfo{title}{Approximate {B}ayesian inference in spatial
  generalized linear mixed models}.
\newblock \bibinfo{journal}{Scand. J. Stat.} \bibinfo{volume}{36},
  \bibinfo{pages}{1--22}.
\bibitem[{Eklund et~al.(2016)Eklund, Nichols and Knutsson}]{eklund2016cluster}
\bibinfo{author}{Eklund, A.}, \bibinfo{author}{Nichols, T.E.},
  \bibinfo{author}{Knutsson, H.}, \bibinfo{year}{2016}.
\newblock \bibinfo{title}{Cluster failure: Why {fMRI} inferences for spatial
  extent have inflated false-positive rates}.
\newblock \bibinfo{journal}{Proc. natl. acad. sci.} \bibinfo{volume}{113},
  \bibinfo{pages}{7900--7905}.
\bibitem[{Erisman and Tinney(1975)}]{art358}
\bibinfo{author}{Erisman, A.M.}, \bibinfo{author}{Tinney, W.F.},
  \bibinfo{year}{1975}.
\newblock \bibinfo{title}{On computing certain elements of the inverse of a
  sparse matrix}.
\newblock \bibinfo{journal}{Commun. {ACM}} \bibinfo{volume}{18},
  \bibinfo{pages}{177--179}.
\bibitem[{Fecchio et~al.()Fecchio, Clark, Bell, Skeen, Lutz, {De La Torre},
  Vaughan, Tkach, Schunck, Ferreira, Braga, Lugarini, Wamiti, Dispoto, Galen,
  Kirchgatter, Sagario, Cueto, {Gonz\'alez-Acu\~na}, Inumaru, Sato, Schumm,
  Quillfeldt, Pellegrino, Dharmarajan, Gupta, Robin, Ciloglu, Yildirim, Huang,
  {Chapa-Vargas}, {\'Alvarez-Mendiz\'abal}, {Santiago-Alarcon}, Drovetski,
  Hellgren, Voelker, Ricklefs, Hackett, Collins, Weckstein and Wells}]{bg:33}
\bibinfo{author}{Fecchio, A.}, \bibinfo{author}{Clark, N.J.},
  \bibinfo{author}{Bell, J.A.}, \bibinfo{author}{Skeen, H.R.},
  \bibinfo{author}{Lutz, H.L.}, \bibinfo{author}{{De La Torre}, G.M.},
  \bibinfo{author}{Vaughan, J.A.}, \bibinfo{author}{Tkach, V.V.},
  \bibinfo{author}{Schunck, F.}, \bibinfo{author}{Ferreira, F.C.},
  \bibinfo{author}{Braga, E.M.}, \bibinfo{author}{Lugarini, C.},
  \bibinfo{author}{Wamiti, W.}, \bibinfo{author}{Dispoto, J.H.},
  \bibinfo{author}{Galen, S.C.}, \bibinfo{author}{Kirchgatter, K.},
  \bibinfo{author}{Sagario, M.C.}, \bibinfo{author}{Cueto, V.R.},
  \bibinfo{author}{{Gonz\'alez-Acu\~na}, D.}, \bibinfo{author}{Inumaru, M.},
  \bibinfo{author}{Sato, Y.}, \bibinfo{author}{Schumm, Y.R.},
  \bibinfo{author}{Quillfeldt, P.}, \bibinfo{author}{Pellegrino, I.},
  \bibinfo{author}{Dharmarajan, G.}, \bibinfo{author}{Gupta, P.},
  \bibinfo{author}{Robin, V.V.}, \bibinfo{author}{Ciloglu, A.},
  \bibinfo{author}{Yildirim, A.}, \bibinfo{author}{Huang, X.},
  \bibinfo{author}{{Chapa-Vargas}, L.},
  \bibinfo{author}{{\'Alvarez-Mendiz\'abal}, P.},
  \bibinfo{author}{{Santiago-Alarcon}, D.}, \bibinfo{author}{Drovetski, S.V.},
  \bibinfo{author}{Hellgren, O.}, \bibinfo{author}{Voelker, G.},
  \bibinfo{author}{Ricklefs, R.E.}, \bibinfo{author}{Hackett, S.J.},
  \bibinfo{author}{Collins, M.D.}, \bibinfo{author}{Weckstein, J.D.},
  \bibinfo{author}{Wells, K.}, .
\newblock \bibinfo{title}{Global drivers of avian haemosporidian infections
  vary across zoogeographical regions}.
\newblock \bibinfo{journal}{Global Ecology and Biogeography}
  \DOIprefix\doi{10.1111/geb.13390}.
\bibitem[{Ferkingstad et~al.(2017)Ferkingstad, Held and Rue}]{art638}
\bibinfo{author}{Ferkingstad, E.}, \bibinfo{author}{Held, L.},
  \bibinfo{author}{Rue, H.}, \bibinfo{year}{2017}.
\newblock \bibinfo{title}{Fast and accurate {B}ayesian model criticism and
  conflict diagnostics using {R-INLA}}.
\newblock \bibinfo{journal}{Stat} \bibinfo{volume}{6},
  \bibinfo{pages}{331--344}.
\newblock \DOIprefix\doi{10.1002/sta4.163}.
\bibitem[{Fischl(2012)}]{fischl2012freesurfer}
\bibinfo{author}{Fischl, B.}, \bibinfo{year}{2012}.
\newblock \bibinfo{title}{Freesurfer}.
\newblock \bibinfo{journal}{Neuroimage} \bibinfo{volume}{62},
  \bibinfo{pages}{774--781}.
\bibitem[{Flor\^encio et~al.(2021)Flor\^encio, Alves, {Lansac-T\^oha}, Silveira
  and Thomaz}]{bg:21}
\bibinfo{author}{Flor\^encio, F.M.}, \bibinfo{author}{Alves, D.C.},
  \bibinfo{author}{{Lansac-T\^oha}, F.M.}, \bibinfo{author}{Silveira, M.J.},
  \bibinfo{author}{Thomaz, S.M.}, \bibinfo{year}{2021}.
\newblock \bibinfo{title}{The success of the invasive macrophyte {Hydrilla}
  verticillata and its interactions with the native {Egeria} najas in response
  to environmental factors and plant abundance in a subtropical reservoir}.
\newblock \bibinfo{journal}{Aquatic Botany} \bibinfo{volume}{175},
  \bibinfo{pages}{103432}.
\newblock \DOIprefix\doi{10.1016/j.aquabot.2021.103432}.
\bibitem[{Friston et~al.(1994)Friston, Holmes, Worsley, Poline, Frith and
  Frackowiak}]{friston1994statistical}
\bibinfo{author}{Friston, K.J.}, \bibinfo{author}{Holmes, A.P.},
  \bibinfo{author}{Worsley, K.J.}, \bibinfo{author}{Poline, J.P.},
  \bibinfo{author}{Frith, C.D.}, \bibinfo{author}{Frackowiak, R.S.},
  \bibinfo{year}{1994}.
\newblock \bibinfo{title}{Statistical parametric maps in functional imaging: a
  general linear approach}.
\newblock \bibinfo{journal}{Human brain mapping} \bibinfo{volume}{2},
  \bibinfo{pages}{189--210}.
\bibitem[{Fuglstad et~al.(2015a)Fuglstad, Lindgren, Simpson and
  Rue}]{fuglstad_exploring_2015}
\bibinfo{author}{Fuglstad, G.A.}, \bibinfo{author}{Lindgren, F.},
  \bibinfo{author}{Simpson, D.}, \bibinfo{author}{Rue, H.},
  \bibinfo{year}{2015}a.
\newblock \bibinfo{title}{Exploring a {New} {Class} of {Non}-stationary
  {Spatial} \{{Gaussian}\} random fields with varying local anisotropy}.
\newblock \bibinfo{journal}{Stat. Sin.} \bibinfo{volume}{25},
  \bibinfo{pages}{115--133}.
\newblock \URLprefix \url{https://www.jstor.org/stable/24311007}.
\bibitem[{Fuglstad et~al.(2015b)Fuglstad, Simpson, Lindgren and
  Rue}]{fuglstad2015non-stationary}
\bibinfo{author}{Fuglstad, G.A.}, \bibinfo{author}{Simpson, D.},
  \bibinfo{author}{Lindgren, F.}, \bibinfo{author}{Rue, H.},
  \bibinfo{year}{2015}b.
\newblock \bibinfo{title}{Does non-stationary spatial data always require
  non-stationary random fields?}
\newblock \bibinfo{journal}{Spat.\ Stat.} \bibinfo{volume}{14},
  \bibinfo{pages}{505--531}.
\newblock \DOIprefix\doi{10.1016/j.spasta.2015.10.001}.
\bibitem[{Fuglstad et~al.(2019)Fuglstad, Simpson, Lindgren and
  Rue}]{fuglstad_constructing_2019}
\bibinfo{author}{Fuglstad, G.A.}, \bibinfo{author}{Simpson, D.},
  \bibinfo{author}{Lindgren, F.}, \bibinfo{author}{Rue, H.},
  \bibinfo{year}{2019}.
\newblock \bibinfo{title}{Constructing {Priors} that {Penalize} the
  {Complexity} of {Gaussian} {Random} {Fields}}.
\newblock \bibinfo{journal}{J. Amer. Statist. Assoc.} \bibinfo{volume}{114},
  \bibinfo{pages}{445--452}.
\newblock \DOIprefix\doi{10.1080/01621459.2017.1415907}.
\bibitem[{Galassi(2018)}]{galassi2018scientific}
\bibinfo{author}{Galassi, M.e.a.}, \bibinfo{year}{2018}.
\newblock \bibinfo{title}{{GNU} {Scientific} {Library} reference manual}.
\newblock \URLprefix \url{https://www.gnu.org/software/gsl/}.
\bibitem[{Ghattas and Willcox(2021)}]{bg:9}
\bibinfo{author}{Ghattas, O.}, \bibinfo{author}{Willcox, K.},
  \bibinfo{year}{2021}.
\newblock \bibinfo{title}{Learning physics-based models from data: perspectives
  from inverse problems and model reduction}.
\newblock \bibinfo{journal}{Acta Numerica} \bibinfo{volume}{30},
  \bibinfo{pages}{445–554}.
\newblock \DOIprefix\doi{10.1017/S0962492921000064}.
\bibitem[{Gneiting and Raftery(2007)}]{gneiting_strictly_2007}
\bibinfo{author}{Gneiting, T.}, \bibinfo{author}{Raftery, A.E.},
  \bibinfo{year}{2007}.
\newblock \bibinfo{title}{Strictly {Proper} {Scoring} {Rules}, {Prediction},
  and {Estimation}}.
\newblock \bibinfo{journal}{J. Amer. Statist. Assoc.} \bibinfo{volume}{102},
  \bibinfo{pages}{359--378}.
\newblock \DOIprefix\doi{10.1198/016214506000001437}.
\bibitem[{G{\'o}mez-Catas{\'u}s et~al.(2021)G{\'o}mez-Catas{\'u}s, Barrero,
  Reverter, Bustillo-de~la Rosa, P{\'e}rez-Granados and Traba}]{bg:18}
\bibinfo{author}{G{\'o}mez-Catas{\'u}s, J.}, \bibinfo{author}{Barrero, A.},
  \bibinfo{author}{Reverter, M.}, \bibinfo{author}{Bustillo-de~la Rosa, D.},
  \bibinfo{author}{P{\'e}rez-Granados, C.}, \bibinfo{author}{Traba, J.},
  \bibinfo{year}{2021}.
\newblock \bibinfo{title}{Landscape features associated to wind farms increase
  mammalian predator abundance and ground-nest predation}.
\newblock \bibinfo{journal}{Biodivers. Conserv.} \bibinfo{volume}{30},
  \bibinfo{pages}{2581--2604}.
\newblock \DOIprefix\doi{10.1007/s10531-021-02212-9}.
\bibitem[{Griffiths and {Lezama-Ochoa}(2021)}]{bg:47}
\bibinfo{author}{Griffiths, S.P.}, \bibinfo{author}{{Lezama-Ochoa}, N.},
  \bibinfo{year}{2021}.
\newblock \bibinfo{title}{A 40-year chronology of the vulnerability of
  spinetail devil ray ({M}obula mobular) to eastern {P}acific tuna fisheries
  and options for future conservation and management}.
\newblock \bibinfo{journal}{Aquat. Conserv.} \DOIprefix\doi{10.1002/aqc.3667}.
\bibitem[{Guinness and Fuentes(2016)}]{guinness_isotropic_2016}
\bibinfo{author}{Guinness, J.}, \bibinfo{author}{Fuentes, M.},
  \bibinfo{year}{2016}.
\newblock \bibinfo{title}{Isotropic covariance functions on spheres: {Some}
  properties and modeling considerations}.
\newblock \bibinfo{journal}{J. Multivar. Anal.} \bibinfo{volume}{143},
  \bibinfo{pages}{143--152}.
\newblock \DOIprefix\doi{10.1016/j.jmva.2015.08.018}.
\bibitem[{Hankin(2006)}]{hankin_r-gsl_2006}
\bibinfo{author}{Hankin, R.K.S.}, \bibinfo{year}{2006}.
\newblock \bibinfo{title}{Special functions in {R}: introducing the gsl
  package}.
\newblock \bibinfo{journal}{R News} \bibinfo{volume}{6}.
\bibitem[{Harbrecht et~al.(2021)Harbrecht, Herrmann, Kirchner and
  Schwab}]{Harbrecht2021}
\bibinfo{author}{Harbrecht, H.}, \bibinfo{author}{Herrmann, L.},
  \bibinfo{author}{Kirchner, K.}, \bibinfo{author}{Schwab, C.},
  \bibinfo{year}{2021}.
\newblock \bibinfo{title}{Multilevel approximation of {Gaussian} random fields:
  Covariance compression, estimation and spatial prediction}.
\newblock \bibinfo{note}{Preprint, arXiv:2103.04424}.
\bibitem[{Herrmann et~al.(2020)Herrmann, Kirchner and Schwab}]{Herrmann2020}
\bibinfo{author}{Herrmann, L.}, \bibinfo{author}{Kirchner, K.},
  \bibinfo{author}{Schwab, C.}, \bibinfo{year}{2020}.
\newblock \bibinfo{title}{Multilevel approximation of {G}aussian random fields:
  fast simulation}.
\newblock \bibinfo{journal}{Math. Models Methods Appl. Sci.}
  \bibinfo{volume}{30}, \bibinfo{pages}{181--223}.
\bibitem[{Hildeman et~al.(2021)Hildeman, Bolin and Rychlik}]{Hildeman2020}
\bibinfo{author}{Hildeman, A.}, \bibinfo{author}{Bolin, D.},
  \bibinfo{author}{Rychlik, I.}, \bibinfo{year}{2021}.
\newblock \bibinfo{title}{Deformed {SPDE} models with an application to spatial
  modeling of significant wave height}.
\newblock \bibinfo{journal}{Spat.\ Stat.} \bibinfo{volume}{42},
  \bibinfo{pages}{100449}.
\bibitem[{Hough et~al.(2021)Hough, Sarafian, Shtein, Zhou, Lepeule and
  Kloog}]{bg:23}
\bibinfo{author}{Hough, I.}, \bibinfo{author}{Sarafian, R.},
  \bibinfo{author}{Shtein, A.}, \bibinfo{author}{Zhou, B.},
  \bibinfo{author}{Lepeule, J.}, \bibinfo{author}{Kloog, I.},
  \bibinfo{year}{2021}.
\newblock \bibinfo{title}{{Gaussian} {Markov} random fields improve ensemble
  predictions of daily 1 km {PM2.5} and p{M10} across {France}}.
\newblock \bibinfo{journal}{Atmos. Environ.} \bibinfo{volume}{264},
  \bibinfo{pages}{118693}.
\newblock \DOIprefix\doi{10.1016/j.atmosenv.2021.118693}.
\bibitem[{Humphreys et~al.()Humphreys, Douglas, Ramey, Mullinax, Soos, Link,
  Walther and Prosser}]{bg:31}
\bibinfo{author}{Humphreys, J.M.}, \bibinfo{author}{Douglas, D.C.},
  \bibinfo{author}{Ramey, A.M.}, \bibinfo{author}{Mullinax, J.M.},
  \bibinfo{author}{Soos, C.}, \bibinfo{author}{Link, P.},
  \bibinfo{author}{Walther, P.}, \bibinfo{author}{Prosser, D.J.}, .
\newblock \bibinfo{title}{The spatial–temporal relationship of blue-winged
  teal to domestic poultry: Movement state modelling of a highly mobile avian
  influenza host}.
\newblock \bibinfo{journal}{J. Appl. Ecol.}
  \DOIprefix\doi{10.1111/1365-2664.13963}.
\bibitem[{Ingebrigtsen et~al.(2014)Ingebrigtsen, Lindgren and
  Steinsland}]{ingebrigtsen_spatial_2014}
\bibinfo{author}{Ingebrigtsen, R.}, \bibinfo{author}{Lindgren, F.},
  \bibinfo{author}{Steinsland, I.}, \bibinfo{year}{2014}.
\newblock \bibinfo{title}{Spatial models with explanatory variables in the
  dependence structure}.
\newblock \bibinfo{journal}{Spat.\ Stat.} \bibinfo{volume}{8},
  \bibinfo{pages}{20--38}.
\newblock \DOIprefix\doi{10.1016/j.spasta.2013.06.002}.
\bibitem[{Jarvis et~al.(2021)Jarvis, McKenna and Rahseed}]{bg:40}
\bibinfo{author}{Jarvis, J.C.}, \bibinfo{author}{McKenna, S.A.},
  \bibinfo{author}{Rahseed, M.A.}, \bibinfo{year}{2021}.
\newblock \bibinfo{title}{Seagrass seed bank spatial structure and function
  following a large-scale decline}.
\newblock \bibinfo{journal}{Mar. Ecol. Prog. Ser.} \bibinfo{volume}{665},
  \bibinfo{pages}{75--87}.
\newblock \DOIprefix\doi{10.3354/meps13668}.
\bibitem[{Katzfuss(2017)}]{katzfuss_multi-resolution_2017}
\bibinfo{author}{Katzfuss, M.}, \bibinfo{year}{2017}.
\newblock \bibinfo{title}{A {Multi}-{Resolution} {Approximation} for {Massive}
  {Spatial} {Datasets}}.
\newblock \bibinfo{journal}{J. Amer. Statist. Assoc.} \bibinfo{volume}{112},
  \bibinfo{pages}{201--214}.
\newblock \DOIprefix\doi{10.1080/01621459.2015.1123632}.
\bibitem[{Khristenko et~al.(2019)Khristenko, Scarabosio, Swierczynski, Ullmann
  and Wohlmuth}]{Ullmann2019}
\bibinfo{author}{Khristenko, U.}, \bibinfo{author}{Scarabosio, L.},
  \bibinfo{author}{Swierczynski, P.}, \bibinfo{author}{Ullmann, E.},
  \bibinfo{author}{Wohlmuth, B.}, \bibinfo{year}{2019}.
\newblock \bibinfo{title}{Analysis of boundary effects on {PDE}-based sampling
  of {W}hittle-{M}at\'{e}rn random fields}.
\newblock \bibinfo{journal}{SIAM/ASA J. Uncertain. Quantif.}
  \bibinfo{volume}{7}, \bibinfo{pages}{948--974}.
\newblock \DOIprefix\doi{10.1137/18M1215700}.
\bibitem[{Kimeldorf and Wahba(1970)}]{kimeldorf_correspondence_1970}
\bibinfo{author}{Kimeldorf, G.S.}, \bibinfo{author}{Wahba, G.},
  \bibinfo{year}{1970}.
\newblock \bibinfo{title}{A {Correspondence} {Between} {Bayesian} {Estimation}
  on {Stochastic} {Processes} and {Smoothing} by {Splines}}.
\newblock \bibinfo{journal}{Ann.\ Math.\ Stat.} \bibinfo{volume}{41},
  \bibinfo{pages}{495--502}.
\newblock \URLprefix \url{https://www.jstor.org/stable/2239347}.
\bibitem[{Kirchner and Bolin(2021)}]{kb-kriging}
\bibinfo{author}{Kirchner, K.}, \bibinfo{author}{Bolin, D.},
  \bibinfo{year}{2021}.
\newblock \bibinfo{title}{Necessary and sufficient conditions for
  asymptotically optimal linear prediction of random fields on compact metric
  spaces}.
\newblock \bibinfo{journal}{Ann.\ Stat.} \bibinfo{note}{To appear}.
\bibitem[{Knorr-Held and Rue(2002)}]{art196}
\bibinfo{author}{Knorr-Held, L.}, \bibinfo{author}{Rue, H.},
  \bibinfo{year}{2002}.
\newblock \bibinfo{title}{On block updating in {M}arkov random field models for
  disease mapping}.
\newblock \bibinfo{journal}{Scand. J. Stat.} \bibinfo{volume}{29},
  \bibinfo{pages}{597--614}.
\bibitem[{Krainski et~al.(2018)Krainski, G{\'o}mez-Rubio, Bakka, Lenzi,
  Castro-Camilio, Simpson, Lindgren and Rue}]{book126}
\bibinfo{author}{Krainski, E.T.}, \bibinfo{author}{G{\'o}mez-Rubio, V.},
  \bibinfo{author}{Bakka, H.}, \bibinfo{author}{Lenzi, A.},
  \bibinfo{author}{Castro-Camilio, D.}, \bibinfo{author}{Simpson, D.},
  \bibinfo{author}{Lindgren, F.}, \bibinfo{author}{Rue, H.},
  \bibinfo{year}{2018}.
\newblock \bibinfo{title}{Advanced Spatial Modeling with Stochastic Partial
  Differential Equations using {R} and {INLA}}.
\newblock \bibinfo{publisher}{CRC press}.
\newblock \bibinfo{note}{Gitbook version
  https://becarioprecario.bitbucket.io/spde-gitbook/}.
\bibitem[{Lang and Pereira(2021)}]{bg:11}
\bibinfo{author}{Lang, A.}, \bibinfo{author}{Pereira, M.},
  \bibinfo{year}{2021}.
\newblock \bibinfo{title}{{Galerkin}--{Chebyshev} approximation of {Gaussian}
  random fields on compact {Riemannian} manifolds}.
\newblock \href{http://arxiv.org/abs/2107.02667}{{\tt arXiv:2107.02667}}.
\bibitem[{Lang and Schwab(2015)}]{Lang2015}
\bibinfo{author}{Lang, A.}, \bibinfo{author}{Schwab, C.}, \bibinfo{year}{2015}.
\newblock \bibinfo{title}{Isotropic {G}aussian random fields on the sphere:
  regularity, fast simulation and stochastic partial differential equations}.
\newblock \bibinfo{journal}{Ann. Appl. Probab.} \bibinfo{volume}{25},
  \bibinfo{pages}{3047--3094}.
\newblock \DOIprefix\doi{10.1214/14-AAP1067}.
\bibitem[{Lauritzen(1996)}]{book24}
\bibinfo{author}{Lauritzen, S.L.}, \bibinfo{year}{1996}.
\newblock \bibinfo{title}{Graphical Models}. volume~\bibinfo{volume}{17} of
  \textit{\bibinfo{series}{Oxford Statistical Science Series}}.
\newblock \bibinfo{publisher}{The Clarendon Press Oxford University Press},
  \bibinfo{address}{New York}.
\newblock \bibinfo{note}{Oxford Science Publications}.
\bibitem[{Lee et~al.(2021)Lee, Arkhipkin and Randhawa}]{bg:45}
\bibinfo{author}{Lee, B.}, \bibinfo{author}{Arkhipkin, A.},
  \bibinfo{author}{Randhawa, H.S.}, \bibinfo{year}{2021}.
\newblock \bibinfo{title}{Environmental drivers of {P}atagonian toothfish
  ({D}issostichus eleginoides) spatial-temporal patterns during an ontogenetic
  migration on the {P}atagonian {S}helf}.
\newblock \bibinfo{journal}{Estuar. Coast. Shelf Sci.} \bibinfo{volume}{259},
  \bibinfo{pages}{107473}.
\newblock \DOIprefix\doi{10.1016/j.ecss.2021.107473}.
\bibitem[{Lee and Haran(2021)}]{lee_picar_2021}
\bibinfo{author}{Lee, B.S.}, \bibinfo{author}{Haran, M.}, \bibinfo{year}{2021}.
\newblock \bibinfo{title}{{PICAR}: {An} efficient extendable approach for
  fitting hierarchical spatial models}.
\newblock \bibinfo{journal}{Technometrics} \bibinfo{volume}{0},
  \bibinfo{pages}{1--12}.
\newblock \DOIprefix\doi{10.1080/00401706.2021.1933596}.
\bibitem[{Levis et~al.(2021)Levis, Lee, Tropp, Gammie and Bouman}]{bg:1}
\bibinfo{author}{Levis, A.}, \bibinfo{author}{Lee, D.}, \bibinfo{author}{Tropp,
  J.A.}, \bibinfo{author}{Gammie, C.F.}, \bibinfo{author}{Bouman, K.L.},
  \bibinfo{year}{2021}.
\newblock \bibinfo{title}{Inference of black hole fluid-dynamics from sparse
  interferometric measurements}.
\newblock \bibinfo{journal}{Proceedings of IEEE International Conference on
  Computer Vision 2021} \bibinfo{note}{To appear}.
\bibitem[{Li(2021)}]{bg:34}
\bibinfo{author}{Li, D.}, \bibinfo{year}{2021}.
\newblock \bibinfo{title}{Urban planning image feature enhancement and
  simulation based on partial differential equation method}.
\newblock \bibinfo{journal}{Adv. Math. Phys.} \bibinfo{volume}{2021},
  \bibinfo{pages}{1700287}.
\newblock \DOIprefix\doi{10.1155/2021/1700287}.
\bibitem[{Lindenmayer et~al.()Lindenmayer, Taylor and Blanchard}]{bg:37}
\bibinfo{author}{Lindenmayer, D.}, \bibinfo{author}{Taylor, C.},
  \bibinfo{author}{Blanchard, W.}, .
\newblock \bibinfo{title}{Empirical analyses of the factors influencing fire
  severity in southeastern {A}ustralia}.
\newblock \bibinfo{journal}{Ecosphere} \bibinfo{volume}{12},
  \bibinfo{pages}{e03721}.
\newblock \DOIprefix\doi{10.1002/ecs2.3721}.
\bibitem[{Lindgren and Rue(2008)}]{art435}
\bibinfo{author}{Lindgren, F.}, \bibinfo{author}{Rue, H.},
  \bibinfo{year}{2008}.
\newblock \bibinfo{title}{A note on the second order random walk model for
  irregular locations}.
\newblock \bibinfo{journal}{Scand. J. Stat.} \bibinfo{volume}{35},
  \bibinfo{pages}{691--700}.
\bibitem[{Lindgren and Rue(2015)}]{art527}
\bibinfo{author}{Lindgren, F.}, \bibinfo{author}{Rue, H.},
  \bibinfo{year}{2015}.
\newblock \bibinfo{title}{Bayesian spatial modelling with {R-INLA}}.
\newblock \bibinfo{journal}{J. Stat. Softw.} \bibinfo{volume}{63},
  \bibinfo{pages}{1--25}.
\bibitem[{Lindgren et~al.(2011)Lindgren, Rue and Lindstr\"{o}m}]{lindgren11}
\bibinfo{author}{Lindgren, F.}, \bibinfo{author}{Rue, H.},
  \bibinfo{author}{Lindstr\"{o}m, J.}, \bibinfo{year}{2011}.
\newblock \bibinfo{title}{An explicit link between {G}aussian fields and
  {G}aussian {M}arkov random fields: the stochastic partial differential
  equation approach}.
\newblock \bibinfo{journal}{J.\ R.\ Stat.\ Soc.\ Ser.\ B Stat.\ Methodol.}
  \bibinfo{volume}{73}, \bibinfo{pages}{423--498}.
\newblock \bibinfo{note}{With discussion and a reply by the authors}.
\bibitem[{Lindgren(2012)}]{LindgrenStocProc}
\bibinfo{author}{Lindgren, G.}, \bibinfo{year}{2012}.
\newblock \bibinfo{title}{Stationary Stochastic Processes: Theory and
  Applications}.
\newblock \bibinfo{publisher}{CRC Press, Chapman and Hall}.
\bibitem[{Liu et~al.(2016)Liu, Guillas and Lai}]{liu_efficient_2016}
\bibinfo{author}{Liu, X.}, \bibinfo{author}{Guillas, S.}, \bibinfo{author}{Lai,
  M.J.}, \bibinfo{year}{2016}.
\newblock \bibinfo{title}{Efficient {Spatial} {Modeling} {Using} the {SPDE}
  {Approach} {With} {Bivariate} {Splines}}.
\newblock \bibinfo{journal}{J. Comp. Graph. Stat.} \bibinfo{volume}{25},
  \bibinfo{pages}{1176--1194}.
\newblock \DOIprefix\doi{10.1080/10618600.2015.1081597}.
\bibitem[{Mannseth et~al.(2021)Mannseth, Berentsen, Skaug, Lie and
  Moster}]{bg:2}
\bibinfo{author}{Mannseth, J.}, \bibinfo{author}{Berentsen, G.D.},
  \bibinfo{author}{Skaug, H.J.}, \bibinfo{author}{Lie, R.T.},
  \bibinfo{author}{Moster, D.}, \bibinfo{year}{2021}.
\newblock \bibinfo{title}{Variation in use of {C}aesarean section in {N}orway:
  An application of spatio-temporal {Gaussian} random fields}.
\newblock \bibinfo{journal}{Scand. J. Public Health}
  \DOIprefix\doi{10.1177/140349482008579}.
\bibitem[{Martino et~al.(2021)Martino, Pace, Moro, Casoli, Ventura, Frachea,
  Silvestri, Arcangeli, Giacomini, Ardizzone and Lasinio}]{bg:27}
\bibinfo{author}{Martino, S.}, \bibinfo{author}{Pace, D.S.},
  \bibinfo{author}{Moro, S.}, \bibinfo{author}{Casoli, E.},
  \bibinfo{author}{Ventura, D.}, \bibinfo{author}{Frachea, A.},
  \bibinfo{author}{Silvestri, M.}, \bibinfo{author}{Arcangeli, A.},
  \bibinfo{author}{Giacomini, G.}, \bibinfo{author}{Ardizzone, G.},
  \bibinfo{author}{Lasinio, G.J.}, \bibinfo{year}{2021}.
\newblock \bibinfo{title}{Integration of presence-only data from several
  sources. a case study on dolphins' spatial distribution}.
\newblock \href{http://arxiv.org/abs/2103.16125}{{\tt arXiv:2103.16125}}.
\bibitem[{Martins et~al.(2013)Martins, Simpson, Lindgren and Rue}]{art522}
\bibinfo{author}{Martins, T.G.}, \bibinfo{author}{Simpson, D.},
  \bibinfo{author}{Lindgren, F.}, \bibinfo{author}{Rue, H.},
  \bibinfo{year}{2013}.
\newblock \bibinfo{title}{Bayesian computing with {INLA}: {N}ew features}.
\newblock \bibinfo{journal}{Comput. Statist. Data Anal.} \bibinfo{volume}{67},
  \bibinfo{pages}{68--83}.
\bibitem[{Mat\'{e}rn(1960)}]{matern60}
\bibinfo{author}{Mat\'{e}rn, B.}, \bibinfo{year}{1960}.
\newblock \bibinfo{title}{Spatial variation: {S}tochastic models and their
  application to some problems in forest surveys and other sampling
  investigations}.
\newblock \bibinfo{publisher}{Meddelanden Fr{\aa}n Statens
  Skogsforskningsinstitut, Band 49, Nr.\ 5, Stockholm}.
\bibitem[{Maynou et~al.(2021)Maynou, Monfort, Morley and {Ord\'o\~nez}}]{bg:25}
\bibinfo{author}{Maynou, L.}, \bibinfo{author}{Monfort, M.},
  \bibinfo{author}{Morley, B.}, \bibinfo{author}{{Ord\'o\~nez}, J.},
  \bibinfo{year}{2021}.
\newblock \bibinfo{title}{Club convergence in {European} housing prices: The
  role of macroeconomic and housing market fundamentals}.
\newblock \bibinfo{journal}{Economic Modelling} \bibinfo{volume}{103},
  \bibinfo{pages}{105595}.
\newblock \DOIprefix\doi{10.1016/j.econmod.2021.105595}.
\bibitem[{Mejia et~al.(2020a)Mejia, Bolin, Yue, Wang, Caffo and
  Nebel}]{mejia2020spatial}
\bibinfo{author}{Mejia, A.F.}, \bibinfo{author}{Bolin, D.},
  \bibinfo{author}{Yue, Y.R.}, \bibinfo{author}{Wang, J.},
  \bibinfo{author}{Caffo, B.S.}, \bibinfo{author}{Nebel, M.B.},
  \bibinfo{year}{2020}a.
\newblock \bibinfo{title}{A spatial template independent component analysis
  model for subject-level brain network estimation and inference}.
\newblock \bibinfo{journal}{arXiv preprint arXiv:2005.13388} .
\bibitem[{Mejia et~al.(2020b)Mejia, Yue, Bolin, Lindgren and
  Lindquist}]{Mejia2020fmri}
\bibinfo{author}{Mejia, A.F.}, \bibinfo{author}{Yue, Y.},
  \bibinfo{author}{Bolin, D.}, \bibinfo{author}{Lindgren, F.},
  \bibinfo{author}{Lindquist, M.A.}, \bibinfo{year}{2020}b.
\newblock \bibinfo{title}{A {B}ayesian general linear modeling approach to
  cortical surface f{MRI} data analysis}.
\newblock \bibinfo{journal}{J. Amer. Statist. Assoc.} \bibinfo{volume}{115},
  \bibinfo{pages}{501--520}.
\newblock \DOIprefix\doi{10.1080/01621459.2019.1611582}.
\bibitem[{Miller et~al.(2020)Miller, Glennie and
  Seaton}]{miller_understanding_2020}
\bibinfo{author}{Miller, D.L.}, \bibinfo{author}{Glennie, R.},
  \bibinfo{author}{Seaton, A.E.}, \bibinfo{year}{2020}.
\newblock \bibinfo{title}{Understanding the {Stochastic} {Partial}
  {Differential} {Equation} {Approach} to {Smoothing}}.
\newblock \bibinfo{journal}{J. Agric. Biol. Environ. Stat.}
  \bibinfo{volume}{25}, \bibinfo{pages}{1--16}.
\newblock \DOIprefix\doi{10.1007/s13253-019-00377-z}.
\bibitem[{Miller et~al.(2021)Miller, Rexstad, Burt, Bravington and
  Hedley.}]{miller_dsm_2021}
\bibinfo{author}{Miller, D.L.}, \bibinfo{author}{Rexstad, E.},
  \bibinfo{author}{Burt, L.}, \bibinfo{author}{Bravington, M.V.},
  \bibinfo{author}{Hedley., S.}, \bibinfo{year}{2021}.
\newblock \bibinfo{title}{dsm: Density Surface Modelling of Distance Sampling
  Data}.
\newblock \URLprefix \url{https://CRAN.R-project.org/package=dsm}.
  \bibinfo{note}{{R} package version 2.3.1}.
\bibitem[{Monnahan et~al.(2021)Monnahan, Thorson, Kotwicki, Lauffenburger,
  Ianelli and Punt}]{bg:42}
\bibinfo{author}{Monnahan, C.C.}, \bibinfo{author}{Thorson, J.T.},
  \bibinfo{author}{Kotwicki, S.}, \bibinfo{author}{Lauffenburger, N.},
  \bibinfo{author}{Ianelli, J.N.}, \bibinfo{author}{Punt, A.E.},
  \bibinfo{year}{2021}.
\newblock \bibinfo{title}{Incorporating vertical distribution in index
  standardization accounts for spatiotemporal availability to acoustic and
  bottom trawl gear for semi-pelagic species}.
\newblock \bibinfo{journal}{ICES J. Mar. Sci.} \bibinfo{volume}{78},
  \bibinfo{pages}{1826--1839}.
\newblock \DOIprefix\doi{10.1093/icesjms/fsab085}.
\bibitem[{Moraga et~al.(2021)Moraga, Dean, Inoue, Morawiecki, Noureen and
  Wang}]{bg:6}
\bibinfo{author}{Moraga, P.}, \bibinfo{author}{Dean, C.},
  \bibinfo{author}{Inoue, J.}, \bibinfo{author}{Morawiecki, P.},
  \bibinfo{author}{Noureen, S.R.}, \bibinfo{author}{Wang, F.},
  \bibinfo{year}{2021}.
\newblock \bibinfo{title}{Bayesian spatial modelling of geostatistical data
  using {INLA} and {SPDE} methods: A case study predicting malaria risk in
  {Mozambique}}.
\newblock \bibinfo{journal}{Spat. Spatio-temporal Epidemiol.}
  \bibinfo{volume}{39}, \bibinfo{pages}{100440}.
\newblock \DOIprefix\doi{10.1016/j.sste.2021.100440}.
\bibitem[{Morales and Laurini(2021)}]{bg:24}
\bibinfo{author}{Morales, A.B.}, \bibinfo{author}{Laurini, M.P.},
  \bibinfo{year}{2021}.
\newblock \bibinfo{title}{Firm location: A spatial point process approach}.
\newblock \bibinfo{journal}{Appl. Spat. Anal. Policy}
  \DOIprefix\doi{10.1007/s12061-021-09419-x}.
\bibitem[{Moses et~al.(2021)Moses, Aheto and Dagne}]{bg:4}
\bibinfo{author}{Moses, J.}, \bibinfo{author}{Aheto, K.},
  \bibinfo{author}{Dagne, G.A.}, \bibinfo{year}{2021}.
\newblock \bibinfo{title}{Geostatistical analysis, web-based mapping, and
  environmental determinants of under-5 stunting: evidence from the 2014
  {Ghana} {Demographic} and {Health} {Survey}}.
\newblock \bibinfo{journal}{The Lancet Planetary Health} \bibinfo{volume}{5},
  \bibinfo{pages}{e347--e355}.
\newblock \DOIprefix\doi{10.1016/S2542-5196(21)00080-2}.
\bibitem[{Novomestky(2013)}]{novomestky_orthopolynom_2013}
\bibinfo{author}{Novomestky, F.}, \bibinfo{year}{2013}.
\newblock \bibinfo{title}{orthopolynom: Collection of functions for orthogonal
  and orthonormal polynomials}.
\newblock \URLprefix \url{https://CRAN.R-project.org/package=orthopolynom}.
  \bibinfo{note}{{R} package version 1.0-5}.
\bibitem[{Nychka et~al.(2016)Nychka, Hammerling, Sain and
  Lenssen}]{nychka_latticekrig_2016}
\bibinfo{author}{Nychka, D.}, \bibinfo{author}{Hammerling, D.},
  \bibinfo{author}{Sain, S.}, \bibinfo{author}{Lenssen, N.},
  \bibinfo{year}{2016}.
\newblock \bibinfo{title}{{LatticeKrig}: Multiresolution {Kriging} based on
  {Markov} random fields}.
\newblock \DOIprefix\doi{10.5065/D6HD7T1R}. \bibinfo{note}{{R} package version
  8.4}.
\bibitem[{Peruzzi et~al.(2020)Peruzzi, Banerjee and
  Finley}]{peruzzi_highly_2020}
\bibinfo{author}{Peruzzi, M.}, \bibinfo{author}{Banerjee, S.},
  \bibinfo{author}{Finley, A.O.}, \bibinfo{year}{2020}.
\newblock \bibinfo{title}{Highly {Scalable} {Bayesian} {Geostatistical}
  {Modeling} via {Meshed} {Gaussian} {Processes} on {Partitioned} {Domains}}.
\newblock \bibinfo{journal}{J. Amer. Statist. Assoc.} \bibinfo{volume}{0},
  \bibinfo{pages}{1--14}.
\newblock \DOIprefix\doi{10.1080/01621459.2020.1833889}.
\bibitem[{Porcu et~al.(2016)Porcu, Bevilacqua and
  Genton}]{porcu_spatio-temporal_2016}
\bibinfo{author}{Porcu, E.}, \bibinfo{author}{Bevilacqua, M.},
  \bibinfo{author}{Genton, M.G.}, \bibinfo{year}{2016}.
\newblock \bibinfo{title}{Spatio-{Temporal} {Covariance} and
  {Cross}-{Covariance} {Functions} of the {Great} {Circle} {Distance} on a
  {Sphere}}.
\newblock \bibinfo{journal}{J. Amer. Statist. Assoc.} \bibinfo{volume}{111},
  \bibinfo{pages}{888--898}.
\newblock \DOIprefix\doi{10.1080/01621459.2015.1072541}.
\bibitem[{Quiroz et~al.(2021)Quiroz, Prates, Dey and Rue}]{art681}
\bibinfo{author}{Quiroz, Z.C.}, \bibinfo{author}{Prates, M.O.},
  \bibinfo{author}{Dey, D.K.}, \bibinfo{author}{Rue, H.}, \bibinfo{year}{2021}.
\newblock \bibinfo{title}{Fast {B}ayesian inference of block nearest neighbor
  {G}aussian process for large data}.
\newblock \href{http://arxiv.org/abs/1908.06437}{{\tt arXiv:1908.06437}}.
\bibitem[{Rayner et~al.(2020)Rayner, Auchmann, Bessembinder, Brönnimann,
  Brugnara, Capponi, Carrea, Dodd, Ghent, Good, Høyer, Kennedy, Kent, Killick,
  Linden, Lindgren, Madsen, Merchant, Mitchelson, Morice, Nielsen-Englyst,
  Ortiz, Remedios, Schrier, Squintu, Stephens, Thorne, Tonboe, Trent, Veal,
  Waterfall, Winfield, Winn and Woolway}]{rayner_eustace_2020}
\bibinfo{author}{Rayner, N.A.}, \bibinfo{author}{Auchmann, R.},
  \bibinfo{author}{Bessembinder, J.}, \bibinfo{author}{Brönnimann, S.},
  \bibinfo{author}{Brugnara, Y.}, \bibinfo{author}{Capponi, F.},
  \bibinfo{author}{Carrea, L.}, \bibinfo{author}{Dodd, E.M.A.},
  \bibinfo{author}{Ghent, D.}, \bibinfo{author}{Good, E.},
  \bibinfo{author}{Høyer, J.L.}, \bibinfo{author}{Kennedy, J.J.},
  \bibinfo{author}{Kent, E.C.}, \bibinfo{author}{Killick, R.E.},
  \bibinfo{author}{Linden, P.v.d.}, \bibinfo{author}{Lindgren, F.},
  \bibinfo{author}{Madsen, K.S.}, \bibinfo{author}{Merchant, C.J.},
  \bibinfo{author}{Mitchelson, J.R.}, \bibinfo{author}{Morice, C.P.},
  \bibinfo{author}{Nielsen-Englyst, P.}, \bibinfo{author}{Ortiz, P.F.},
  \bibinfo{author}{Remedios, J.J.}, \bibinfo{author}{Schrier, G.v.d.},
  \bibinfo{author}{Squintu, A.A.}, \bibinfo{author}{Stephens, A.},
  \bibinfo{author}{Thorne, P.W.}, \bibinfo{author}{Tonboe, R.T.},
  \bibinfo{author}{Trent, T.}, \bibinfo{author}{Veal, K.L.},
  \bibinfo{author}{Waterfall, A.M.}, \bibinfo{author}{Winfield, K.},
  \bibinfo{author}{Winn, J.}, \bibinfo{author}{Woolway, R.I.},
  \bibinfo{year}{2020}.
\newblock \bibinfo{title}{The {EUSTACE} {Project}: {Delivering} {Global},
  {Daily} {Information} on {Surface} {Air} {Temperature}}.
\newblock \bibinfo{journal}{Bull. Am. Meteorol. Soc.} \bibinfo{volume}{101},
  \bibinfo{pages}{E1924--E1947}.
\newblock \DOIprefix\doi{10.1175/BAMS-D-19-0095.1}.
\bibitem[{Roininen et~al.(2019)Roininen, Girolami, Lasanen and
  Markkanen}]{Roininen2019}
\bibinfo{author}{Roininen, L.}, \bibinfo{author}{Girolami, M.},
  \bibinfo{author}{Lasanen, S.}, \bibinfo{author}{Markkanen, M.},
  \bibinfo{year}{2019}.
\newblock \bibinfo{title}{Hyperpriors for {M}at\'{e}rn fields with applications
  in {B}ayesian inversion}.
\newblock \bibinfo{journal}{Inverse Probl. Imaging} \bibinfo{volume}{13},
  \bibinfo{pages}{1--29}.
\newblock \DOIprefix\doi{10.3934/ipi.2019001}.
\bibitem[{Roininen et~al.(2014)Roininen, Huttunen and Lasanen}]{Roininen2014}
\bibinfo{author}{Roininen, L.}, \bibinfo{author}{Huttunen, J.M.J.},
  \bibinfo{author}{Lasanen, S.}, \bibinfo{year}{2014}.
\newblock \bibinfo{title}{Whittle-{M}at\'{e}rn priors for {B}ayesian
  statistical inversion with applications in electrical impedance tomography}.
\newblock \bibinfo{journal}{Inverse Probl. Imaging} \bibinfo{volume}{8},
  \bibinfo{pages}{561--586}.
\newblock \DOIprefix\doi{10.3934/ipi.2014.8.561}.
\bibitem[{Roininen et~al.(2018)Roininen, Lasanen, Orispää and
  Särkkä}]{roininen_sparse_2018}
\bibinfo{author}{Roininen, L.}, \bibinfo{author}{Lasanen, S.},
  \bibinfo{author}{Orispää, M.}, \bibinfo{author}{Särkkä, S.},
  \bibinfo{year}{2018}.
\newblock \bibinfo{title}{Sparse {Approximations} of {Fractional} {Matérn}
  {Fields}}.
\newblock \bibinfo{journal}{Scand. J. Stat.} \bibinfo{volume}{45},
  \bibinfo{pages}{194--216}.
\newblock \DOIprefix\doi{10.1111/sjos.12297}.
\bibitem[{Roksv{\aa}g et~al.(2021)Roksv{\aa}g, Steinsland and Engeland}]{bg:13}
\bibinfo{author}{Roksv{\aa}g, T.}, \bibinfo{author}{Steinsland, I.},
  \bibinfo{author}{Engeland, K.}, \bibinfo{year}{2021}.
\newblock \bibinfo{title}{A two-field geostatistical model combining point and
  areal observations--{A} case study of annual runoff predictions in the {Voss}
  area}.
\newblock \bibinfo{journal}{J. R. Stat. Soc. Ser. C. Appl. Stat.}
  \bibinfo{volume}{70}, \bibinfo{pages}{934--960}.
\newblock \DOIprefix\doi{10.1111/rssc.12492}.
\bibitem[{{Roksv{\r{a}}g} et~al.(2021){Roksv{\r{a}}g}, {Steinsland} and
  {Engeland}}]{bg:12}
\bibinfo{author}{{Roksv{\r{a}}g}, T.}, \bibinfo{author}{{Steinsland}, I.},
  \bibinfo{author}{{Engeland}, K.}, \bibinfo{year}{2021}.
\newblock \bibinfo{title}{Estimating mean annual runoff by using a
  geostatistical spatially varying coefficient model that incorporates
  process-based simulations and short records}, in: \bibinfo{booktitle}{EGU
  General Assembly Conference Abstracts}, pp. \bibinfo{pages}{EGU21--4233}.
\bibitem[{Rozanov(1977)}]{rozanov1977markov}
\bibinfo{author}{Rozanov, J.A.}, \bibinfo{year}{1977}.
\newblock \bibinfo{title}{{Markov} random fields and stochastic partial
  differential equations}.
\newblock \bibinfo{journal}{Sbornik: Mathematics} \bibinfo{volume}{32},
  \bibinfo{pages}{515--534}.
\bibitem[{Rue and Held(2005)}]{rue2005gaussian}
\bibinfo{author}{Rue, H.}, \bibinfo{author}{Held, L.}, \bibinfo{year}{2005}.
\newblock \bibinfo{title}{Gaussian {M}arkov Random Fields: {T}heory and
  Applications}. volume \bibinfo{volume}{104} of
  \textit{\bibinfo{series}{Monographs on Statistics and Applied Probability}}.
\newblock \bibinfo{publisher}{Chapman \& Hall}, \bibinfo{address}{London}.
\newblock \DOIprefix\doi{10.1201/9780203492024}.
\bibitem[{Rue and Held(2010)}]{col27}
\bibinfo{author}{Rue, H.}, \bibinfo{author}{Held, L.}, \bibinfo{year}{2010}.
\newblock \bibinfo{title}{Markov random fields}, in: \bibinfo{editor}{Gelfand,
  A.}, \bibinfo{editor}{Diggle, P.}, \bibinfo{editor}{Fuentes, M.},
  \bibinfo{editor}{Guttorp, P.} (Eds.), \bibinfo{booktitle}{Handbook of Spatial
  Statistics}. \bibinfo{publisher}{CRC/Chapman \& Hall}, \bibinfo{address}{Boca
  Raton, FL}, pp. \bibinfo{pages}{171--200}.
\bibitem[{Rue and Martino(2007)}]{art375}
\bibinfo{author}{Rue, H.}, \bibinfo{author}{Martino, S.}, \bibinfo{year}{2007}.
\newblock \bibinfo{title}{Approximate {B}ayesian inference for hierarchical
  {G}aussian {M}arkov random fields models}.
\newblock \bibinfo{journal}{J. Stat. Plan. Inference} \bibinfo{volume}{137},
  \bibinfo{pages}{3177--3192}.
\newblock \bibinfo{note}{Special Issue: {B}ayesian Inference for Stochastic
  Processes}.
\bibitem[{Rue et~al.(2009)Rue, Martino and Chopin}]{art451}
\bibinfo{author}{Rue, H.}, \bibinfo{author}{Martino, S.},
  \bibinfo{author}{Chopin, N.}, \bibinfo{year}{2009}.
\newblock \bibinfo{title}{Approximate {B}ayesian inference for latent
  {G}aussian models using integrated nested {L}aplace approximations}.
\newblock \bibinfo{journal}{J.\ R.\ Stat.\ Soc.\ Ser.\ B Stat.\ Methodol.}
  \bibinfo{volume}{71}, \bibinfo{pages}{319--392}.
\newblock \bibinfo{note}{With discussion and a reply by the authors}.
\bibitem[{Rue et~al.(2017)Rue, Riebler, S{\o}rbye, Illian, Simpson and
  Lindgren}]{art632}
\bibinfo{author}{Rue, H.}, \bibinfo{author}{Riebler, A.},
  \bibinfo{author}{S{\o}rbye, S.H.}, \bibinfo{author}{Illian, J.B.},
  \bibinfo{author}{Simpson, D.P.}, \bibinfo{author}{Lindgren, F.K.},
  \bibinfo{year}{2017}.
\newblock \bibinfo{title}{Bayesian computing with {INLA}: {A} review}.
\newblock \bibinfo{journal}{Annu. Rev. Stat. Appl.} \bibinfo{volume}{4},
  \bibinfo{pages}{395--421}.
\newblock \DOIprefix\doi{10.1146/annurev-statistics-060116-054045}.
\bibitem[{Sampson and Guttorp(1992)}]{sampson_nonparametric_1992}
\bibinfo{author}{Sampson, P.D.}, \bibinfo{author}{Guttorp, P.},
  \bibinfo{year}{1992}.
\newblock \bibinfo{title}{Nonparametric {Estimation} of {Nonstationary}
  {Spatial} {Covariance} {Structure}}.
\newblock \bibinfo{journal}{J. Amer. Statist. Assoc.} \bibinfo{volume}{87},
  \bibinfo{pages}{108--119}.
\newblock \DOIprefix\doi{10.2307/2290458}.
\bibitem[{Sangalli(2021)}]{sangalli_spatial_2021}
\bibinfo{author}{Sangalli, L.M.}, \bibinfo{year}{2021}.
\newblock \bibinfo{title}{Spatial regression with partial differential equation
  regularisation}.
\newblock \bibinfo{journal}{Int. Stat. Rev.}
  \DOIprefix\doi{10.1111/insr.12444}.
\bibitem[{Sangalli et~al.(2013)Sangalli, Ramsay and
  Ramsay}]{sangalli_spatial_2013}
\bibinfo{author}{Sangalli, L.M.}, \bibinfo{author}{Ramsay, J.O.},
  \bibinfo{author}{Ramsay, T.O.}, \bibinfo{year}{2013}.
\newblock \bibinfo{title}{Spatial spline regression models}.
\newblock \bibinfo{journal}{J. R. Stat. Soc. Ser. B. Stat. Methodol.}
  \bibinfo{volume}{75}, \bibinfo{pages}{681--703}.
\newblock \DOIprefix\doi{10.1111/rssb.12009}.
\bibitem[{Sanz-Alonso and Yang(2021a)}]{bg:10}
\bibinfo{author}{Sanz-Alonso, D.}, \bibinfo{author}{Yang, R.},
  \bibinfo{year}{2021}a.
\newblock \bibinfo{title}{Finite element representations of {Gaussian}
  processes: Balancing numerical and statistical accuracy}.
\newblock \href{http://arxiv.org/abs/2109.02777}{{\tt arXiv:2109.02777}}.
\bibitem[{Sanz-Alonso and Yang(2021b)}]{Alonso2021}
\bibinfo{author}{Sanz-Alonso, D.}, \bibinfo{author}{Yang, R.},
  \bibinfo{year}{2021}b.
\newblock \bibinfo{title}{The {SPDE} approach to {Matérn} fields: Graph
  representations}.
\newblock \bibinfo{journal}{Stat.\ Sci.} ,
  \bibinfo{pages}{1--53}\bibinfo{note}{Accepted}.
\bibitem[{Schoenberg(1942)}]{schoenberg_positive_1942}
\bibinfo{author}{Schoenberg, I.J.}, \bibinfo{year}{1942}.
\newblock \bibinfo{title}{Positive definite functions on spheres}.
\newblock \bibinfo{journal}{Duke Math. J.} \bibinfo{volume}{9},
  \bibinfo{pages}{96--108}.
\newblock \DOIprefix\doi{10.1215/S0012-7094-42-00908-6}.
\bibitem[{Scott(2021)}]{bg:3}
\bibinfo{author}{Scott, R.P.}, \bibinfo{year}{2021}.
\newblock \bibinfo{title}{Shared streets, park closures and environmental
  justice during a pandemic emergency in {D}enver, {C}olorado}.
\newblock \bibinfo{journal}{J. Trans. Health} \bibinfo{volume}{21},
  \bibinfo{pages}{101075}.
\newblock \DOIprefix\doi{10.1016/j.jth.2021.101075}.
\bibitem[{{Sicacha-Parada} et~al.(2021){Sicacha-Parada}, {Pavon-Jordan},
  Steinsland, May, Stokke and {\O}ien}]{bg:28}
\bibinfo{author}{{Sicacha-Parada}, J.}, \bibinfo{author}{{Pavon-Jordan}, D.},
  \bibinfo{author}{Steinsland, I.}, \bibinfo{author}{May, R.},
  \bibinfo{author}{Stokke, B.}, \bibinfo{author}{{\O}ien, I.J.},
  \bibinfo{year}{2021}.
\newblock \bibinfo{title}{A spatial modeling framework for monitoring surveys
  with different sampling protocols with a case study for bird populations in
  mid-{S}candinavia}.
\newblock \href{http://arxiv.org/abs/2104.05751}{{\tt arXiv:2104.05751}}.
\bibitem[{Sidén et~al.(2021)Sidén, Lindgren, Bolin, Eklund and
  Villani}]{Siden2021}
\bibinfo{author}{Sidén, P.}, \bibinfo{author}{Lindgren, F.},
  \bibinfo{author}{Bolin, D.}, \bibinfo{author}{Eklund, A.},
  \bibinfo{author}{Villani, M.}, \bibinfo{year}{2021}.
\newblock \bibinfo{title}{{Spatial} {3D} {Matérn} priors for fast whole-brain
  {fMRI} analysis}.
\newblock \bibinfo{journal}{Bayesian Anal.} ,
  \bibinfo{pages}{1--28}\DOIprefix\doi{10.1214/21-BA1283}.
\bibitem[{Simpson et~al.(2016)Simpson, Illian, Lindgren, S{\o}rbye and
  Rue}]{Simpson2016}
\bibinfo{author}{Simpson, D.}, \bibinfo{author}{Illian, J.B.},
  \bibinfo{author}{Lindgren, F.}, \bibinfo{author}{S{\o}rbye, S.H.},
  \bibinfo{author}{Rue, H.}, \bibinfo{year}{2016}.
\newblock \bibinfo{title}{Going off grid: computationally efficient inference
  for log-{G}aussian {C}ox processes}.
\newblock \bibinfo{journal}{Biometrika} \bibinfo{volume}{103},
  \bibinfo{pages}{49--70}.
\newblock \DOIprefix\doi{10.1093/biomet/asv064}.
\bibitem[{Simpson et~al.(2012a)Simpson, Lindgren and Rue}]{simpson_order_2012}
\bibinfo{author}{Simpson, D.}, \bibinfo{author}{Lindgren, F.},
  \bibinfo{author}{Rue, H.}, \bibinfo{year}{2012}a.
\newblock \bibinfo{title}{In order to make spatial statistics computationally
  feasible, we need to forget about the covariance function}.
\newblock \bibinfo{journal}{Environmetrics} \bibinfo{volume}{23},
  \bibinfo{pages}{65--74}.
\newblock \DOIprefix\doi{10.1002/env.1137}.
\bibitem[{Simpson et~al.(2012b)Simpson, Lindgren and Rue}]{art512}
\bibinfo{author}{Simpson, D.}, \bibinfo{author}{Lindgren, F.},
  \bibinfo{author}{Rue, H.}, \bibinfo{year}{2012}b.
\newblock \bibinfo{title}{Think continuous: {Markovian} {Gaussian} models in
  spatial statistics}.
\newblock \bibinfo{journal}{Spat.\ Stat.} \bibinfo{volume}{1},
  \bibinfo{pages}{16--29}.
\newblock \DOIprefix\doi{10.1016/j.spasta.2012.02.003}.
\bibitem[{Simpson et~al.(2017)Simpson, Rue, Riebler, Martins and
  S{\o}rbye}]{art631}
\bibinfo{author}{Simpson, D.P.}, \bibinfo{author}{Rue, H.},
  \bibinfo{author}{Riebler, A.}, \bibinfo{author}{Martins, T.G.},
  \bibinfo{author}{S{\o}rbye, S.H.}, \bibinfo{year}{2017}.
\newblock \bibinfo{title}{Penalising model component complexity: A principled,
  practical approach to constructing priors (with discussion)}.
\newblock \bibinfo{journal}{Stat.\ Sci.} \bibinfo{volume}{32},
  \bibinfo{pages}{1--28}.
\bibitem[{Solin and Särkkä(2020)}]{solin_hilbert_2020}
\bibinfo{author}{Solin, A.}, \bibinfo{author}{Särkkä, S.},
  \bibinfo{year}{2020}.
\newblock \bibinfo{title}{Hilbert space methods for reduced-rank {Gaussian}
  process regression}.
\newblock \bibinfo{journal}{Stat. Comput.} \bibinfo{volume}{30},
  \bibinfo{pages}{419--446}.
\newblock \DOIprefix\doi{10.1007/s11222-019-09886-w}.
\bibitem[{Spencer et~al.(2021)Spencer, Yue, Bolin, Ryan and
  Mejia}]{spencer2021spatial}
\bibinfo{author}{Spencer, D.}, \bibinfo{author}{Yue, Y.R.},
  \bibinfo{author}{Bolin, D.}, \bibinfo{author}{Ryan, S.},
  \bibinfo{author}{Mejia, A.F.}, \bibinfo{year}{2021}.
\newblock \bibinfo{title}{Spatial {Bayesian} {GLM} on the cortical surface
  produces reliable task activations in individuals and groups}.
\newblock \href{http://arxiv.org/abs/2106.06669}{{\tt arXiv:2106.06669}}.
\bibitem[{Stein(1999)}]{stein99}
\bibinfo{author}{Stein, M.L.}, \bibinfo{year}{1999}.
\newblock \bibinfo{title}{Interpolation of {S}patial {D}ata: {S}ome {T}heory
  for {K}riging}.
\newblock Springer Series in Statistics, \bibinfo{publisher}{Springer-Verlag,
  New York}.
\bibitem[{Stein(2005)}]{stein2005space}
\bibinfo{author}{Stein, M.L.}, \bibinfo{year}{2005}.
\newblock \bibinfo{title}{Space--time covariance functions}.
\newblock \bibinfo{journal}{J. Amer. Statist. Assoc.} \bibinfo{volume}{100},
  \bibinfo{pages}{310--321}.
\bibitem[{Särkkä et~al.(2013)Särkkä, Solin and
  Hartikainen}]{sarkka_spatiotemporal_2013}
\bibinfo{author}{Särkkä, S.}, \bibinfo{author}{Solin, A.},
  \bibinfo{author}{Hartikainen, J.}, \bibinfo{year}{2013}.
\newblock \bibinfo{title}{Spatiotemporal {Learning} via
  {Infinite}-{Dimensional} {Bayesian} {Filtering} and {Smoothing}: {A} {Look}
  at {Gaussian} {Process} {Regression} {Through} {Kalman} {Filtering}}.
\newblock \bibinfo{journal}{IEEE Signal Process. Mag.} \bibinfo{volume}{30},
  \bibinfo{pages}{51--61}.
\newblock \DOIprefix\doi{10.1109/MSP.2013.2246292}.
\bibitem[{Sørbye et~al.(2019)Sørbye, Myrvoll-Nilsen and
  Rue}]{sorbye_approximate_2019}
\bibinfo{author}{Sørbye, S.H.}, \bibinfo{author}{Myrvoll-Nilsen, E.},
  \bibinfo{author}{Rue, H.}, \bibinfo{year}{2019}.
\newblock \bibinfo{title}{An approximate fractional {Gaussian} noise model with
  {O}(n) computational cost}.
\newblock \bibinfo{journal}{Stat. Comput.} \bibinfo{volume}{29},
  \bibinfo{pages}{821--833}.
\newblock \DOIprefix\doi{10.1007/s11222-018-9843-1}.
\bibitem[{Takahashi et~al.(1973)Takahashi, Fagan and Chen}]{pro20}
\bibinfo{author}{Takahashi, K.}, \bibinfo{author}{Fagan, J.},
  \bibinfo{author}{Chen, M.S.}, \bibinfo{year}{1973}.
\newblock \bibinfo{title}{Formation of a sparse bus impedance matrix and its
  application to short circuit study}, in: \bibinfo{booktitle}{8th PICA
  Conference proceedings}, \bibinfo{organization}{IEEE Power Engineering
  Society}. pp. \bibinfo{pages}{63--69}.
\newblock \bibinfo{note}{Papers presented at the $1973$ Power Industry Computer
  Application Conference in Minneapolis, Minnesota}.
\bibitem[{Taylor et~al.(2021)Taylor, Blanchard and Lindenmayer}]{bg:36}
\bibinfo{author}{Taylor, C.}, \bibinfo{author}{Blanchard, W.},
  \bibinfo{author}{Lindenmayer, D.B.}, \bibinfo{year}{2021}.
\newblock \bibinfo{title}{What are the associations between thinning and fire
  severity?}
\newblock \bibinfo{journal}{Austral Ecology} \bibinfo{volume}{46},
  \bibinfo{pages}{145--1439}.
\newblock \DOIprefix\doi{10.1111/aec.13096}.
\bibitem[{Thorson et~al.(2021)Thorson, Barbeaux, Goethel, Kearney, Laman,
  Nielsen, Siskey, Siwicke and Thompson}]{bg:46}
\bibinfo{author}{Thorson, J.T.}, \bibinfo{author}{Barbeaux, S.J.},
  \bibinfo{author}{Goethel, D.R.}, \bibinfo{author}{Kearney, K.A.},
  \bibinfo{author}{Laman, E.A.}, \bibinfo{author}{Nielsen, J.K.},
  \bibinfo{author}{Siskey, M.R.}, \bibinfo{author}{Siwicke, K.},
  \bibinfo{author}{Thompson, G.G.}, \bibinfo{year}{2021}.
\newblock \bibinfo{title}{Estimating fine-scale movement rates and habitat
  preferences using multiple data sources}.
\newblock \bibinfo{journal}{Fish and Fisheries} \bibinfo{volume}{22},
  \bibinfo{pages}{1359--1376}.
\newblock \DOIprefix\doi{10.1111/faf.12592}.
\bibitem[{Valente and Laurini(2021a)}]{bg:22}
\bibinfo{author}{Valente, F.}, \bibinfo{author}{Laurini, M.},
  \bibinfo{year}{2021}a.
\newblock \bibinfo{title}{Pre-harvest sugarcane burning: {A} statistical
  analysis of the environmental impacts of a regulatory change in the energy
  sector}.
\newblock \bibinfo{journal}{Cleaner Engineering and Technology}
  \bibinfo{volume}{4}, \bibinfo{pages}{100255}.
\newblock \DOIprefix\doi{10.1016/j.clet.2021.100255}.
\bibitem[{Valente and Laurini(2021b)}]{bg:19}
\bibinfo{author}{Valente, F.}, \bibinfo{author}{Laurini, M.},
  \bibinfo{year}{2021}b.
\newblock \bibinfo{title}{Spatio-temporal analysis of fire occurrence in
  {Australia}}.
\newblock \bibinfo{journal}{Stochastic Environmental Research and Risk
  Assessment} \bibinfo{volume}{35}, \bibinfo{pages}{1759--1770}.
\newblock \DOIprefix\doi{10.1007/s00477-021-02043-8}.
\bibitem[{{van Niekerk} et~al.(2021){van Niekerk}, Bakka, Rue and
  Schenk}]{art664}
\bibinfo{author}{{van Niekerk}, J.}, \bibinfo{author}{Bakka, H.},
  \bibinfo{author}{Rue, H.}, \bibinfo{author}{Schenk, O.},
  \bibinfo{year}{2021}.
\newblock \bibinfo{title}{New frontiers in {B}ayesian modeling using the {INLA}
  package in {R}}.
\newblock \bibinfo{journal}{J. Stat. Softw.} \bibinfo{volume}{100},
  \bibinfo{pages}{1--28}.
\newblock \DOIprefix\doi{10.18637/jss.v100.i02}.
\bibitem[{{van Woesik} and Cacciapaglia(2021)}]{bg:39}
\bibinfo{author}{{van Woesik}, R.}, \bibinfo{author}{Cacciapaglia, C.W.},
  \bibinfo{year}{2021}.
\newblock \bibinfo{title}{Thermal stress jeopardizes carbonate production of
  coral reefs across the western and central {P}acific {O}cean}.
\newblock \bibinfo{journal}{PLOS ONE} \bibinfo{volume}{16},
  \bibinfo{pages}{1--18}.
\newblock \DOIprefix\doi{10.1371/journal.pone.0249008}.
\bibitem[{Vandeskog et~al.(2021a)Vandeskog, Martino, {Castro-Camilo} and
  Rue}]{bg:15}
\bibinfo{author}{Vandeskog, S.M.}, \bibinfo{author}{Martino, S.},
  \bibinfo{author}{{Castro-Camilo}, D.}, \bibinfo{author}{Rue, H.},
  \bibinfo{year}{2021}a.
\newblock \bibinfo{title}{Modelling short-term precipitation extremes with the
  blended generalised extreme value distribution}.
\newblock \href{http://arxiv.org/abs/2105.09062}{{\tt arXiv:2105.09062}}.
\bibitem[{Vandeskog et~al.(2021b)Vandeskog, Thorarinsdottir, Steinsland and
  Lindgren}]{vandeskog_quantile_2021}
\bibinfo{author}{Vandeskog, S.M.}, \bibinfo{author}{Thorarinsdottir, T.L.},
  \bibinfo{author}{Steinsland, I.}, \bibinfo{author}{Lindgren, F.},
  \bibinfo{year}{2021}b.
\newblock \bibinfo{title}{Quantile based modelling of diurnal temperature range
  with the five-parameter lambda distribution}.
\newblock \href{http://arxiv.org/abs/2109.11180}{{\tt arXiv:2109.11180}}.
\bibitem[{Vehtari et~al.(2017)Vehtari, Gelman and
  Gabry}]{vehtari_practical_2017}
\bibinfo{author}{Vehtari, A.}, \bibinfo{author}{Gelman, A.},
  \bibinfo{author}{Gabry, J.}, \bibinfo{year}{2017}.
\newblock \bibinfo{title}{Practical {Bayesian} model evaluation using
  leave-one-out cross-validation and {WAIC}}.
\newblock \bibinfo{journal}{Stat. Comput.} \bibinfo{volume}{27},
  \bibinfo{pages}{1413--1432}.
\newblock \DOIprefix\doi{10.1007/s11222-016-9696-4}.
\bibitem[{Vogel et~al.(2021)Vogel, Biuw, Blanchet, Jonsen, Mul, Johnsen,
  Hj{\o}llo, Olsen, Dietz and Rikardsen}]{EF2021}
\bibinfo{author}{Vogel, E.F.}, \bibinfo{author}{Biuw, M.},
  \bibinfo{author}{Blanchet, M.A.}, \bibinfo{author}{Jonsen, I.D.},
  \bibinfo{author}{Mul, E.}, \bibinfo{author}{Johnsen, E.},
  \bibinfo{author}{Hj{\o}llo, S.S.}, \bibinfo{author}{Olsen, M.T.},
  \bibinfo{author}{Dietz, R.}, \bibinfo{author}{Rikardsen, A.},
  \bibinfo{year}{2021}.
\newblock \bibinfo{title}{Killer whale movements on the {N}orwegian shelf are
  associated with herring density}.
\newblock \bibinfo{journal}{Marine Ecology Progress Series}
  \bibinfo{volume}{665}, \bibinfo{pages}{217--231}.
\newblock \DOIprefix\doi{10.3354/meps13685}.
\bibitem[{Wahba(1981)}]{wahba_spline_1981}
\bibinfo{author}{Wahba, G.}, \bibinfo{year}{1981}.
\newblock \bibinfo{title}{Spline {Interpolation} and {Smoothing} on the
  {Sphere}}.
\newblock \bibinfo{journal}{SIAM J. Sci. Statist. Comput.} \bibinfo{volume}{2},
  \bibinfo{pages}{5--16}.
\newblock \DOIprefix\doi{10.1137/0902002}.
\bibitem[{Walder and Hanks(2020)}]{Walder2020}
\bibinfo{author}{Walder, A.}, \bibinfo{author}{Hanks, E.M.},
  \bibinfo{year}{2020}.
\newblock \bibinfo{title}{Bayesian analysis of spatial generalized linear mixed
  models with {L}aplace moving average random fields}.
\newblock \bibinfo{journal}{Comput. Statist. Data Anal.} \bibinfo{volume}{144},
  \bibinfo{pages}{106861, 13}.
\newblock \DOIprefix\doi{10.1016/j.csda.2019.106861}.
\bibitem[{Wallin and Bolin(2015)}]{Wallin15}
\bibinfo{author}{Wallin, J.}, \bibinfo{author}{Bolin, D.},
  \bibinfo{year}{2015}.
\newblock \bibinfo{title}{Geostatistical modelling using non-{G}aussian
  {M}at\'{e}rn fields}.
\newblock \bibinfo{journal}{Scand. J. Stat.} \bibinfo{volume}{42},
  \bibinfo{pages}{872--890}.
\newblock \DOIprefix\doi{10.1111/sjos.12141}.
\bibitem[{Wang and Zuo(2021)}]{bg:16}
\bibinfo{author}{Wang, J.}, \bibinfo{author}{Zuo, R.}, \bibinfo{year}{2021}.
\newblock \bibinfo{title}{Spatial modelling of hydrothermal
  mineralization-related geochemical patterns using {INLA}+{SPDE} and local
  singularity analysis}.
\newblock \bibinfo{journal}{Comput. Geosci.} \bibinfo{volume}{154},
  \bibinfo{pages}{104822}.
\newblock \DOIprefix\doi{10.1016/j.cageo.2021.104822}.
\bibitem[{Whittle(1954)}]{whittle54}
\bibinfo{author}{Whittle, P.}, \bibinfo{year}{1954}.
\newblock \bibinfo{title}{On stationary processes in the plane}.
\newblock \bibinfo{journal}{Biometrika} \bibinfo{volume}{41},
  \bibinfo{pages}{434--449}.
\newblock \DOIprefix\doi{10.1093/biomet/41.3-4.434}.
\bibitem[{Whittle(1963)}]{whittle63}
\bibinfo{author}{Whittle, P.}, \bibinfo{year}{1963}.
\newblock \bibinfo{title}{Stochastic processes in several dimensions}.
\newblock \bibinfo{journal}{Bull.\ Internat.\ Statist.\ Inst.}
  \bibinfo{volume}{40}, \bibinfo{pages}{974--994}.
\bibitem[{Wikle et~al.(2019)Wikle, Zammit-Mangion and
  Cressie}]{wikle_spatio-temporal-with-R_2019}
\bibinfo{author}{Wikle, C.K.}, \bibinfo{author}{Zammit-Mangion, A.},
  \bibinfo{author}{Cressie, N.}, \bibinfo{year}{2019}.
\newblock \bibinfo{title}{Spatio-Temporal Statistics with R}.
\newblock \bibinfo{publisher}{Chapman \& Hall/CRC, Boca Raton, FL}.
\bibitem[{Williamson et~al.(2021)Williamson, Scott, Laxton, Bachl, Illian,
  Brookes and Thompson}]{bg:29}
\bibinfo{author}{Williamson, L.D.}, \bibinfo{author}{Scott, B.E.},
  \bibinfo{author}{Laxton, M.R.}, \bibinfo{author}{Bachl, F.E.},
  \bibinfo{author}{Illian, J.B.}, \bibinfo{author}{Brookes, K.L.},
  \bibinfo{author}{Thompson, P.M.}, \bibinfo{year}{2021}.
\newblock \bibinfo{title}{Spatiotemporal variation in harbor porpoise
  distribution and foraging across a landscape of fear}.
\newblock \bibinfo{journal}{Marine Mammal Science} ,
  \bibinfo{pages}{1--16}\DOIprefix\doi{10.1111/mms.12839}.
\bibitem[{Wright et~al.(2021)Wright, Newell, Lam, Kurmi, Chen and
  Kartsonaki}]{bg:17}
\bibinfo{author}{Wright, N.}, \bibinfo{author}{Newell, K.},
  \bibinfo{author}{Lam, K.B.H.}, \bibinfo{author}{Kurmi, O.},
  \bibinfo{author}{Chen, Z.}, \bibinfo{author}{Kartsonaki, C.},
  \bibinfo{year}{2021}.
\newblock \bibinfo{title}{Estimating ambient air pollutant levels in {Suzhou}
  through the {SPDE} approach with {R-INLA}}.
\newblock \bibinfo{journal}{Int. J. Hyg. Environ. Health}
  \bibinfo{volume}{235}, \bibinfo{pages}{113766}.
\newblock \DOIprefix\doi{10.1016/j.ijheh.2021.113766}.
\bibitem[{Xi et~al.(2021)Xi, Wu, Qian, Liu and Wang}]{bg:32}
\bibinfo{author}{Xi, C.}, \bibinfo{author}{Wu, Z.}, \bibinfo{author}{Qian, T.},
  \bibinfo{author}{Liu, L.}, \bibinfo{author}{Wang, J.}, \bibinfo{year}{2021}.
\newblock \bibinfo{title}{A {B}ayesian model for estimating the effects of
  human disturbance on wildlife habitats based on nighttime light data and
  {INLA}-{SPDE}}.
\newblock \bibinfo{journal}{Appl. Spat. Anal. Policy}
  \DOIprefix\doi{10.1007/s12061-021-09402-6}.
\bibitem[{Yuan et~al.(2017)Yuan, Bachl, Lindgren, Borchers, Illian, Buckland,
  Rue and Gerrodette}]{yuan_point_2017}
\bibinfo{author}{Yuan, Y.}, \bibinfo{author}{Bachl, F.E.},
  \bibinfo{author}{Lindgren, F.}, \bibinfo{author}{Borchers, D.L.},
  \bibinfo{author}{Illian, J.B.}, \bibinfo{author}{Buckland, S.T.},
  \bibinfo{author}{Rue, H.}, \bibinfo{author}{Gerrodette, T.},
  \bibinfo{year}{2017}.
\newblock \bibinfo{title}{Point process models for spatio-temporal distance
  sampling data from a large-scale survey of blue whales}.
\newblock \bibinfo{journal}{Ann.\ Appl.\ Stat.} \bibinfo{volume}{11},
  \bibinfo{pages}{2270--2297}.
\newblock \DOIprefix\doi{10.1214/17-AOAS1078}.
\bibitem[{Yue et~al.(2014)Yue, Simpson, Lindgren and Rue}]{art532}
\bibinfo{author}{Yue, Y.R.}, \bibinfo{author}{Simpson, D.},
  \bibinfo{author}{Lindgren, F.}, \bibinfo{author}{Rue, H.},
  \bibinfo{year}{2014}.
\newblock \bibinfo{title}{Bayesian adaptive smoothing spline using stochastic
  differential equations}.
\newblock \bibinfo{journal}{Bayesian Anal.} \bibinfo{volume}{9},
  \bibinfo{pages}{397--424}.
\bibitem[{Zhang(2004)}]{Zhang2004}
\bibinfo{author}{Zhang, H.}, \bibinfo{year}{2004}.
\newblock \bibinfo{title}{Inconsistent estimation and asymptotically equal
  interpolations in model-based geostatistics}.
\newblock \bibinfo{journal}{J.\ Amer.\ Statist.\ Assoc.} \bibinfo{volume}{99},
  \bibinfo{pages}{250--261}.
\bibitem[{Zhang et~al.(2021)Zhang, Guilleminot and Gomez}]{bg:8}
\bibinfo{author}{Zhang, H.}, \bibinfo{author}{Guilleminot, J.},
  \bibinfo{author}{Gomez, L.J.}, \bibinfo{year}{2021}.
\newblock \bibinfo{title}{Stochastic modeling of geometrical uncertainties on
  complex domains, with application to additive manufacturing and brain
  interface geometries}.
\newblock \bibinfo{journal}{Comput. Methods Appl. Mech. Eng.}
  \bibinfo{volume}{385}, \bibinfo{pages}{114014}.
\newblock \DOIprefix\doi{10.1016/j.cma.2021.114014}.
\bibitem[{Zhang et~al.(2016)Zhang, Czado and Sigloch}]{zhang_bayesian_2016}
\bibinfo{author}{Zhang, R.}, \bibinfo{author}{Czado, C.},
  \bibinfo{author}{Sigloch, K.}, \bibinfo{year}{2016}.
\newblock \bibinfo{title}{Bayesian spatial modelling for high dimensional
  seismic inverse problems}.
\newblock \bibinfo{journal}{J. R. Stat. Soc. Ser. C. Appl. Stat.}
  \bibinfo{volume}{65}, \bibinfo{pages}{187--213}.
\newblock \URLprefix \url{https://www.jstor.org/stable/24772417}.

\end{thebibliography}

\appendix

\section{Reproducing kernel Hilbert space connection details}
\label{app:derivations}

As in Section~\ref{sec:precisions}, let $\adj{\cL}$ be the adjoint of
$\cL$, i.e.\ an operator such that
$\scal{\adj{\cL}f}{g}=\scal{f}{\cL g}$ on the manifold domain $\cD$,
and also assume that $\cL$ is invertible. Also let $\cW(\s)$ be a
standard Gaussian white noise process on $\cD$, and let $u(\s)$ be a
solution to $\cL u=\cW$ with covariance function $\cov(\s,\s')$. Then,
by construction,
\begin{align}
  \label{eq:deriv-R}
  \scal{f}{g}
  &=
    \cR_{\cW}(f,g)
    =
    \cR_{\cL u}(f,g)
    =
    \cR_{u}(\adj{\cL}f,\adj{\cL}g),
\end{align}
so that
\begin{align*}
  \scal{f}{g}
  &=
    \iint (\adj{\cL} f)(\s)\cov(\s,\s')(\adj{\cL} g)(\s') \md\s\md\s'
  \\
  &=
    \scal{\adj{\cL}_\s f(\s)}{\scal{\cov(\s,\cdot)}{\adj{\cL} g(\cdot)}}
  \\
  &=
    \iint f(\s)g(\s')\cL_\s\cL_{\s'}\cov(\s,\s') \md\s\md\s',
\end{align*}
where the subscript notation $\cL_{\s}^*$ and $\cL_{\s}$ is used to
indicate which spatial variable is operated on. Since the equality
holds for all admissible $f$ and $g$, this shows that
\begin{align*}
  \cL_\s\cL_{\s'}\cov(\s,\s') &= \delta_{\s}(\s'),
\end{align*}
where $\delta_\s(\cdot)$ is a Dirac delta. Define
$\cQ_u(f,g) = \scal{\cL f}{\cL g}$. Then, for all $\s\in\cD$ and all
suitable $g(\cdot)$,
\begin{align}
  \label{eq:deriv-Q}
  \cQ_u\{\cov(\s,\cdot),g(\cdot)\}
  &=
    \scal{\cL \cov(\s,\cdot)}{\cL g}
    =
    \scal{\cL_\s^{-1} \cL_\s \cL \cov(\s,\cdot)}{\cL g}
    =
    \scal{\cL_\s^{-1} \delta_{\s}(\cdot)}{\cL g}
  \\\nonumber
  &=
    \cL_\s^{-1} \scal{\delta_{\s}(\cdot)}{\cL g}
    =
    \cL_\s^{-1} (\cL_\s g)(\s)
    =
    g(\s) ,
\end{align}
which shows that this $\cQ_u(\cdot,\cdot)$ is the inner product for
the RKHS for the covariance $\cov(\s,\s')$ that is associated with the
covariance product $\cR_u(\cdot,\cdot)$ in \eqref{eq:deriv-R}. By
taking $g(\cdot)=\cov(\cdot,\s')$ we obtain the kernel reproducing
property $\cQ_u\{\cov(\s,\cdot),\cov(\cdot,\s')\}=\cov(\s,\s')$. The
derivation \eqref{eq:deriv-Q} directly shows that
$\scal{\adj{\cL}\cL\cov(\s,\cdot)}{g}=g(\s)$, which means that the
covariance function is a Green's function of the precision operator
$\cL^*\cL$. After a change of test functions in \eqref{eq:deriv-R}, we
can also write
$ \cR_{u}(f,g) = \scal{(\adj{\cL})^{-1} f}{(\adj{\cL})^{-1} g} $.


\end{document}